 \documentclass{cambridge6A}

\usepackage{hyperref}
  \usepackage{makeidx}
  \makeindex
    \usepackage{color,empheq}
    \usepackage{amsmath}
   \usepackage{amssymb}
       \usepackage{amsthm}
  \usepackage{natbib}
  \usepackage{rotating}
  \usepackage{floatpag}
  \rotfloatpagestyle{empty}

\usepackage{bm}
   \usepackage{graphicx}

  \theoremstyle{plain}

  \theoremstyle{definition}

  \theoremstyle{remark}

\usepackage{dharamvir}

\begin{document}
 
\allowdisplaybreaks

  		\title{{\textcolor{black}{\textsc{Mass dimension one fermions}}}
\\
   \textcolor{red}{-----------------------------------}\vspace{2pt}  }
 \vspace{101pt}
  	\author{{\sc\Large \textcolor{black}{Dharam Vir Ahluwalia}}  \\[1\baselineskip] 
  	\ ~~\\
 	{{  \textsc \small
	Centre for the Studies of the Glass Bead Game
	}}
 	\\ [1\baselineskip] 
}
 \frontmatter
 \maketitle
.\vspace{201pt}

 \hfill \texttt{dedicated to the Journeyers to the East}

\hfill \texttt{in the tradition of Hermann Hesse's The Glass Bead Game}
\newpage

 \tableofcontents



\newcommand{\gdual}[1]{\overset{\:{}^{{}^{\boldsymbol{\sim}}}}{\smash[t]{#1}}} 
\newcommand{\dual}[1]{\overset{{}^{{}^{\boldsymbol{\sim}}}}{\smash[t]{#1}}} 
\newcommand{\gdualn}[1]{\overset{\:{}^{{}^{\boldsymbol{\neg}}}}{\smash[t]{#1}}} 
\newcommand{\dualn}[1]{\overset{{}^{{}^{\boldsymbol{\neg}}}}{\smash[t]{#1}}} 

\def\bmfj{\mbox{\boldmath$\displaystyle\boldsymbol{\mathfrak{J}}$}}
\def\bmfk{\mbox{\boldmath$\displaystyle\boldsymbol{\mathfrak{K}}$}}

\def\bmfa{\mbox{\boldmath$\displaystyle\boldsymbol{\mathfrak{A}}$}}
\def\bmfb{\mbox{\boldmath$\displaystyle\boldsymbol{\mathfrak{B}}$}}

\def\0{\mbox{\boldmath$\displaystyle\mathbb{O}$}}
\def\C{\mbox{\boldmath$\displaystyle\mathbb{C}$}}
\def\R{\mbox{\boldmath$\displaystyle\mathbb{R}$}}
\def\bpi{\mbox{\boldmath$\displaystyle\boldsymbol{\pi}$}}
\def\j{\mbox{\boldmath$\displaystyle\boldsymbol{j}$}}
\def\J{\mbox{\boldmath$\displaystyle\boldsymbol{J}$}}
\def\vp{\mbox{\boldmath$\displaystyle\boldsymbol{\varphi}$}}
\def\vt{\mbox{\boldmath$\displaystyle\boldsymbol{\vartheta}$}}

\def\K{\mbox{\boldmath$\displaystyle\boldsymbol{K}$}}
\def\L{\mbox{\boldmath$\displaystyle\boldsymbol{L}$}}
\def\x{\mbox{\boldmath$\displaystyle\boldsymbol{x}$}}
\def\n{\mbox{\boldmath$\displaystyle\boldsymbol{n}$}}
\def\kb{\mbox{\boldmath$\displaystyle\boldsymbol{\kappa}$}}
\def\bz{\mbox{\boldmath$\displaystyle\boldsymbol{\zeta}$}}

\def\A{\mbox{\boldmath$\displaystyle\boldsymbol{A}$}}
\def\B{\mbox{\boldmath$\displaystyle\boldsymbol{B}$}}

\def\bn{\mbox{\boldmath$\displaystyle\boldsymbol{\nabla}$}}

\def\I{\openone}
\def\s{\mbox{\boldmath$\displaystyle\boldsymbol{\sigma}$}}
\def\p{\mbox{\boldmath$\displaystyle\boldsymbol{p}$}}
\def\q{\mbox{\boldmath$\displaystyle\boldsymbol{q}$}}
\def\k{\mbox{\boldmath$\displaystyle\boldsymbol{k}$}}
\def\e{\rm e}
\def\openone{\mathbb I}

\def\beq{\begin{equation}}
\def\eeq{\end{equation}}

\def\bea{\begin{align}}
\def\ena{\end{align}}


\mainmatter
\chapter*{Preface}

\vspace{7pt}
\begin{itemize}
\item[\empty]\textit{We report an unexpected theoretical discovery of a spin one half matter field with mass dimension one.}
\item[\empty]
\hfill {\small 
\cite{Ahluwalia:2004sz}}
\end{itemize}

\begin{itemize}
\item[\empty]\textit{We provide the first details on the unexpected theoretical discovery of a spin-one-half matter field with mass dimension one.}
\item[\empty]
\hfill {\small 
\cite{Ahluwalia:2004ab}}
\end{itemize}
\vspace{21pt}

With these opening lines Daniel Grumiller and I introduced an entirely new class of fermions. They carry   
 mass dimension one. That is, they do not satisfy Dirac equation, but only the spinorial Klein-Gordon equation for spin one half. In the 
 intervening decade and a half, the issues of non-locality, and Lorentz symmetry violation, have been completely resolved. But a self contained and an \textit{ab initio} treatment of such an unexpected theoretical discovery is missing.
It is therefore necessary to lift a logical version of this development from the pages of various journals to a monograph.

   I present here what we know of the subject at the present moment (late 2018). In making this selection I have  strictly confined, with a minor exception,   to that part of the existing literature 
 which has passed through my own pen and paper. This is not to negate the 
 contributions of my collaborators, and many others who have worked on the subject,  but to 
  take full personal responsibility for the presented formalism.\footnote{Of the reader it is assumed that she is at home with the theory of special relativity, and first few chapters of books on the theory of quantum 
  fields. With that background she would be ready for the journey through this monograph}
  
     Not unexpectedly, 
there is a group of physicists who have siezed upon the new construct and based much of their careers on exploiting the physical consequences
 and studying the underlying mathematical  structure of the new theoretical discovery. This is evident from some one hundred papers, and several doctoral thesis, that are entirely devoted to the new spinors and the associated fermions.
 
 Then there is a group of physicists who simply dismiss the subject as an impossibility. For the latter, I can only suggest that they first construct the eigenspinors of the $(1/2,0)\oplus(0,1/2)$ charge conjugation operator -- with eigenvalues $\pm 1$ -- and show that neither
$\left(\gamma_\mu p^\mu + m\I_4\right)$ nor $\left(\gamma_\mu p^\mu - m\I_4\right)$
annihilates these spinors. Having done this preliminary exercise carefully, without falling into the temptation of Grassmann-isaton of the new spinors, 
they may start with Chapter~\ref{ch5} -- returning to the earlier chapters only for notational details -- and come to the end of chapter~\ref{ch11}. At that stage, they would have enough information to develop their own theory and to see if their calculations produce something similar to what follows in the remainder of the monograph. 

The rest of the Preface provides a brief scientific journey of the author.
It provides a context in which the reported results were obtained. What follows may thus be seen as a brief scientific autobiography\index{Scientific autobiography} that excludes, 
with the exception of the next paragraph,
 large parts of my work on the interface of the gravitational and quantum realms, and on neutrino oscillations.

 \centerline{--------------}

Sometime in the early 1980's, on the banks of Charles river in Boston, my quantum mechanics teacher was scheduled to give three fifty-minutes lectures, thrice a week. Instead, to my pleasure, I was exposed to five lectures a week, each of three hours duration with a five minutes water break half the way through each of the lectures.  And these continued for three trimesters. I learned many things, among them, significance of phases in the quantum description of reality. 
Years later,
it was, in part, for that reason that a day after Christmas of 1995,
 seeing falling snow flakes on a road trip by car from the 
Maulbronn Monastery,
Germany, to the French border, that I asked myself as to what is the difference between  classical snow and quantum snow.\index{Snow: classical versus quantum} I realised that each of the snow flakes had a different mass. Each flake thus picked up a different gravitationally induced phase.\footnote{I now realise that this inference requires a   revision  due to the inevitable modification of  the  wave particle duality in the Planck realm~\citep{Kempf:1994su,Ahluwalia:2000iw}: going from a snow flake to the neutrino mass eigenstates one goes from  the Planck-scale induced gravitational modifications of the wave particle duality to the low energy realm of quantum mechanics where the de Broglie's wave particle duality holds to a great accuracy \citep{PhysRevLett.91.090408}. For $C_{60} F_{48}$ molecule the experiment finds fringe visibility lower than expected. It may  be indicative that the modification to the de Broglie wave particle duality may become significant at a much lower energy.}
Soon, within minutes, I was thinking of solar neutrinos instead of snow flakes and a back of the envelope calculation rolled through me. It became clear that neutrino oscillations provide a set of flavour oscillation clocks and that these clocks redshift according to the general relativistic expectation, and suffer Zeeman like splitting in the oscillation frequencies for generalised flavour oscillations clocks. In the process I came to realise that there are instances when the gravitationally induced forces may be zero, but not the gravitationally induced phases. All this, in collaboration with Christoph Burgard, led to a shared 1996 First Prize from the Gravity Research Foundation (GRF), a Fourth Prize for 1997, and a series of other publications that inspired a few hundred papers devoted to the interface of the gravitational and quantum realms and won the third, in 2004, and the fifth in 2000, prizes from GRF.

When my quantum mechanics teacher moved from the east coast to the mid west, I returned with him to my original host university (which has been utterly flexible and graceful to me),
I considered him as a natural advisor for my doctoral thesis.
Gradually it dawned on me that my questions in Physics were different from  his.
For my teacher one must start a conversation with a Lagrangian, while for me 
I needed a more systematic approach to arrive at Lagrangians -- be they be for the Maxwell field, or the Dirac, or any other matter or gauge field.
In fact one cannot even fully formulate gauge covariance without knowing the kinematical description of the matter fields. 

I now know that once the Lagrangian is given then the principles of quantum mechanics and inhomogeneous Lorentz symmetries -- coupled with the operations of parity, time reversal, and charge conjugation  -- intermingle to make a theory predictable in the resulting S-matrix 
formalism.\footnote{This is the quantum field theoretic formalism presented in~\citep{Weinberg:1995mt,Weinberg:1996kr}. Its origins go back to~\citep{Weinberg:1964cn,Weinberg:1964ev,Weinberg:1969di}.
 Its most celebrated offspring is the standard model of high energy physics.} 
My aim was to understand  the opening chapters of any quantum field theory text better. 
 Given a representation space associated with the Lorentz algebra,  and the behaviour under  discrete symmetries, my desire was to derive Lagrangians for the objects inhabiting these spaces. I expected nothing more than to arrive at the standard results, but in my own way. This is already done by Steven Weinberg in his classic on quantum fields~\citep{Weinberg:1995mt,Weinberg:1996kr}

During the year I started working on my doctoral thesis, Lewis Ryder's book on quantum field theory arrived on the scene. It provided a derivation of Dirac equation.\footnote{Dirac equation, as originally introduced and as now understood are two very different things. The original acted $(i\gamma_\mu\partial^\mu - m\I_4)$ on a spinor, while that of the standard model of high energy physics the same very operator acts on a spinorial quantum field.}
 I could easily extend his derivation to obtain Maxwell equations. So, I was happy for sometime that I could make progress in obtaining the kinematical structure of matters fields, and understand gauge fields from my own perspective. At this juncture, I arranged for a series of breakfast/lunch conversations with a nuclear physicist who had been my teacher for a wonderfully-taught course from Jackson's electrodynamics. As a result of these conversations,
I realised that I could  be useful to the nuclear physics community by formulating a pragmatic approach to dealing with higher spin baryonic and mesonic resonances.
These were copiously produced at the then new Continuous Electron Beam Accelerator Facility  in Virginia.\footnote{Now called
Thomas Jefferson National Accelerator Facility.}

As soon I submitted my doctoral work to the thesis  clerk, came Christoph Burgard, and declared to me that Ryder's derivation of Dirac equation is `all wrong.'\footnote{To be fair to Lewis Ryder, his derivation, in essence
reproduced the then-existing literature on the subject. For a parallel treatment as Ryder's our reader may consult~\citep{Hladik:1999tt}. It apparently began in the Istanbul lectures in early 1960's with Feza G\"ursey hosting the theoretical physics school.}  
And it turned out that Christoph was correct, as always. This is now explained
in my review of Ryder's book (written a few years later) and by Gaioli and Garcia Alvarez in their  
\textit{Am. J. Phys.} paper.
\footnote{At the time I wrote my book review I was not aware of the analysis of Gaioli and Garcia Alvarez. I only learned of their work when, unexpectedly one day,  they walked into my office at the Los Alamos National Laboratory  (as by then I was a Director's Fellow there) and told me about their publication. Lewis Ryder in the second edition of his book does cite 
Gaioli and Garcia Alvarez, but without attending to the raised concerns.
} To me, it is again a story that weaves important phases in the analysis: for Dirac spinors the right and left transforming components of the particle and antiparticle rest-spinors have opposite relative phases. 
This important fact can be read off from
Steven Weinberg's analysis of the Dirac field. The result follows without invoking Dirac equation.

Soon after my arrival at the Los Alamos Meson Physics Facility\footnote{Now named Los Alamos Neutron Science Center.}  in the July of 1992, I started to look at Majorana neutrinos. There was a significant element of confusion on the subject and I requested LAMPF to obtain for me an English translation of the 1937 paper of Majorana. I thus found out that in the 1937 paper there was no notion of Majorana spinors: The Majorana field was still the Dirac field, expanded in terms of the Dirac spinors, but with particle and antiparticle creation operators identified with each other. 

On the other hand, I found in Pierre Ramond's primer\index{Ramond, Pierre} a systematic development of Majorana spinors:\index{Majorana!spinors} their origin resided in the fact that if $\phi$ transformed as a left-handed Weyl spinor then $\sigma_2 \phi^\ast$ transformed as a right-handed Weyl spinor ($\sigma_2 = \mbox{`second' Pauli matix}$). But $\phi$, and consequently the Majorana spinor, had to be treated as a Grassmann variable. Furthermore, Majorana spinors were looked upon as Weyl spinors in disguise.

I was uncomfortable with both of the mentioned elements. For the Dirac spinors, understood as a direct sum of the right and left transforming Weyl spinors, no such Grassmann-isation was necessary -- at least in the operator formalism of quantum field theory. Another matter that concerned me was that for higher spin generalisation of the 
$\phi$-$\sigma_2\phi^\ast$ argument a replacement of $\sigma_2$ by its higher spin counterparts failed to do the magic of Pauli matrices,\index{Magic of Pauli matrices} as Ramond had called it. For a while, this seemed to make a higher spin generalisation of Majorana spinors untenable.

I resolved these problems by taking note of the fact that for spin one half the charge conjugation operator has four, rather than two, independent eigenspinors. And that the $\phi$-$\sigma_2\phi^\ast$ argument\index{The $\phi$-$\sigma_2\phi^\ast$ argument} works magically well for all spins if $\sigma_2$ was instead recognised, up to a phase, as Wigner's time reversal operator, $\Theta$, for spin one half: if  $\phi$ transforms as an $(0,j)$ object then $\Theta \phi^\ast$ transforms as a $(j,0)$ object -- with $\Theta$ now a spin-$j$ Wigner time reversal operator. 
As a consequence the $(j,0)\oplus(0,j)$ object
\begin{equation}
\left(
\begin{array}{c}
\mbox{a phase} \times \Theta \phi^\ast\\
\phi
\end{array}
\right)
\end{equation}
becomes an eigenvector of the spin-$j$ charge conjugation operator if the indicated phases are chosen correctly to satisfy the self/anti-self conjugacy condition. This generalised Majorana spinors to all spins. And such new objects were not $(j,0)$, or $(0,j)$  objects in disguise: there were $2j$, and not $j$, independent $(j,0)\oplus(0,j)$ vectors. Half of them were self-conjugate under charge conjugation, and other half were anti self-conjugate. Later new names were to be invented to avoid confusion with the newly constructed $(j,0)\oplus(0,j)$ vectors.

During this phase, I also began to suspect that the dynamics associated with these new objects would carry unexpected features. It was as true for spin one half, as for higher spins.

All these observations were essentially a mathematical science fiction.
In addition, I encountered the same problem as Aitchison and Hey\index{Aitchison and Hey} did in their attempt to construct a Lagrangian for the  c-number Majorana spinors. With some caveats, without the Lagrangian the story cannot unfold, cannot come to fruition.

Bringing back the focus to spin one half, I expressed some of my frustration in an unpublished e-print where, towards a resolution of the problem, I initiated a work to construct dual space for the new set of four-component spinors. It allowed to define an adjoint for the quantum field with the new spinors as expansion coefficients. At this stage Daniel Grumiller joined my efforts and we calculated the vacuum expectation value of the time ordered product of the field with the  newly defined adjoint -- that is, the Feynman-Dyson propagator. This gave us a most startling result: the new spin one half fermionic field carried mass dimension one. This we presented as an ``unexpected theoretical discovery'' in JCAP and PRD -- the two 2004 e-prints, were published in 2005, almost back to back due to a long, but constructive, refereeing process.

Soon after the publication of these papers, I moved from Zacatecas in Mexico to Canterbury in New Zealand. There, I formed a very active research group till it dismantled in the aftermath of the Christchurch earthquakes. 
The most active members of this group were my doctoral students: Cheng-Yang Lee,\index{Lee, Cheng-Yang}  Sebastian Horvath,\index{Horvath, Sebastian}  Dimitri Schritt,\index{Schritt, Dimitri} and 
Tom Watson.\index{Watson, Tom} Though Tom stayed in the group only for a short time, he made an interesting contribution. The most important of these was that Ryder's definition of the Dirac quantum field, as was the case with many other authors, was not consistent with the construction of quantum fields formulated by Steven Weinberg. This fact, along with what I'll later describe as the IUCAA breakthrough\index{The IUCAA breakthrough} in Section \ref{Sec:IUCAA} of this monograph led to evaporating the problems of non-locality and Lorentz-symmetry violation. 

After the Canterbury earthquakes, I took a two-year detour to Campinas in Brasil, and returned to India more or less permanently.
Thus by 2017, in India I had on my hands a spin one half fermionic field that was local and did not suffer from the violation of the Lorentz symmetry. 

\centerline{--------------}


With this background, this monograph presents the new theory at a level that should be easily accessible to any good graduate student. An outstanding question that still remains is to reformulate Weinberg's construction of quantum fields so as to accommodate this new field. My preliminary thoughts on evading the no-go result of Weinberg can be found in my latest  paper in
Europhysics Letters (EPL) written under the title, ``Evading Weinberg's no-go theorem to construct mass dimension one fermions: Constructing darkness.''


\chapter*{Acknowledgements}
\label{ch16}

Projects like these often evolve over years if not decades. In the process one's scientific style takes birth, often in a merging of one's own genius and an inspiration owed to least one great teacher.
The latter for me was Dick Arnowitt. I am immensely grateful to him for teaching me many things of the quantum realm and for his absolute accessibility. Steven Weinberg's books are another source of my inspiration, as is Dirac's classic on quantum mechanics. While for Arnowitt a story begins with the Lagrangian density, for me once the Lagragian density is given one gives essentially the whole story. And so my quest was for the logical path that leads to Lagrangian densities. This monograph is a reflection of that quest, and that path taken. Arnowitt, in the very first lecture I attended by him, told us all that Dirac's  classic was the best book written in one hundred years~\citep{Dirac:1930pam}. Weinberg, in my opinion does for quantum field theory what Dirac did for quantum mechanics. I am grateful to these scholars. I took Arnowitt's emphasis on the importance of phases in quantum mechanics to heart and there is perhaps not a single publication of mine where this is not apparent.

My gratefulness also includes numerous referees and one in particular. He is 
Louis Michel. I urge the reader to read my indebtedness to him published as an acknowledgement~\citep{Ahluwalia:1995ur}. I am thankful to Peter Herczeg for bringing certain sentiments of Louis Michel to me and for his friendship and scholarship during my 1992-1998 stay at Los Alamos.

Very special thanks go to Daniel Grumiller for joining my seed efforts in a 2003 preprint~\citep{Ahluwalia:2003jt} and evolving them collaboratively into an `unexpected theoretical discovery'  reported in~\citep{Ahluwalia:2004sz,Ahluwalia:2004ab}. My students, Cheng-Yang Lee, Sebastian Horvath, and Dimitri Schritt, became my close friends and collaborators and contributed immensely in creating a warm scholarly ambiance in our research group at the University of Canterbury (Christchurch, New Zealand) and in developing the formalism of mass dimension one fermions. I am grateful to them, and to numerous other students who either attended my lectures at Canterbury and Zacatecas (Mexico) or/and worked under my supervision for projects or thesis.

For securing a continuing academic position at the University of Canterbury I am grateful to Matt Visser and to David Wiltshire and equally thankful for making that decade a very productive and pleasant one. For my two-year long detour to Brasil I am grateful to Marco Dias,  Saulo Pereira, Julio Marny Hoff da Silva, Alberto Saa, and to Rold\~ao da Rocha for their friendship and many insightful discussions on subject of this monograph.

Zacatecas is a beautiful small city in northern Mexico at roughly two thousand and five hundred meters. I very much enjoyed my  tenure at Universidad Aut\'onoma de Zacatecas and the city. The  papers with Daniel Grumiller were published from there. Gema Mercado, then Director of the Department of Mathematics, not only invited me back from India to Zacatecas but she also provided an inspired scholarly ambiance, and a friendship and leadership of unprecedented selflessness. I am utterly grateful to Gema for that and for allowing me to pursue my work without hindrance and with encouragement and support.

 I thank Llohann 
Speran\c{c}a for discussions in the initial stages of this manuscript at Unicamp (S\~ao Paulo, Brasil).
The breakthrough on the Lorentz symmetry and locality presented here began in late 2015 during a  three-month long visit to the Inter-University Centre for Astronomy and Astrophysics (IUCAA, Pune) where I gave a series of lectures on mass dimension one fermions. For their insightful questions and the ensuing discussions, I thank  the participants of those lectures and in particular Sourav Bhattachaya, Sumanta Chakraborty, Swagat Mishra, Karthik Rajeev, and Krishnamohan Parattu and my host Thanu Padmanabhan.
Raghu Rangarajan (Physical Research Laboratory, Allahabad) carefully read the entire first draft of an important manuscript~\citep{Ahluwalia:2016rwl} and provided many insightful suggestions. I am grateful to him for
his generosity and for engaging in long insightful discussions. The calculations that led to the reported results were done at Centre for the Studies of the Glass Bead Game and Physical Research Laboratory. 

Chia-Ren Hu and George Kattawar brought me to Texas A\&M University (College Station, USA) for me to pursue my doctoral degree and supported me through my entire 1983-1991 stay there. I am immensely indebted to them for their conviction that a man could enter a Ph.D. program at age 31 and take his time reflecting to secure a Ph.D. at just a little shy of his 39th birthday. My gratefulness also goes to my supervisor for the Ph.D. degree, Dave Ernst. My tenure at Texas A\&M was made particularly meaningful by the inspired friendship and collaboration with Christoph Burgard. I treasured and treasure his warmth, his insights, and many things \textit{zimpoic}.

At the Los Alamos National Laboratory, Terry Goldman, Peter Herczeg, Mikkel Johnson,  Hewyl White, among so many other friends like Cy Hoffman, George Glass and Nu Xu, kept their faith in my studies and supported my independence with warmth and scholarship. I am grateful to them, and those others who know who they  are. Who sat down on mesas or walked with me, and shared their insights and wisdom.

My return to India has been warmly supported by Pankaj Jain through an Institute Fellowship at the Indian Institute of Technology Kanpur, and through similar grants and invitations by T. P. Singh (Tata Institute of Fundamental Research, Mumbai), Sudhakar Panda (Institute of Physics, Bhubaneshwar), Thanu Padmanbhan (IUCAA), Mohammad Sami (Jamia Milia Islamia, New Delhi).  I thank them for welcoming me back and for opening doors for me. 

The scene where the reported work took place changed from Los Alamos National Laboratory in the States, to 
Universidad Aut\'onoma de Zacatecas (UAZ) in Mexico, to the University of Canterbury in New Zealand, to Universidade Estadual de Campinas (Unicamp) in Brasil, to Inter-University Centre for Astronomy and Astrophysics (IUCAA) and to the Centre for the Studies of the Glass Bead Game in Bir, Himachal Pradesh.  I am indebted to these institutions for reasons too many to enumerate.

Yeluripati Rohin and Suresh Chand  read the final draft of the manuscript and caught several typos, and helped me improve the presentation.  I thank them both. I thank Shreyas Tiruvaskar for his suggestions on an earlier draft.

I am utterly thankful to my editor, Simon Capelin, and Roisin Munnelly, and Sarah Lambert in the role of  of Editorial Assistants. Without their patience and advice, and without the personal invitation of Simon, this monograph would simply not have come to exist.

Last, but not least my warm thanks go to my children Jugnu, Vikram, Shanti, and Wellner, and to my father Bikram Singh Ahluwalia.
They all have been sages of wisdom and affection to me. Karan and Bobby, my brothers and their families, provided a home when I had none, for that and their warmth I am grateful. Sangeetha Siddheswaran is a constant source  of affection and encouragement, and I am equally grateful to her for helping me bring this monograph to completion.

While writing this monograph works of J. M. Coetzee, Hermann Hesse and Carl Jung provided the poetic and moral 
 background. The monograph is dedicated to Hermann Hesse's 
Journeyers to the East.

During writing of this work Sweta Sarmah became an inspired inspiration in the ways of the Golu Molu. Many butter scotch ice creams to her.

\chapter{Introduction}

In a broad brush, the grand metamorphosis\index{The grand metamorphosis} that has created the astrophysical and cosmic structures arise from an interplay of 
(a) Wigner's work on unitary representations of the inhomogeneous Lorentz group including reflections, (b) Yang-Mills-Higgs framework for understanding interactions, and (c) an expanding universe governed by Einstein's theory of general relativity.
The wood\index{The wood} is provided by the spacetime symmetries. These tell us whatever matter exists, it must be one representation or the other of the extended Lorentz symmetries~\citep{Wigner:1939cj,Wigner:1962ep}.
The fire and the glue\index{The fire and the glue} 
is provided by  the principle of local gauge symmetries \emph{\`a la} Yang and Mills~\citep{Yang:1954ek} and by general relativistic gravity.
 The Wigner-Yang-Mills framework, as implemented in the standard model of the high energy physics, then provides the right hand side of the Einstein's field equations to determine  the evolution of that very spacetime which these matter and gauge fields give birth to in a mutuality yet not completely formulated to its quantum completion. Dark energy, thought to be needed for the accelerated expansion of the universe, and dark matter, required by 
 data on the velocities of the stars in galaxies, the motion of galaxies in galactic 
 clusters, and cosmic structure formation, keeps  roughly ninety five percent of the universe dark, and of yet to be understood origin.
 
 It is nothing less than the most sublime poetry and primal magic that this picture can explain the rise of mountains and flow of water in the rivers, 
and go as deep as to invoke metamorphosis of light into the water and the mountains, the stars and galaxies. 
 Where a scientist knows where and how the water first came to 
be~\citep{Hogerheijde:2011pq,                                                                                                                                                                                                                                                                                                                                                                                                                                                                                                                                                                                                                                                                                                                                                                                                                                                                                                  Podio:2013wnh}, 
and a poet asks if there was thirst when the water first rose.   The origin of biological structures remains an inspiring open subject. 

It is within this framework, that we wish to add a new chapter and show how to construct dark matter and understand its darkness from first principles. It is thus, in this monograph we present  an unexpected theoretical discovery of new fermions of spin one half. The fermions of the standard model, be they leptons or quarks,  carry mass dimension three halves. The mass dimension of the new 
fermions is one.  Their quartic self interaction, despite being fermions, is a mass dimension four operator as is their interaction with the Higgs. Their interaction with the standard model fermions is suppressed by one power of the Planck scale and because they couple to Higgs, quantum corrections can bring about tiny magnetic moments for the new fermions. These aspects make them  natural dark matter candidate and can provide tiny interaction between matter and gauge fields of the standard model  -- something that is already suggested observationally~\citep{Barkana:2018lgd}.
Studies in cosmology hint that the quantum field associated with the new particles may also play an important role in inflation and accelerated expansion of the universe~\citep{Boehmer:2007dh,Boehmer:2010ma,Basak:2012sn,Basak:2014qea,Pereira:2014wta,Pereira:2017bvq,Pereira:2016emd,Rogerio:2017gvr}.


We thus weave a story of how the non-locality of the first effort evaporated~\citep{Ahluwalia:2004sz,Ahluwalia:2004ab,Ahluwalia:2016jwz,Ahluwalia:2016rwl}. 
We tell of the evaporation of the violation of Lorentz symmetry. In the process, we construct a quantum field that is local and fermionic.  It finds not its description in the Dirac formalism, but in a new formalism appropriate for its own nature. Beyond the immediate focus it makes explicit many insights,  otherwise hidden in the work of Weinberg~\citep{Weinberg:1995mt}.

The meandering path from non-locality to locality, from Lorentz symmetry violation to preserving Lorentz symmetry, owes its existence to  certain wide-spread errors and misconceptions in most textbook presentations of quantum field theory (and we had to learn, and correct these), and the eventual breakthrough to certain phases that affect locality and to a construction of a theory of duals and adjoints.




In the first volume, in chapters 2 to 5 of~\citep{Weinberg:1995mt} Steven Weinberg proves what may be called  a no-go theorem:  a Lorentz and parity covariant local theory of spin half  fermions  must be based on a field expanded in terms of the eigenspinors of the parity operator, that is Dirac spinors -- and nothing else. Furthermore, these expansion coefficients must come with certain relative phases. And in addition, there must be a specific pairing between the expansion coefficients and the annihilation and creation operators satisfying fermionic statistics.

A reader who finds these remarks mysterious, may undertake the exercise of comparing ``coefficient functions at zero momentum'' which Weinberg arrives at in his equations (5.5.35) and (5.5.36) with their counterparts written by some of the other popular authors, for example~\citep{Ryder:1985wq,Folland:2008zz,Schwartz:2014md}. The book by Srednicki avoids these  errors with profound consequences for the consistency of the theory with Lorentz symmetry and locality~\citep{Srednicki:2007qs}.

In arriving at the canonical spin one half fermionic field, Weinberg does not use or invoke Dirac equation, or the Dirac Lagrangian density. These follow by evaluating the vacuum expectation value of the time ordered product of the field and its adjoint at two spacetime points $(x,x^\prime)$. The resulting Feynman-Dyson propagator determines  the mass dimensionality of the field to be 
three halves.\footnote{See chapter 12 of Weinberg's cited monograph for a rigorous definition of mass dimensionality of a quantum field.}

Thus, information about the mass dimensionality of a quantum field is spread over two objects: the field, and its adjoint. Weinberg first derives the field from general quantum mechanical considerations consistent with spacetime symmetries, cluster decomposition principle, and then as just indicated, uses this field to arrive at the Lagrangian density through evaluating the vacuum expectation value of the time ordered product of the field and its adjoint at two spacetime points $(x,x^\prime)$. The powers of spacetime derivatives that enter the Lagrangian density is not assumed, but it is determined by the representation space, and the mentioned formalism,  in which the field resides. 
The broad brush lesson is: Given a spin, it is naive to propose a Lorentz covariant Lagrangian density. It must be derived \emph{\`a la} Weinberg.
The expansion coefficient, $f_\alpha$ and $f^\prime_\alpha$, of a quantum field, $\psi$, are determined by an appropriate finite dimensional representation of the Lorentz algebra and the symmetry of spacetime translation
\begin{equation}
\psi = \sum_\alpha \big[f_\alpha \textbf{a}_\alpha + f^\prime_\alpha \textbf{b}^\dagger_\alpha\big]\nonumber
\end{equation}
where $\textbf{a}_\alpha$ and  $\textbf{b}_\alpha$ satisfy canonical fermionic or bosonic commutators or anticommutators. For simplicity of our argument we have suppressed the usual integration on four momentum, and it may be considered absorbed in the summation sign. As is clear from Weinberg's work, though not explicitly stated by him, if $f_\alpha$ and $f^\prime_\alpha$ 
satisfy a wave equation, so do $ u_\alpha = e^{i \zeta_\alpha} f_\alpha$ and $v_\alpha = e^{i \xi_\alpha} f^\prime_\alpha$, with $\zeta_\alpha,\xi_\alpha\in \Re$. If the field $\psi$ has to respect Lorentz covariance, locality, and certain discrete symmetries then the phases,
$ e^{i \zeta_\alpha}$ and $ e^{i \xi_\alpha}$ cannot be arbitrary,  but must acquire certain values. Up to an overall phase factor,  these  are determined uniquely in the Weinberg formalism. Furthermore, the pairing of the $u_\alpha$ and $v_\alpha$ with the annihilation and creation operators is also not arbitrary. A concrete example of all this can be found in~\citep{Ahluwalia:2016jwz}. The second subtle element is: how to define dual of $u_\alpha$ and $v_\alpha$, and the adjoint of $\psi$ (see below). We develop a general theory of these elements in this monograph suspecting that mathematicians may have already addressed this issue in one form or another -- that said, a tourist guide by a mathematician has missed the issues that we point out~\citep{Folland:2008zz}.

Once these observations are taken into account, 
if one were to envisage a new fermionic field of spin one half and evade Weinberg's no-go theorem then something non-trivial has to be done. 
Our approach would be to combine elements  of Weinberg's approach  and that of a naive one indicated above. We shall take the $f_\alpha$ and $f^\prime_\alpha$ not to be complete set of eigenspinors of the spinorial parity operator but that of the spin one half charge conjugation operator. 
We shall fix the phases $e^{i \zeta_\alpha}$ and $e^{i \xi_\alpha}$ to control the covariance under various symmetries, and to satisfy locality.
We will find that each of the eigenspinors of the charge conjugation operator has a zero norm under the canonical Dirac dual. This would lead us to an \textit{ab initio} analysis of constructing duals and adjoints. 
In the process, we  find that if the eigenspinors of the parity operators in the Dirac field are replaced by a complete set of the eigenspinors of the charge conjugation operator, and one chooses appropriate relative phases between the ``coefficient functions at zero momentum,'' and follows a Weinberg analogue of pairing of the expansion coefficients with the annihilation and creation operators, then the resulting field on evaluating the vacuum expectation value of the time ordered product of the field and its adjoint at two spacetime points $(x,x^\prime)$ is found to be endowed with mass dimension one. Thus,  giving a fundamentally new fermionic field of spin one half.
\centerline{--------------}

One of my younger friends, and a physicist in his own right, explains to me the new fermions with the following 
wisdom~\citep{Swagat:2017pc}, ``Why should Parity get all the privilege? Charge Conjugation has equal rights." We will see here that he captures the essence of one of the main results of this monograph.

The monograph may also be seen as chapters envisaged by a referee of a 2006 Marsden Funding Application to the Royal Society of New Zealand. The referee report read, in part~\citep{RefereeMarsden:2006nz}:

\begin{itemize}\item[\empty]
\textrm{The problem has fueled intense debates in recent years and is generally considered
fundamental for the advancement in the field. As for the proposed
solution [by Ahluwalia], I find the approach advocated in the project
a very solid one, and, remarkably, devoid of speculative excesses
common in the field; the whole program is firmly rooted in quantum
field theoretic fundamentals, and can potentially contribute to
them. If \textit{Elko} and its siblings can be shown to account for
dark matter, it will be a major theoretical advancement that will
necessitate the rewriting of the first few chapters in any textbook
in quantum field theory.  If not, the enterprise will still have
served its purpose in elucidating the role of all representations of
the extended Poincar\'e group.}
\end{itemize}

Thus this monograph presents the first long chapter envisaged by the referee and contains much that has been  discovered since.

From time to time, a junior reader would come across a remark that is not immediately obvious. For example, after equation (\ref{eq:primordial-generators}), there suddenly appears 
a paragraph reading, ``Without the existence of two, rather than one, representations for each $\bmfj$ one would not be able to respect causality in quantum field theoretic formalism respecting Poincar\'e symmetries, or have antiparticles required to avoid causal paradoxes.'' In such an instance our reader may simply go past such matters and continue. The chances are in the course of her studies, she will come to appreciate the insight, or perhaps disagree with it. I hope such liberties  shall serve their purpose in the spirit of Hermann  Hesse's journeyers to the east, to whom this monograph is dedicated.

 
 \chapter{A trinity of duplexities} \index{Duplexity! Dirac} 


A view is presented in which dark matter is seen as a continuation of historical emergence of spin and antiparticles.

\section{From emergence of spin, to antiparticles, to dark matter}

Arguing for ``some incompleteness'' in the earlier works of  Darwin and Pauli,   Dirac confronts a ``duplexity'' phenomena:~a discrepancy that the observed number of stationary states of an electron in an atom being twice the number given by the then-existing  theory~\citep{Dirac:1928hu,Darwin:1927du,Pauli:1927,Uhlenbeck:1925ge,Uhlenbeck:1926ge}.\footnote{For the contribution of Otto Stern to this story, see \citep{Pakvasa:2018xlz}.} 
The  solution he proposed, with the subsequent development of the theory of  quantum 
fields, not only resolved the discrepancy but it also introduced a new unexpected\index{Unexpected duplexity}
duplexity~\citep{Tomonaga:1946zz,PhysRev.76.749,PhysRev.75.1736,PhysRev.82.914,Weinberg:1964cn,tHooft:1973mfk,Weinberg:1995mt}: 
For each spin one half particle, the Dirac theory predicted an antiparticle. Associated with this prediction was the charge conjugation symmetry\index{Charge conjugation symmetry} -- a notion that soon afterwards was generalised to all spins. This symmetry shall play a pivotal role in this monograph.

The doubling of the degrees of freedom, for  spin one half fermions of Dirac, can be traced to the parity covariance built into the formalism. This symmetry requires not only the left-handed Weyl spinors but also the right-handed Weyl spinors. In the process for a spin one half particle we are forced to deal with four, rather than two, degrees of freedom. The antiparticles of the Dirac formalism may be interpreted as a consequence of this doubling.

\noindent 
Parenthetic remarks:

\begin{itemize}
\item The existence of antiparticles is not confined to spin 
one half as is beautifully argued by Feynman in his 1986 Dirac memorial 
lecture delivered under the title, ``The reason for antiparticles'' 
~\citep{Feynman:1987gs}. It is based on a calculation of amplitudes for sources with strictly 
positive energy superpositions.
A related argument by Weinberg emphasises that antiparticles are required to 
avoid causal paradoxes~\citep[chap. 2, Sec. 13]{Weinberg:1972gc}.
\index{Antiparticles! causality and parity covariance} 
\item For each spin, Lorentz algebra  provides two separate representation spaces. These transmute into each other under the operation of parity. The existence of two separate representation spaces is important for the existence of antiparticles. It allows the doubling in the degrees of freedom required by the existence of antiparticles. That in turn allows to build a causal theory.\footnote{A departure from this remark must be made when dealing with massless particles, and parity violation.}

\end{itemize}

Fast forward a few decades, with the intervening years placing Dirac's work on a more systematic footing,
the new astrophysical and cosmological observations have now introduced a new duplexity.\index{A new duplexity} With the exception of interaction with gravity, these observations strongly hint that there exists a new form of matter which carries no, or limited, interactions with the matter and gauge fields of the standard model of high energy physics~\citep{Bertone:2016nfn}.  To distinguish it from the matter fields of the standard model of high energy physics the new form of matter has come to be called dark matter.\index{A new duplexity!dark matter} \index{Dark matter}

For some decades now supersymmetry\index{Supersymmetry} was thought to provide precisely such a duplexity  in a natural manner by introducing a symmetry that transmuted mass dimensionality \textit{and} statistics of particles~\citep{Coleman:1967ad,Haag:1974qh}. However, at the date of this writing, despite intense searches there is no observational evidence for its existence.

Here I suggest that its origin instead lies in  
a new duplexity. And that dark matter is simply not   yet another familiar  particle of the types  found in the standard model of the high energy physics and the standard general relativistic cosmology.
The new duplexity,\index{A new duplexity} I suggest, is provided by mass dimension one fermions. Unlike supersymmetry I suspect that there exists a new symmetry\index{A new symmetry} that transmutes only the mass dimension of the fermions, and \textit{not} the statistics.


\begin{table}\label{tab:trinity}
\begin{minipage}{350pt}
\caption{A trinity of duplexities}\index{A trinity of duplexities}
\label{table2}
\addtolength\tabcolsep{2pt}
\begin{tabular}{@{}c@{\hspace{30pt}}lll@{\hspace{10pt}}}
\hline\hline
Duplexity & Phenomena & Consequence\\
\hline
1 & Doubling of states of an electron & Spin \\ 
\empty & in an atom & \empty \\ \\
2 & Doubling of degrees of freedom  & Antiparticles\\
\empty & for spin $\frac12$ particles ($m\ne 0$) & \empty\\ \\
3\footnote{Here, perhaps interchanging the entries in the second and third columns may better reflect the logical order. As we proceed the reader would discover why mass dimension one fermions are a first principle candidate for dark matter.}& Doubling of types of matter fields by & Dark matter \\
\empty & introducing fermions of mass dimension one & \empty\\
\hline\hline
\end{tabular}
\end{minipage}
\end{table}

The new fermions  carry a foundational importance for the representations of the Lorentz symmetries and the particle content contained in them. The limited interactions of the new fermions with the standard model particles is an ineluctable consequence of its nature.\index{Duplexity! trinity of}
The unexpected theoretical discovery of the new fermions thus provides a natural dark matter candidate.
We outline the logical thread leading to the new proposal in Table 1.1 \citep{Ahluwalia:2016jwz}.

The identification of dark matter  with the mass dimension one fermions is consistent with the conjecture that whatever dark matter it must still be one representation or the other of the Lorentz\footnote{Note, we say Lorentz symmetry  rather than spacetime symmetry. For the latter is just one representation arising out of the Lorentz algebra. See, 
section~\ref{sec:constructing-xt} for further remarks.} symmetries~\citep{Wigner:1939cj}. 
It also suggests a possible existence of  a new symmetry~\textendash~yet to be 
discovered~\textendash~ that mutates the mass dimensionality of fermions, without affecting the statistics. In that event the no-go theorems resulting from the works of Wigner, Weinberg, Lee and 
Wick may no longer apply~\citep{Wigner:1962ep,Weinberg:1964cn,Weinberg:1964ev,Weinberg:1995mt,PhysRev.148.1385} and open up truly new physics systematically constructed on well-known and experimentally verified symmetries.


The new fermions do not allow the usual local gauge symmetries of the standard model. Concurrently, their mass dimension is in mismatch with mass dimension of the standard model fermions. It  prevents them from entering the standard model doublets. The new fermions have a natural quartic self interaction with a dimensionless coupling constant, something that cannot occur for the Dirac/Majorana fermions. They also have a natural coupling with the Higgs and gravity. Additional interactions arise from quantum corrections.

As this monograph was composed a new unexpected aspect of the new fermions under rotation came to attention. It has important cosmological consequences. This is now the subject of  Chapter~\ref{ch10}.

Like the Majorana fermions the symmetry of charge conjugation plays a central role for the new fermions: while for the Majorana field the coefficient functions are eigenspinors of the parity operator, the field itself equals its charge conjugate.  For the new fermions the field is expanded in terms of the eigenspinors of the charge conjugation operator. 
Once that is done, one may choose to impose the Majorana condition, but it is not mandated.

Thus the new formalism, in a parallel with the Dirac formalism, allows for darkly-charged fermions, and Majorana-like neutral fields.

To avoid possible confusion we remind the reader that both the  Dirac and Majorana quantum field are  expanded in terms of Dirac spinors. These are eigenspinors of the parity operator: $m^{-1} \gamma_\mu p^\mu$, see Chapter~\ref{ch5} below, and ~\citep{Speranca:2013hqa}. Eigenspinors of the charge conjugation operator are thought to provide no Lagrangian description in a quantum 
field theoretic construct~\citep[App. P]{Aitchison:2004cs}. Thus placing the parity and charge conjugation symmetries on an asymmetric footing~\textendash~that is, as far as their roles in constructing spin one half quantum fields are concerned. Here I show that the Aitchison and Hey claim is in error (it seems to be edited out in later editions). It has remained hidden in a lack of full appreciation as to how one is to construct duals for spinors, and the associated adjoints~\textendash~that is, in the mathematics underlying the definition of the dual spinors via
$\overline\psi(\p) = \psi(\p)^\dagger \gamma_0$~\citep{Ahluwalia:2016jwz,Ahluwalia:2016rwl}. 
The resolution occurs through a generalisation of the Dirac dual presented here in chapter \ref{ch12}. 

To develop the physics hinted above we are forced to complete the development of this mathematics. Taken to its logical conclusion
it leads to the doubling of the fundamental form of matter fields.  One form of matter is described by the Dirac formalism, while the other, that of the dark sector, by the new fermions reported here. For each sector the needed matter fields 
 require a complete set of four, four-component spinors. For the former these are eigenspinors of the  parity operator while for the latter these are eigenspinors of the  charge conjugation operator (Elko)\index{Elko!acronym}.\footnote{As noted earlier: Elko is a German acronym for \textbf{E}igenspinoren 
 des \textbf{L}adungs\textbf{k}onjugations\textbf{o}perators introduced in~\citep{Ahluwalia:2004sz,Ahluwalia:2004ab}. In English, it translates to  eigenspinors of the charge conjugation operator.} 
 
 Global phases associated with these eigenspinors, and the pairing of these eigenspinors with the creation and annihilation operators influence the Lorentz covariance and locality of the fields (as already discussed at some length in the Preface). This last observation, often ignored in textbooks, when coupled with the 
discovery of a freedom in defining spinorial duals 
 accounts for the removal of the non-locality and restores the Lorentz symmetry for the mass dimension one fermions~\citep{Ahluwalia:2016rwl,Ahluwalia:2016jwz}.



\chapter{From elements of Lie symmetries to Lorentz algebra}\label{ch3}

Our first exposure to Lorentz algebra often happens in the context of some course on the theory of special relativity.
Because of historical reasons one often thinks of the two in the same breath. On some planet endowed with individuals who reflect on their origins one may arrive at Lorentz algebra by looking at the spectrum of the  hydrogen atom. At another planet, observations on light may lead thinking beings to arrive at Maxwell equations, and then Lorentz algebra  -- and may be even at conformal algebra. Thus, Lorentz algebra is a unifying theme underlying all attempts to understand associated  matter and gauge fields, and the very spacetime in which the associated quanta propagate. Somewhat poetically, that which walks and that in which it walks are determined by each other (this thread continues further in  chapter~\ref{ch15}). With this being so, we here take the view that Lorentz algebra is deeper than the symmetries of the Minkowski space (where the additional symmetry of spacetime translations exist). Its different solutions furnish different representation spaces. Minkowski spacetime being just one of them.

This chapter develops the needed concepts in a pedagogic manner. Its pace is to the point, and leisurely, and exploits the opportunity to present a point of view that to my knowledge contains several novel points of view. This chapter is written for an advanced undergraduate student to provide her a simple entry into the subject at hand.


\section{Introduction}

At Texas A\&M University, I did not begin work on my doctoral thesis till such time I came to know of Eugene Wigner's 1939 work and of Steven Weinberg's 1964 papers~\citep{Wigner:1939cj,Weinberg:1964cn,Weinberg:1964ev}. Not that I instantly understood them, or even appreciated their depth in the first encounter. But I  realised that the Dirac equation in its 1928 form and Maxwell equations of classical electrodynamics are so utterly simple to arrive at. Modulo a careful handling of the discrete symmetries, all one needs is the unifying theme provided by the Loretnz algebra.\footnote{In a quantum field theoretic context significantly more structure needs to be introduced to arrive at the Dirac and Maxwell equation for the fields~\citep{Weinberg:1995mt}.}
And if the latter were to change, say at the Planck scale, so would these equations. Assertions such as `let us take a Lagrangian density that depends only on one or two spacetime derivatives of the fields,' as I was to learn in the process, often led to mysterious pathologies for once the representation space to which a field belonged to was specified, one lost the freedom to invoke simplicity and convenience to choose how many spacetime derivatives entered the Lagrangian density. Many, if not all, such pathologies disappear if one is careful at the starting point: the Lagrangian density.

In other words, and as already alluded to above, my trouble with the quantum field theory courses as a doctoral student   was that one had to provide a Lagrangian density, and even at the free field level it amounted to invoking the genius of Dirac or that of Maxwell coupled with a set of associated experiments and data. But theoretical physics to me was to be a logical exercise which depended on as few experiments as possible and explained all the relevant phenomena.
 This approach, if it existed, would then 
provide the unifying thread from which wave equations and Lagrangian densities would emerge. The hosts of experimental results would then simply be a logical consequence of this unifying theme and these Lagrangian densities. Additional formal structure would then be built from additional principles, such as that of local gauge invariance and the technology of the $S$-matrix theory would help extract many of the observables.\footnote{Unknown to me for sometime, similar questions were being asked by Steven Weinberg in 1960's. In 1980's when I was to take quantum field theory courses, not too far from Texas A\&M, he was offering quantum field theory courses at the University of Texas at Austin. Had I attended those courses this monograph would not have come to be written. The beauty of his formalism, how he viewed and views quantum field theory, and its seduction was too intense to think that a couple places needed to be explored deeper leading to the results presented in this monograph.} 

Roughly two decades later I have a better understanding as to from where  do the Lagrangian densities come from and how one may evade certain no-go theorems of Weinberg~\citep{Ahluwalia:2016jwz}.  This is the story I want to develop. In the process we shall arrive at several known, and some totally unexpected, results. Very often the manner in which we arrive at the known results shall be significantly different from their presentation elsewhere in the physics literature. The role played by various symmetries shall be transparent, or so I hope. I make significant effort to keep the presentation at a level that makes the presented material easily accessible to the advanced undergraduate students and  the beginning doctoral students of Physics. Some mathematicians may frown upon it but even for them it is hoped that there is, at times, new mathematics, and a lesson in the importance of phase factors in physics.

The first thing, then, is to introduce the notion of symmetry generators for a Lie algebra towards the eventual aim to facilitate bringing Lorentz algebra on the scene. To introduce the said notion it suffices to begin with the rotational symmetry in the familiar landscape of three spatial dimensions. With rotational symmetry in our focus we would not forgo the opportunity to emphasise the unavoidable inevitability  of the quantum structure of the physical reality -- or at least, make it more plausible.

 \section{Generator of a Lie symmetry}
 
 By a Lie symmetry\index{Lie symmetry} we mean a symmetry that depends on a continuous parameter. 
 These are to be distinguished from the discrete symmetries, such as that of parity.
The familiar rotational symmetry in the ordinary space is one simple and important example.  We will use it to introduce the notion of a generator\index{Generator(s)!notion of}
 and make quantum aspect of reality essentially unavoidable.  
 
 Consider a rotation of a frame of reference, or the vector itself (keeping the frame of reference fixed), with the following effect on a vector $\x =(x,y,z)$
 \begin{equation}
 \left(
 \begin{array}{c}
 x^\prime\\
 y^\prime\\
 z^\prime
 \end{array}
 \right) =  \left(\begin{array}{ccc}
 \cos\vartheta & \sin\vartheta & 0\\
 -\sin\vartheta & \cos\vartheta & 0 \\
 0 & 0 & 1
 \end{array}
 \right)  \left(\begin{array}{c}
 x\\
 y\\
 z
 \end{array}\right)
 \end{equation}
 with $0\le \vartheta\le 2 \pi$.
 Denoting the $3\times 3$ matrix that appears above by $R_z(\vartheta)$, \index{Generator(s)!of rotations} we define a generator of rotation $J_z$  through the relation
 \begin{equation}
R_z(\theta) = {\e}^{i J_z \vartheta} \label{eq:def-of-Jz}
\end{equation}
 with $J_z$ a $3\times3$ matrix.
  The knowledge of $R_z(\vartheta)$ uniquely determines
 \begin{equation}
 J_z =\frac{1}{i} \frac{\partial R_z(\vartheta)}{\partial\vartheta}\bigg\vert_{\vartheta = 0} =  i \left( 
 \begin{array}{ccc}
 0 & -1 & 0\\  
 1 &0 &0\\
 0 & 0 & 0
 \end{array}
 \right). \label{eq:jz}
 \end{equation}

The factor of $i$ in the exponential of the definition (\ref{eq:def-of-Jz}) is often omitted by mathematicians. We shall keep the factor of $i$. It makes $J_z$ hermitian. In the process it becomes a candidate for an observable in the quantum formalism.

The interval $dr^2 = dx^2+dy^2+dz^2$, besides other symmetry transformations of the Galilean group, is invariant  not only under the transformation $R_z(\theta)$ but also under two additional transformations associated with rotations
\begin{equation}
R_y(\psi) = 
\left(
\begin{array}{ccc}
\cos\psi & 0 & -\sin\psi \\
0 & 1 & 0\\
\sin\psi & 0 & \cos\psi
\end{array}
\right) \stackrel{\textrm{\textrm{def} \textrm{$J_y$}}}{=}  {\e}^{i J_y\psi}
\end{equation}
and
\begin{equation}
R_x(\phi) = 
\left(
\begin{array}{ccc}
1 & 0 & 0 \\
0 &\cos\phi &\sin\phi \\
0 & -\sin\phi & \cos\phi
\end{array}
\right) \stackrel{\textrm{def $J_x$}}{=}  {\e}^{i J_x \phi}
\end{equation}
leading to the associated generators of the rotations
 \begin{equation}
 J_y = \frac{1}{i} \frac{\partial R_y(\psi)}{\partial\psi}\bigg\vert_{\psi= 0} = i \left( 
 \begin{array}{ccc}
 0 & 0 & 1\\
 0 &0 &0\\
 -1 & 0 & 0
 \end{array}
 \right)  \label{eq:jy}
 \end{equation}
and 
 \begin{equation}
 J_x =\frac{1}{i} \frac{\partial R_x(\phi)}{\partial\phi}\bigg\vert_{\phi = 0} = i \left( 
 \begin{array}{ccc}
 0 & 0 & 0\\
 0 &0 &-1\\
 0 & 1 & 0
 \end{array}
 \right). \label{eq:jx}
 \end{equation}
 This is very elementary. But it allows us to introduce the important notion of generators of Lie symmetries in a familiar landscape. For the moment we refrain from studying boosts and spacetime translations.

\section{A beauty of abstraction and a hint for the quantum nature of reality}\index{Abstraction!its beauty}\label{sec:quantum}

 Now comes an unreasonable beauty of abstraction.
 Since the matrices do not commute, the generators of rotations given by equations (\ref{eq:jz}),  (\ref{eq:jy}), and  (\ref{eq:jx})  satisfy the algebraic relationship, or simply the algebra (or, Lie algebra)\index{Lie algebra!associated with rotations}
 \begin{equation}
 [J_x,J_y] = i J_z,\quad\mbox{and cyclic permutations}.\label{angular-momentum-algebra-in-3D}
 \end{equation}
 The abstraction, well known to physicists and mathematicians, consists of the following.  Let us momentarily forget how this algebra has arisen, and instead marvel at the infinitely many solutions that exist for this algebra. Each of these solutions is called a representation of the algebra,\index{Representation!of an algebra} and the spaces on which the elements of the representation act are called  representation spaces.\index{Representation!space} 

All of us know that (\ref{angular-momentum-algebra-in-3D}) also follows from the fundamental commutator, but we now ask: Does the fundamental commutator follow from rotational symmetry?

Continuing the thread, not only are there finite-dimensional  matrix representations of (\ref{angular-momentum-algebra-in-3D}), infinitely many of them (this thought continues in chapter~\ref{ch4}),  but there is also  an infinite-dimensional representation that is made of differential operators, with
\begin{align}
&J_x = \frac{1}{i}\left( y\frac{\partial}{\partial{z}} - z\frac{\partial}{\partial{y}} \right),\quad
J_y = \frac{1}{i}\left( z\frac{\partial}{\partial{x}} - x\frac{\partial}{\partial{z}} \right),\label{eq:jxjy}\\
&J_z = \frac{1}{i}\left( x\frac{\partial}{\partial{y}} - y\frac{\partial}{\partial{x}} \right).\label{eq:jzz}
\end{align}
Many students of Physics encounter this result in the context of their first quantum mechanics course. Starting with the Heisenberg fundamental commutator, they are taught that taking the classical definition of the angular momentum and replacing the position and momentum  by their quantum counterparts, one obtains (\ref{angular-momentum-algebra-in-3D}), followed by (\ref{eq:jxjy})  and (\ref{eq:jzz}).

Given our discussion above, we ask the question: Does quantum aspect of reality spring from rotational symmetry and the implicit assumption of a continuous spacetime? 
The answer we adopt would then tell us if the quantum gravity  spacetime would be discrete, not covered by Lorentz algebra and what modifications would the Heisenberg algebra and the deBroglie wave particle duality suffer. This view is supported by the Kempf, Mangano, and Mann~\citep{Kempf:1994su} and by my own 
work~\citep{Ahluwalia:2000iw,Ahluwalia:1993dd}.

 Taking this view has the consequence that quantum mechanical foundations are seen as an inevitable consequence of the rotational symmetry. The primary result is the Heisenberg algebra\index{Heisenberg algebra}, and from that follows the secondary result: the de Broglie wave particle duality
We thus rewrite (\ref{eq:jxjy}) and (\ref{eq:jzz}) as
\begin{align}
&J_x = \frac{1}{\hbar}\times\frac{\hbar}{i}\left( y\frac{\partial}{\partial{z}} - z\frac{\partial}{\partial{y}} \right),\quad
J_y =  \frac{1}{\hbar}\times\frac{\hbar}{i}\left( z\frac{\partial}{\partial{x}} - x\frac{\partial}{\partial{z}} \right),\\
&J_z =  \frac{1}{\hbar}\times\frac{\hbar}{i}\left( x\frac{\partial}{\partial{y}} - y\frac{\partial}{\partial{x}} \right).
\end{align}
with $\hbar = h/2\pi$ as the reduced Planck's constant. Identifying 
\begin{equation}
x\quad \mbox{and} \quad \frac{\hbar}{i}\frac{\partial}{\partial x},\quad
y\quad \mbox{and} \quad \frac{\hbar}{i}\frac{\partial}{\partial y},\quad
z\quad \mbox{and} \quad \frac{\hbar}{i}\frac{\partial}{\partial z}\quad
\end{equation}
as position and momentum operators naturally leads to the Heisenberg's fundamental commutators
\begin{equation}
\left[x,p_x\right] = i \hbar,\quad \left[y,p_y\right] = i \hbar,\quad \left[z,p_z\right] = i \hbar
\label{eq:fundamentalC}
\end{equation}
with $[x,y]=0$, etc. Now the  eigenfunctions of the momentum operator 
\begin{equation}
\p = \frac{\hbar}{i}\,\nabla
\end{equation}
have the form 
\begin{equation}
\exp\left(i\, \frac{\p^\prime\cdot \x^\prime}{\hbar}\right)\label{eq:eigenf}
\end{equation}
where $\p^\prime$ and $\x^\prime$ denote eigenvalues of the operators $\p$ and $\x$. We find this Dirac-Schwinger notation useful but only invoke it when an ambiguity is likely to arise.
The eigenfunctions~(\ref{eq:eigenf}) have the spatial periodicity
\begin{equation}
\lambda =\frac{h}{p}
\end{equation}
with $p=\vert\p^\prime\vert$. We are thus required that we  associate a wave length $\lambda$ with momentum $p$ \`a la de Broglie.\footnote{We shall often set $\hbar$ and the speed of light $c$ to be unity.} 

In this interpretation the foundations of classical mechanics are at odds with the rotational symmetry and echo considerations related to the  lack of stability of the algebra underling classical mechanics~\citep{Flato:1982yu,Faddeev:1989LD,VilelaMendes:1994zg,Chryssomalakos:2004gk}.

To emphasise  we repeat that the argument of the conventional  courses
on quantum mechanics begins with the fundamental commutator and results in the `angular momentum commutators.' It is generally not realised that the de Broglie's $\lambda = h/p$ is a direct consequence of the Heisenberg's fundamental commutator. Whereas here we reverse that argument and see the fundamental commutator, and hence the wave particle duality, as a consequence of rotational symmetry. It makes it clear, or opens a discussion, that the classical description of reality is incompatible with the rotational symmetry. 
Quantum aspect of realty thus seem to spring from the rotational symmetry of the space in which events occur. 

 The presence of the momentum operator in the Heisenberg algebra   introduces kinetic energy in the measurement process. It induces inevitable gravitational effects through the  modification of the local curvature. These effects make position measurements 
 non-commutative~\citep{Ahluwalia:1993dd,Doplicher:1994zv}.
 The implicit assumption of the commuting position operators of the arrived at quantum nature of reality must, therefore, undergo  a modification at the Planck scale where gravitational effects become important. But rotational symmetry seems to make the fundamental commutator inevitable. This paradoxical circumstance can be averted if  we entertain the possibility that quantum-gravity requires a non-commutative spacetime accompanied by a modified Heisenberg algebra.

In the argument above we can easily take $\hbar$ as some unknown constant with the dimension of angular momentum and then obtain its chosen identification by, say, looking at the spectrum of a system governed by the 
 hamiltonian 
$
H = {\p^2}/{2 m} + \left({1}/{2}\right) m \omega^2 \x^2
$.
The fundamental commutator yields the energy spectrum of such systems to be equidistant lines with separation $\hbar\omega$, and a zero point energy of 
$\frac{1}{2} \hbar\omega$.


 \section{A unification of the microscopic and the macroscopic} 
 
 All this is not mathematical science fiction.\index{Earth!its stability}
 Because of its simplicity and its manifest importance, it is good to recall that the stability of  the Earth beneath us and Avogadro number of primal entities in a palmful of water speak of it. Take the simplest of the simple systems, the hydrogen atom. Classically, if one minimises the energy, $E=\left(p^2/2m\right) - e^2/r$,
 of the hydrogen atom
 then one immediately sees that the energy is minimised at $r=0$. The atom collapses to size zero. In contrast the fundamental commutator in (\ref{eq:fundamentalC}) requires that $E$ be minimised to the constraint $r\, p\sim\hbar$. To implement this constraint one may set $r \sim \hbar/p$  and get
 $E = \left(p^2/2m\right) - e^2 p/\hbar$.
 Setting, $\partial{E}/{\partial p} =0$, then gives the $E$-minimising  $p$,
 $p_0 \sim m e^2/\hbar$,
 for which $E$ takes its minimum value
 $E_0 \sim - m e^4/2 \hbar^2 \approx -13.6 ~\mbox{eV}$.
The $r$ instead of collapsing to zero, is now constrained to
$
r_0 =\hbar^2/{m e^2} = 0.5 \times 10^{-8} ~\textrm{cm}
$. 
 It is a good measure of the size of most atoms.  The reason being that for heavier nuclei with atomic number $Z$, for the outermost electron the $(Z-1)$ electrons screen the nucleus in such a way that the effective nuclear charge remains  $e$~\citep{Weinberg:2012qm}.
 In this simple manner we not only understand the origin of the 
 ionisation energy of the hydrogen atom but we also obtain the 
 order of magnitude for the
Avagadro's number\index{Avogadro number} ($\sim (1/r_0)^3$).

These simple argument unify the microscopic, the hydrogen atom, with the macroscopic, the Earth. Beyond the stability of the planets, and burning of the stars, quantum nature of reality  leaves its imprints in the entire cosmos and to possible physics beyond. Beyond, where rotational symmetry either ceases to be, or takes a new -- not yet known -- form in discreteness of spacetime~\citep{Padmanabhan:2016eld}.

\section{Lorentz algebra}\label{sec:LorentzAlgebra}

On the planet Earth, the Lorentz algebra is usually arrived at by considering the transformations of the spatial and temporal specifications of events.
Elsewhere in the cosmos
one may arrive at the very same algebra by studying the spectrum of hydrogen atom, instead through the null result of the terrestrially famous 1887 experiment of Michelson and 
Morley~\citep{Michelson:1887zz}.

The folklore that the 1887 experiment requires Lorentz symmetries is, strictly speaking, not true~\citep{Cohen:2006ky}. The Lorentz symmetries follow only if one requires in addition any of the four discrete symmetries. One of these discrete  symmetries is  Parity~\textendash~a symmetry known to be violated in the electroweak interactions~\citep{Lee:1956qn,Wu:1957my}.
Remaining three are: time reversal, and charge conjugation conjugated with parity, or time reversal. 

The most familiar way to arrive at the Lorentz algebra is to simply note that absolute space and absolute time are now empirically untenable. While a rotation, say about the $z$-axis, \textit{does not} mix time and space (units: speed of light is now taken as unity)
\begin{equation}
 \left(
 \begin{array}{c}
 t^\prime\\
 x^\prime\\
 y^\prime\\
 z^\prime
 \end{array}
 \right) =  \left(\begin{array}{cccc}
 1& 0 & 0 & 0\\
0&  \cos\vartheta & \sin\vartheta & 0\\
0&  -\sin\vartheta & \cos\vartheta & 0 \\
0&  0 & 0 & 1
 \end{array}
 \right)  \left(\begin{array}{c}
 t\\
 x\\
 y\\
 z
 \end{array}\right).
 \end{equation}
 a boost, say along the $x$-axis \textit{does}
 \begin{equation}
 \left(
 \begin{array}{c}
 t^\prime\\
 x^\prime\\
 y^\prime\\
 z^\prime
 \end{array}
 \right) =  \left(\begin{array}{cccc}
 \cosh\varphi&\sinh\varphi& 0 & 0\\
\sinh\varphi&  \cosh\varphi & 0 & 0\\
0 & 0 & 1&0 \\
0&  0 & 0 & 1
 \end{array}
 \right)  \left(\begin{array}{c}
 t\\
 x\\
 y\\
 z
 \end{array}\right)\label{eq:boost-x}
 \end{equation}
 where the rapidity parameter,\index{Rapidity!parameter} $\vp = \varphi\,\widehat \p$, is defined as\index{Rapidity parameter}
 \begin{align}
 &\cosh\varphi = E/m = \gamma = 1/\sqrt{1-v^2}\nonumber\\
 &\sinh\varphi = p/m = \gamma v\label{eq:rapidity-parameter}
 \end{align}
 with all symbols carrying their usual meaning. Denoting the $4\times 4$ boost matrix in (\ref{eq:boost-x}) by $B_x(\varphi)$. The generator of the boost along the $x$-axis is thus\index{Generator(s)!of boosts and rotations}
 \begin{equation}
 K_x= \frac{1}{i}\frac{\partial B_x(\varphi)}{\partial\varphi}\bigg\vert_{\varphi=0} = 
 - i \left(
 \begin{array}{cccc}
 0 & 1 & 0 & 0\\
 1 & 0 & 0 & 0\\
 0 & 0 & 0 &0\\
 0 & 0 & 0 & 0
 \end{array}
 \right).\label{eq:kx}
 \end{equation}
 This is complemented by the remaining two generators of the boosts 
  \begin{equation}
 K_y = -i \left(
 \begin{array}{cccc}
 0 & 0 & 1 & 0\\
 0 & 0 & 0 & 0\\
 1 & 0 & 0 &0\\
 0 & 0 & 0 & 0
 \end{array}
 \right),\quad K_z = -i \left(
 \begin{array}{cccc}
 0 & 0 & 0 & 1\\
 0 & 0 & 0 & 0\\
 0 & 0 & 0 &0\\
 1 & 0 & 0 & 0
 \end{array}
 \right)\label{eq:kykz}
 \end{equation}
 and the three generators of the rotations.  These are directly read off from our work earlier with the added observation that under rotations $t$ and $t^\prime$ are identical
 \begin{equation}
 J_x= - i \left(
 \begin{array}{cccc}
 0 & 0 & 0 & 0\\
 0 & 0 & 0 & 0\\
 0 & 0 & 0 &1\\
 0 & 0 & -1 & 0
 \end{array}
 \right),\label{eq:jx-new}
 \end{equation}
  \begin{equation}
 J_y = -i \left(
 \begin{array}{cccc}
 0 & 0 & 0 & 0\\
 0 & 0 & 0 & -1\\
 0 & 0 & 0 &0\\
 0 & 1 & 0 & 0
 \end{array}
 \right),\quad J_z = -i \left(
 \begin{array}{cccc}
 0 & 0 & 0 & 0\\
 0 & 0 & 1 & 0\\
 0 & -1 & 0 &0\\
 0 & 0 & 0 & 0
 \end{array}
 \right).\label{eq:jyjz-new}
 \end{equation}

\noindent
The six generators satisfy the following algebra\index{Lorentz algebra}
\begin{align}
& \left[J_x,J_y\right] = i J_z\quad\textrm{and~cyclic~permutations}\label{eq:rotational-sub-algebra}\\
& \left[K_x,K_y \right] = - i J_z,\quad\textrm{and~cyclic~permutations}\\
& \left[J_x, K_x\right] = 0,\quad\textrm{etc.} \\
&\left[J_x,K_y\right] = i K_z,\quad\textrm{and~cyclic~permutations}.
\end{align}
It is named Lorentz algebra.\index{Lorentz algebra} 

\section[Further abstraction: Un-hinging the Lorentz algebra \ldots]{Further abstraction: Un-hinging the Lorentz algebra from its association with Minkowski spacetime}

To underline the mysteries of nature coded in the algebra just arrived at let us take the liberty of imagining that we forget how we arrived at this algebra. Each civilisation in the cosmos, sooner or later, is likely to arrive at this truth, this reflection of low-energy reality, in one way or another. Some in a  way similar to ours, others in ways different, even ways we have not yet dreamed of. Lorentz algebra is a powerful unifying element  in unearthing the nature of reality. Its various aspects thread through this monograph, with Physics as our primary focus.

We thus symbolically unhinge the Lorentz algebra from spacetime symmetries and express it as an abstract reality expressed through the following abstraction
\beq
\J\rightarrow \mathfrak{J},\quad \K\rightarrow \mathfrak{K}\label{eq:abstractionJ}
\eeq
such that $\mathfrak{J}$ and $\mathfrak{K}$ are no longer confined to being identified with the $4\times 4$ matrices given in (\ref{eq:kx}) to (\ref{eq:jyjz-new}), or with their unitarily transformed expressions, but still satisfy 
\begin{align}
&\left[\mathfrak{J}_x,\mathfrak{J}_y \right]= i \mathfrak{J}_z,\quad\mbox{and cyclic permutations}\label{eq:sub-new}\\
&\left[\mathfrak{K}_x,\mathfrak{K}_y\right] = - i\mathfrak{J}_z,\quad\mbox{and cyclic permutations}\\
&\left[\mathfrak{J}_x,\mathfrak{K}_x \right]=0,  \quad\mbox{etc.}\\
&\left[\mathfrak{J}_x,\mathfrak{K}_y \right]= i \mathfrak{K}_z,\quad\mbox{and cyclic permutations}.
\label{eq:la-new}
\end{align}\index{Lorentz algebra}
The ${\mathfrak{J}}_i$ and $\mathfrak{K}_j$, $i,j=x,y,z$, represent generators of rotations and boosts -- in an abstract space.  Their exponentiations
\begin{equation}
 \exp\left(i\bmfj\cdot\bm{\theta}\right), \quad \exp\left(i\bmfk\cdot\vp\right)
 \label{eq:exponentiation}
\end{equation}
give the group transformations\index{Group transformation} under rotations and boosts for the `vectors' spanning the associated representation space. While the underlying algebraic structure remains the same, each of the group transformations -- chosen by a specific choice of $\bmfj$ and $\bmfk$ in~(\ref{eq:exponentiation}) --
defines a new group. These group transformations act on vectors that inhabit the associated representation spaces. Upto a convention related freedom of a unitary transformation, the transformations of four vectors, to transformations of Dirac spinors, to `other vectors' in infinitely many other representation spaces, are all obtained 
from~(\ref{eq:exponentiation}).


To cast the Lorentz algebra in a manifestly covariant form,  we recall the  definition of the three-dimensional Levi-Civita symbol
\beq
\epsilon_{ijk} = \begin{cases}
+1 & \textrm{if $(ijk)$ is an even permutation of $(123)$} \\
-1 & \textrm{if $(ijk)$ is an odd permutation of $(123)$} \\
0 & \textrm{otherwise}
\end{cases}
\eeq
That is, if any two indices are equal the $\epsilon$ symbol vanishes. If all the indices are unequal, we have
\beq
\epsilon_{ijk} = (-)^p \epsilon_{123}
\eeq
where p, known as the parity of the permutation, is the number of interchanges of indices necessary to transmute $ijk$  into the order $123$. The factor $(-1)^p$ is called the signature of the permutation. Equipped with the  Levi-Civita symbol we can define a completely antisymmetric 
 operator 
 \beq
\mathfrak{J}_{\mu\nu} =  
\begin{cases}
\mathfrak{J}_{ij} =  - \mathfrak{J}_{ji} = \epsilon_{ijk} \mathfrak{J}_k \\
\mathfrak{J}_{i0} =  -\mathfrak{J}_{0i}  = - \mathfrak{K}_i
\end{cases}
\eeq
with the greek indices $\mu$ and $\nu$ taking the values $0,1,2,3$, and the latin indices confined to the values $1,2,3$. We follow the Einstein convention. It assumes repeated indices are summed.

These definitions can be used to cast the Lorentz algebra in the following two equivalent form:\index{Lorentz algebra}
\beq
\left[\mathfrak{J}_{\mu\nu},\mathfrak{J}_{\rho\sigma}\right] = i\left(\eta_{\nu\rho} \mathfrak{J}_{\mu\sigma} 
- \eta_{\mu\rho} \mathfrak{J}_{\nu\sigma} + \eta_{\mu\sigma} \mathfrak{J}_{\nu\rho}
-\eta_{\nu\sigma} \mathfrak{J}_{\mu\rho} 
\right)
\eeq
where $\eta_{\mu\nu} = \textrm{diag}(1,-1,-1,-1)$ is the spacetime metric,\index{Spacetime metric} 
and
\beq
\left[\mathfrak{J}_i,\mathfrak{J}_j\right] = i \epsilon_{ijk} \mathfrak{J}_k,\quad  \left[\mathfrak{J}_i,\mathfrak{K}_j\right]  =  i \epsilon_{ijk} \mathfrak{K}_k,\quad
\left[\mathfrak{K}_i,\mathfrak{K}_j\right] = -i \epsilon_{ijk} \mathfrak{J}_k 
\label{eq:LorentzAlgebra}
\eeq


The $\J$ and $\K$ of section~\ref{sec:LorentzAlgebra} constitute the most familiar representation. The associated representation space for historical reasons is called Minkowski spacetime (or, generally the space of four vectors).




\chapter{Representations of Lorentz  Algebra}\label{ch4}

\section{Poincar\'e algebra, mass, and spin}

For the purposes of a physicist a `representation of an algebra' is simply a specific solution to the symmetry algebra under consideration. 

Thus the six $4\times 4$ matrices given in equations
 (\ref{eq:kx}) to (\ref{eq:jyjz-new}) form a representation of the Lorentz algebra. It is a finite dimensional representation\index{Finite dimensional representation}, and there are infinitely many of them. We will start considering them in the next section.
 
On the other hand, the set of generators
  \begin{align}
&J_x = \frac{1}{i}\left( y\frac{\partial}{\partial{z}} - z\frac{\partial}{\partial{y}} \right),\quad
J_y = \frac{1}{i}\left( z\frac{\partial}{\partial{x}} - x\frac{\partial}{\partial{z}} \right),\label{eq:diffa}\\
&J_z = \frac{1}{i}\left( x\frac{\partial}{\partial{y}} - y\frac{\partial}{\partial{x}} \right)\label{eq:diffb}
\end{align}
 coupled with 
 \begin{align}
&K_x = {i}\left( t\frac{\partial}{\partial{x}} + x\frac{\partial}{\partial{t}} \right),\quad
K_y = {i}\left( t\frac{\partial}{\partial{y}} + y\frac{\partial}{\partial{t}} \right),\label{eq:diffc}\\
&K_z = {i}\left( t\frac{\partial}{\partial{z}} + z\frac{\partial}{\partial{t}} \right).\label{eq:diffd}
\end{align}
 provide   an infinite dimensional representation\index{Infinite dimensional representation} of the same very algebra -- the Lorentz albegra. 

In the context of Minkowski space, Nature also supports the symmetry induced by the
spacetime translations
\begin{equation}
x^\mu\rightarrow x^{\prime\mu} = {\Lambda^\mu}_\nu x^\nu + a^\mu
\end{equation}
generated by
\begin{equation}
P_\mu = i\frac{\partial}{\partial x^\mu} \label{eq:sttr}
\end{equation}
In the above we have defined ${\Lambda^\mu}_\nu$, the transformations of the Lorentz 
group\index{Lorentz group! transformations}, as follows
\begin{align}
{\Lambda^\mu}_\nu =
\begin{cases}
 {\Big[\exp(i\J\cdot \bm{\vt})\Big]^\mu}_\nu, & ~~~~~ \mbox{for rotations}\\ \\
 {\Big[\exp(i\K\cdot \bm{\varphi})\Big]^\mu}_\nu, & ~~~~~\mbox{for boosts}
 \end{cases}
\end{align}
with $\J$ and $\K$ given by equations~(\ref{eq:kx}-\ref{eq:jyjz-new}) and take $a^\mu$ as a constant four vector.

When one adjoins these four generators, encoded in (\ref{eq:sttr}),  
to the six generators of rotations and boosts (\ref{eq:diffa})-(\ref{eq:diffd}) one obtains the 10 generators of the Poincar\'e algebra:\index{Poincar\'e algebra}
\begin{align}
& \left[{J}_{\mu\nu},{J}_{\rho\sigma}\right] = i\left(\eta_{\nu\rho} {J}_{\mu\sigma} 
- \eta_{\mu\rho} {J}_{\nu\sigma} + \eta_{\mu\sigma} {J}_{\nu\rho}
-\eta_{\nu\sigma} {J}_{\mu\rho} 
\right) \\
& \left[P_\mu, J_{\rho\sigma}\right] = i\left( \eta_{\mu\rho} P_\sigma - \eta_{\mu\sigma} P_\rho\right) \\
& \left[P_\mu,P_\nu\right] =0 .
\end{align}
It is remarkable that the fundamental notions of mass and spin arise from these symmetries.
To see this, we introduce two `Casimir' operators\index{Poincar\'e algebra! Casimir invariants} \index{Casimir operators}
\begin{equation}
C_1 = P_\mu P^\mu, \quad C_2 = W_\mu W^\mu
\end{equation}
with the Pauli Luba\'nski pseudovector defined as~\citep{Lubanski:1942jk} 
\begin{equation}
W_\mu \stackrel{\text{def}}{=} \frac{1}{2 m}\epsilon_{\mu\nu\rho\sigma}J^{\nu\rho} P^\sigma
\end{equation} \index{Pauli Luba\'nski pseudovector}
The `extra' factor of $1/m$ in the above definition is consistent with 
 Luba\'nski's original paper. In considering the massless case we may introduce a related operator that is more in keeping with its later re-defintion.
  Now because the commutator $\left[P^\sigma,
 P^\mu \right]$ vanishes, the projection of $W_\mu$ on the generators of spacetime translations vanishes
 \begin{equation}
 W_\mu P^\mu =0
 \end{equation}
 because the $\epsilon$ symbol is antisymmetric under the interchange of the indices $\sigma,\mu$ and $P^\sigma P^\mu$ is symmetric under the interchange of the same indices. It is then a few lines of exercise to 
 show that (see, for example,~\citep{Tung:1985na} or~\citep{Ryder:1985wq})
\begin{equation}
\left[W^\mu,P^\mu\right] = 0
\end{equation}
with the consequence that not only $W_\mu W^\mu$ is invariant under the Lorentz transformation but also under spacetime translations. And thus it commutes with all the ten generators of the Poincar\'e algebra. A parallel of the same argument tells us that $P_\mu P^\mu$ also commutes with all the ten generators of the Poincar'e algebra. 
The eigenvalues of the latter are identified with the square of the mass parameter $m$, and modulo a sign the eigenvalues of the former can be easily identified with the eigenvalues of the square of the generators of rotations: 
$- s(s+1)$.\footnote{In making this identification with $\J^2$ we implicitly assume the abstraction explicitly noted in equation (\ref{eq:abstractionJ}) that allows $s$ to take half integral and integral values.} It is for these reasons that mass and spin\index{Mass and spin} arise in the description of the physical reality. 
Once one inertial observer measures them,
all inertial observers related by Poincar\'e symmetries measure the same value. \index{Poincar\'e algebra!mass and spin}

We generally assume that electron has the same `mass' and same `spin,' here in our solar system as in a far away distant galaxy going back to the dawn of our universe -- despite the fact that in earlier epochs Poincar\'e symmetries may have suffered a modification. But perhaps where and when these departures occur we enter another realm, the realm of massless particles, with hypothetical massless observers. These observers have no rest frame, and to us they lie in a realm where spacetime dimensionality changes, and time as we generally define does not exist. Such a dramatic change arises from singularities of length contraction and time dilation for truly massless particles. 

\noindent
\subsection{A cautionary remark}\index{Spin!a cautionary remark}

 Except in the rest frame, the $s(s+1)$ encountered above is not an eigenvalue of $\bmfj^2$. It is $C_2$ whose eigenvalues remain invariant under Poincar\'e transformations and not the eigenvalues of 
$\bmfj^2$ except  in the rest frame -- or, in an accidental situation to be discussed below.

\section{Representations of Lorentz algebra\label{sec:representations-of-lorentz-algebra}}

To identify various representations of Lorentz algebra one often starts with introducing two $\mathfrak{su\left(2\right)}$ generators:
\begin{equation}
\bmfa =\frac{1}{2}\left(\bmfj+ i \bmfk\right),\quad \bmfb =\frac{1}{2}\left(\bmfj- i \bmfk\right) 
\end{equation}
and studies representations of $\mathfrak{su\left(2\right)}\otimes\mathfrak{su\left(2\right)}$, keeping in mind that
\begin{align}
&\left[\mathfrak{A}_x, \mathfrak{A}_y\right] = i \mathfrak{A}_z,\quad\textrm{and~cyclic~permutations}\\
&\left[\mathfrak{B}_x, \mathfrak{B}_y\right] = i \mathfrak{B}_z,\quad\textrm{and~cyclic~permutations}\\
&\left[\mathfrak{A}_i,\mathfrak{B}_j\right] = 0,\quad i,j=x,y,z. \label{eq:two-su(2)s}
\end{align}
In this way one labels the resulting representation space by two labels $(a,b)$, with $a(a+1)$ and $b(b+1)$ being eigenvalues of $\bmfa^2$ and $\bmfb^2$, respectively. We find this complexification of the generators unnecessary. 
I mention it here because our reader may encounter it in her/his studies often.
If due interpretational caution is not exercised all this can cause confusion as $\bmfj^2$ too is not a Casimir invariant of the Lorentz algebra (as just cautioned above), but only of the rotational subalgebra represented by the first of the equations in~(\ref{eq:LorentzAlgebra}). Introducing two $\mathfrak{su\left(2\right)}$ algebras through complexifications of the generators~\textendash~when one is permitted only real linear combination~\textendash~strictly speaking, takes us away from the Lorentz algebra and it is avoidable.

 \centerline{--------------}

To avoid conceptual confusion we do not follow this potentially misleading traditional approach but instead follow a straight forward method.
For this we note that once a representation of $\bmfj$ is found\footnote{To find explicit form of $\bmfj$ 
a junior reader may consult 
any good book on quantum mechanics ranging from Dirac's classic~\citep{Dirac:1930pam}, to recent lectures by 
Weinberg~\citep{Weinberg:2012qm}, or to~a very pedagogically written two volume set by 
Cohen-Tannoudji et al.~ \citep{Cohen-Tannoudji:1977qm}.}
that satisfies the first part of~(\ref{eq:LorentzAlgebra}) an inspection of the remainder of the Lorentz algebra shows that for each representation of $\bmfj$ there exist \textit{two} independent primordial representations for the boost generators
\begin{equation}
\bmfk = - i \bmfj,\quad \bmfk= + i\bmfj
\end{equation}
The sets 
\begin{equation}
\underbrace{\bmfj, \bmfk= - i\bmfj}_{\mathcal{R}~\textrm{type}}\quad{\textrm{and}}\quad\underbrace{ \bmfj,\bmfk= i\bmfj}_{\mathcal{L}~\textrm{type}} \label{eq:primordial-generators}
\end{equation}
provide two independent representations of the Lorentz algebra. 
As indicated, we call these representaions as $\mathcal{R}$ type and $\mathcal{\mathcal L}$ type, respectively. The reason for this nomenclature shall become apparent in section~\ref{sec:RL} below. 

Without the existence of two, in contrast to one, representations for each $\bmfj$ one would not be able to respect causality in quantum field theoretic formalism respecting Poincar\'e symmetries, or have massive antiparticles required to avoid causal paradoxes. And thus this doubling of representations is more than an oddity, or a coincidence. It is a reflection a deep underlying thread of symmetries, causality, and degrees  of freedom encountered in quantum fields.

In accordance with (\ref{eq:exponentiation}), the associated symmetry transformations are implemented as follows
\begin{equation}
\mathcal{R:}
\begin{cases}
\textrm{Rotations~by}: & \exp[i\bmfj\cdot{\boldmath\boldsymbol{\vt}}] \label{eq:R-type-rb} \\
\textrm{Boosts~by}: & \exp[ i (-i\bmfj)\cdot{\boldmath\boldsymbol{\varphi}}] = \exp[\bmfj\cdot{\boldmath\boldsymbol{\varphi}}] 
\end{cases}
\end{equation}
and 
\begin{equation}
\mathcal{L:}
\begin{cases}
\textrm{Rotations~by}: & \exp[i\bmfj\cdot{\boldmath\boldsymbol{\vartheta}}] \\
\textrm{Boosts~by}: & \exp[ i (i\bmfj)\cdot{\boldmath\boldsymbol{\varphi}}] = \exp[-\bmfj\cdot{\boldmath\boldsymbol{\varphi}}] \label{eq:L-type-rb}
\end{cases} 
\end{equation} 
For these primordial representations the knowledge of $\bmfj$ completely determines the rotation, parameterised  by ${\boldmath\boldsymbol{\vartheta}}$,  and the boost, parameterised by ${\boldmath\boldsymbol{\varphi}}$, transformations.

The boosts do not carry the imaginary $i$ in the exponentiation. It is not that we have suddenly changed to the convention of the mathematics literature but because the $i$ in the exponentiation for the boost operator  when encountered with the $\pm i$ in the expression for $\mathfrak{K}$ in  (\ref{eq:primordial-generators}) results in $\mp 1$.~I make this parenthetic note because again and again in my seminars I encounter a question to this effect.              
\vspace{7pt}

  \subsection{Notational remark}
  
  In the language of the traditional approach of  labelling representation spaces,
   the $\mathcal{R}$ and $\mathcal{L}$ stand for the $(j,0)$ and $(0,j)$ representation spaces, respectively:
   $\mathcal{R}\leftrightarrow (j,0),
   \mathcal{L}\leftrightarrow (0,j).$

 \subsection{Accidental Casimir} \index{Accidental Casimir}
 
For the primordial representations introduced above \textendash~ despite the general remark made above about $\bmfj^2$ after 
equations~(\ref{eq:two-su(2)s})~ \textendash~$\bmfj^2$ does commute with all the generators given in~(\ref{eq:primordial-generators}). For this reason, in the restricted context of the $\mathcal{R}$ and $\mathcal{L}$ representation spaces, we introduce the term `accidental Casimir operator' for $\bmfj^2$. Its invariant eigenvalue $s(s+1)$ coincides with that of the $C_2$ in all inertial frames.

\section{Simplest representations of Lorentz algebra\label{sec:RL}}

With our focus on (\ref{eq:R-type-rb}) and (\ref{eq:L-type-rb}), the simplest non trivial $\bmfj$ that satisfies the Lorentz algebra (\ref{eq:LorentzAlgebra}) is formed from the Pauli  matrices\index{Pauli matrices}
\begin{equation}
\sigma_x=\left(
\begin{array}{cc}
0 & 1 \\
1 &0
\end{array}
\right),\quad 
\sigma_y=\left(
\begin{array}{cc}
0 & -i \\
i &0
\end{array}
\right),\quad
\sigma_z=\left(
\begin{array}{cc}
1 & 0 \\
0 &-1
\end{array}
\right)\label{eq:Pauli-Matrices}
\end{equation}
and is given by
\begin{equation}
\bmfj = \s/2.
\end{equation}
Through~(\ref{eq:primordial-generators}), it introduces two independent  representations spaces of spin one half.

Referring to equations~(\ref{eq:R-type-rb}) and (\ref{eq:L-type-rb}) the group transformations for rotations and boosts for these representation spaces are given by\index{Weyl spinors!rotations and boosts}
\begin{align}
& R_{\mathcal{R},\mathcal{L}}({\boldmath\boldsymbol{\vartheta}})  = \exp\left[{i \frac{{\boldmath\boldsymbol{\sigma}}}{2}\cdot{\boldmath\boldsymbol{\vartheta}} }\right] \label{eq:rotations-for-spin-1/2} \\
 & B_\mathcal{R,L} (\vp)  =\exp\left[\pm {\frac{{\boldmath\boldsymbol{\sigma}}}{2}\cdot{\boldmath\boldsymbol{\varphi}}} \right]
 \label{eq:boosts-for-spin-1/2}
\end{align}
where the upper sign is for the $\mathcal{R}$ type boosts, while the lower sign is for the 
 $\mathcal{L}$ type boosts. The  expression for the transformation for rotation is the same for both type of primordial representations.

Writing ${\boldmath\boldsymbol{\vartheta}} = \vartheta \,{\boldmath\boldsymbol{\widehat{n}}}$, with $\vartheta$ representing the angle of rotation and ${\boldmath\boldsymbol{\widehat{n}}}$ denoting a unit vector along the axis of rotation, the transformation for rotations (\ref{eq:rotations-for-spin-1/2}) takes the form
\begin{align}
R_{\mathcal{R},\mathcal{L}}({\boldmath\boldsymbol{\vartheta}}) & = \I + i \s\cdot{\boldmath\boldsymbol{\widehat{n}}}\frac{\vartheta}{2} -\frac{1}{2!}
\left(\s\cdot{\boldmath\boldsymbol{\widehat{n}}}\right)^2\left(\frac{\vartheta}{2}\right)^2
- i \frac{1}{3!}
\left(\s\cdot{\boldmath\boldsymbol{\widehat{n}}}\right)^3\left(\frac{\vartheta}{2}\right)^3 \nonumber\\
& \hspace{21pt} +\frac{1}{4!}
\left(\s\cdot{\boldmath\boldsymbol{\widehat{n}}}\right)^4\left(\frac{\vartheta}{2}\right)^4 
+ i\frac{1}{5!}
\left(\s\cdot{\boldmath\boldsymbol{\widehat{n}}}\right)^5\left(\frac{\vartheta}{2}\right)^5 + \ldots 
\end{align}
Using the identity $\left(\s\cdot{\boldmath\boldsymbol{\widehat{n}}}\right)^2 = \I$, the above expansion reduces to 
\begin{align}
& R_{\mathcal{R},\mathcal{L}}({\boldmath\boldsymbol{\vartheta}})  =\bigg[1-  \frac{1}{2!} \left(\frac{\vartheta}{2}\right)^2
+\frac{1}{4!} \left(\frac{\vartheta}{2}\right)^4 -\ldots
\bigg]\I \nonumber\\
 &\hspace{51pt}+ i \s\cdot{\boldmath\boldsymbol{\hat{n}}} 
\bigg[ \frac{\vartheta}{2} -\frac{1}{3!}\left( \frac{\vartheta}{2} \right)^3
+
\frac{1}{5!}
\left( \frac{\vartheta}{2} \right)^5-\ldots
\bigg]
\end{align}
and simplifies to yield
\begin{equation}
R_{\mathcal{R},\mathcal{L}}({\boldmath\boldsymbol{\vartheta}}) =
 \cos\left(\frac{\vartheta}{2}\right)\I + i \s\cdot{\boldmath\boldsymbol{\widehat{n}}} \sin\left(\frac{\vartheta}{2}\right)\label{eq:rotation-for-Weyl-spinors}
\end{equation} \index{Weyl spinors!rotations and boosts}

 An exactly similar calculation as above yields the 
boosts noted in (\ref{eq:boosts-for-spin-1/2}) to be given by
\begin{align}
B_\mathcal{R,L} (\vp) &= \cosh\left(\frac{\varphi}{2}\right)\I 
\pm \s\cdot{\boldmath\boldsymbol{\widehat{p}}}\,
\sinh\left(\frac{\varphi}{2}\right) \\ 
&= \cosh\left(\frac{\varphi}{2}\right)\left[\I\pm\s\cdot {\boldmath\boldsymbol{\widehat{p}}} 
\,\tanh\left(\frac{\varphi}{2}\right)\right] \label{eq:R-L-boost}
\end{align}
where the the upper sign on the righthand side of the above result holds for the $\mathcal{R}$ type representations, and the lower sign holds for the $\mathcal{L}$ type representations. To further simplify (\ref{eq:R-L-boost}) we use the identities
\begin{align}
&\cosh\left(\frac{\varphi}{2}\right) = \sqrt{ \frac{\cosh\varphi+1}{2}}  =\sqrt{\frac{E+m}{2 m}}\\
&\tanh\left(\frac{\varphi}{2}\right) =\frac{\sinh\varphi}{\cosh\varphi+1} = \frac{p}{E+m}
\end{align}\label{eq:boosts-for-R-L-spinors}
for the rapidity parameter defined in 
(\ref{eq:rapidity-parameter})
and rewrite (\ref{eq:R-L-boost})~as
\begin{align}
B_\mathcal{R,L} (\vp) = \sqrt{\frac{E+m}{2 m}}\left[\I \pm \frac{\s\cdot\p}{E+m}   \right]\label{eq:Weyl-boosts}
\end{align} \index{Weyl spinors!rotations and boosts}
Now note that the spin-$1/2$ helicity operator is defined as
\beq
\mathfrak{h} \stackrel{\textrm{def}}{=}\frac{\s}{2}\cdot\widehat\p\label{eq:helicity}
\eeq
and thus $\s\cdot\p$ that appears in $B_\mathcal{R,L} (\vp)$ can be written as $ 2 p \mathfrak{h}$. This allows us to re-write (\ref{eq:Weyl-boosts}) in the form
\beq
B_\mathcal{R,L} (\vp) = \sqrt{\frac{E+m}{2 m}}\left[\I \pm \frac{2 p \mathfrak{h}}{E+m}   \right]\label{eq:Weyl-boosts-h}
\eeq

The transformations $R_{\mathcal{R},\mathcal{L}}({\boldmath\boldsymbol{\theta}})$ and  $B_{\mathcal{R},\mathcal{L}}
({\boldmath\boldsymbol{\varphi}})$ act on the two-dimensional representation spaces. For historical reasons the `vectors' in these spaces are called  $\mathcal{R}$-type and $\mathcal{L}$-type Weyl spinors. \index{Weyl spinors}
To gain insights into these representation spaces we work out the effect of the boosts and rotations on the eigenspinors of $\mathfrak{h}$. For an arbitrary four-momentum $p^\mu$ these are defined as 
\begin{equation}
\mathfrak{h} \,\phi_\pm^{\mathcal{R},\mathcal{L}}(p^\mu) = \pm \frac{1}{2} \phi_\pm^{\mathcal{R},\mathcal{L}}(p^\mu)\label{eq:Weyl-spinors-h}
\end{equation}
Introducing the standard four momentum vector for massive particles
\begin{equation}
k^\mu = \left(
\begin{array}{c}
m\\
\bm{0}
\end{array}
\right),\quad \bm{0}\stackrel{\textrm{def}}{=}\p \big\vert_{p\to 0}
\label{eq:standard-four-vector}
\end{equation}
the Weyl spinors for a particle at  rest $\phi_\pm^{\mathcal{R},\mathcal{L}}(k^\mu)$ also satisfy
(\ref{eq:Weyl-spinors-h})
\begin{equation}
\mathfrak{h} \,\phi_\pm^{\mathcal{R},\mathcal{L}}(k^\mu) = \pm \frac{1}{2} \phi_\pm^{\mathcal{R},\mathcal{L}}(k^\mu)
\end{equation}
Once $\phi_\pm^{\mathcal{R},\mathcal{L}}(k^\mu)$ are chosen, the boosted $\phi_\pm^{\mathcal{R},\mathcal{L}}(p^\mu)$ follow immediately by the application of the  boost operators (\ref{eq:Weyl-boosts-h}). 
To avoid keeping track of two many $\pm$ and $\mp$ signs, consider $\mathcal{R}$ type Weyl spinors first
\begin{align}
\phi_\pm^{\mathcal{R}}(p^\mu) & = B_\mathcal{R} (\vp) 
\phi_\pm^{\mathcal{R}}(k^\mu) \\
& =\sqrt{\frac{E+m}{2 m}}\left[\I \pm \frac{p}{E+m}   \I\right] \phi_\pm^{\mathcal{R}}(k^\mu) 
\end{align}
 To explore the high-energy limit we note that for $p \gg m$,
  \begin{align}
  \frac{p}{E+m}  \approx 1-\frac{m}{p}
 \end{align} 
 As a result we have
 \begin{equation}
 B_\mathcal{R} \,(\vp) \phi_\pm^{\mathcal{R}}(k^\mu)  
  \approx \sqrt{\frac{E+m}{2 m}}\left[1\pm\left(1-\frac{m}{p}\right) \right] \phi_\pm^{\mathcal{R}}(k^\mu) 
  \label{eq:interpretation-R}
 \end{equation}
 Implementing the high-energy limit by taking the massless limit one thus arrives at the result that for the $\mathcal{R}$ type Weyl spinors only the positive helicity degree of freedom survives. 
 While for the  $\mathcal{L}$ type Weyl spinors the counterpart of (\ref{eq:interpretation-R}) reads
 \begin{equation}
 B_\mathcal{L} \,(\vp) \phi_\pm^{\mathcal{L}}(k^\mu)  
  \approx \sqrt{\frac{E+m}{2 m}}\left[1\mp\left(1-\frac{m}{p}\right) \right] \phi_\pm^{\mathcal{L}}(k^\mu) 
  \label{eq:interpretation-L}
 \end{equation}
 As a consequence, in the high energy limit only the negative helicity spinors survive. 
 
 The nomenclature of  $\mathcal{R}$ and $\mathcal{L}$ type Weyl spinors arises from these high energy limits.\index{Weyl spinors!$\mathcal{R}$ and $\mathcal{L}$ type}
 For the $m=0$ case the helicity is an invariant under boosts, and from there arose the nomenclature of the right-handed ($\mathcal{R}$ type) and left-handed ($\mathcal{L}$ type) Weyl spinors. But for the massive case helicity is not an invariant and thus one must be careful. The $\mathcal{R}$ and 
 $\mathcal{L}$ must be taken as labels for the representation spaces under consideration unless one wishes to restrict one's calculations to a preferred frame for convenience. Both the $\mathcal{R}$ type and the $\mathcal{L}$ type Weyl spinors support positive helicity and the negative helicity spinors. A similar exercise for higher spins tells us that in the high energy limit only the maximal helicity $(\pm j)$ associated with the $j$-vectors survives. By $j$-vectors we mean $(2j+1)$ column vectors with $(2j+1)$ complex entries 
  \textendash~either in the $\mathcal{R}$ or $\mathcal{L}$ representation space. In this language massive Weyl spinors are $1/2$-vectors.

Equation (\ref{eq:rotation-for-Weyl-spinors}) implies that 
the action of $R_{\mathcal{R},\mathcal{L}}({\boldmath\boldsymbol{\vartheta}}) $ on Weyl spinors for $\vartheta=2\pi$, irrespective of the axis of rotation ${\boldmath\boldsymbol{\widehat{n}}}$,  introduces a phase factor equal to $-1$. It requires a $4\pi$ rotation for a spinor to return to itself.\index{Weyl spinors!rotation by $2\pi$} In general  Weyl spinors pick up a helicity-dependent phase factor
\begin{equation}
\phi(p^\mu) \to  \begin{cases}\e^{+i\vartheta/2} \phi(p^\mu),\hspace{3pt}\mbox{for the positive helicity Weyl spinor}\\
\e^{-i\vartheta/2} \phi(p^\mu),\hspace{3pt}\mbox{for the negative helicity Weyl spinor}
\end{cases}\label{eq:phasefactor}
\end{equation}
We shall discover in Chapter~\ref{ch10} that these phase factors\index{Phase factors under rotation} play a dramatic role for Elko, the eigenspinors of the charge conjugation operator.

\section[Spacetime: Its construction \ldots]{Spacetime: Its construction from the simplest representations of Lorentz algebra}
\label{sec:constructing-xt} \index{Spacetime!construction of}

The simplest representations of the Lorentz algebra provide the basic elements to describe the matter fields that we encounter around us, the basic blocks from which we are constructed from. The gauge fields that keep them interacting, glue them and add fire, carry a `vector' spacetime index. This section is devoted to construct this vector index, in essence by constructing spacetime itself.

Towards this end, using the the $\mathcal{R}$ and $\mathcal{L}$ type generators for spin one half
 \beq
 \Big\{\bz_{\mathcal{R}} = \s/2,\kb_{\mathcal{R}}=-i\s/2\Big\},\quad 
 \Big\{\bz_{\mathcal{L}} = \s/2,\kb_{\mathcal{L}}=i\s/2\Big\}
 \label{eq:rl-generators}
 \eeq
 we introduce the following six generators for the $\mathcal{R}\otimes\mathcal{L}$ representation space
\begin{align}
 &{\mathfrak{J}}_i = \big[(\zeta_{\mathcal{R}})_i \otimes\I \big] + \big[\I\otimes(\zeta_{\mathcal{L}})_i\big] \label{eq:zeta-i}\\
 & {\mathfrak{K}}_i = \big[(\kappa_{\mathcal{R}})_i \otimes\I \big] 
 + \big[\I\otimes(\kappa_{\mathcal{L}})_i\big]  \label{eq:kappa-i}
\end{align}
with $i=x,y,z$. These generators correspond to rotation and boost  transformations given by
\beq
R({\boldmath\boldsymbol{\vartheta}})\stackrel{\textrm{def}}{=}R_\mathcal{R} ({\boldmath\boldsymbol{\vartheta}})\otimes R_\mathcal{L} ({\boldmath\boldsymbol{\vartheta}}),\quad B({\boldmath\boldsymbol{\varphi}})\stackrel{\textrm{def}}{=}B_\mathcal{R} ({\boldmath\boldsymbol{\varphi}}) \otimes B_\mathcal{L} ({\boldmath\boldsymbol{\varphi}}).\label{eq:RB}
\eeq

The definitions~(\ref{eq:RB}), and  (\ref{eq:zeta-i}) and (\ref{eq:kappa-i}) mutually define each other

\beq
\mathfrak{J}_i= \frac{1}{i}\frac{\partial}{\partial{\boldmath\boldsymbol{\vartheta}}} 
R({\boldmath\boldsymbol{\vartheta}})\bigg\vert_{{\boldmath\boldsymbol{\vartheta}}\to 0}
\eeq
Each of the ${\mathfrak{J}}_i$ corresponds to taking ${\boldmath\boldsymbol{\widehat{n}}}$ in the defintion ${\boldmath\boldsymbol{\vartheta}} = \vartheta \,{\boldmath\boldsymbol{\widehat{n}}}$ along each of the three spatial axes. Similarly, 
\beq
 \mathfrak{K}_i= \frac{1}{i}\frac{\partial}{\partial{\boldmath\boldsymbol{\varphi}}} 
B({\boldmath\boldsymbol{\varphi}})\bigg\vert_{{\boldmath\boldsymbol{\varphi}}\to 0}
 \eeq
Each of the $\mathfrak{K}_i$ corresponds to taking ${\boldmath\boldsymbol{\widehat{\p}}}$ in the defintion ${\boldmath\boldsymbol{\varphi}} = \varphi \,{\boldmath\boldsymbol{\widehat{\p}}}$ along each of the three spatial axes. 

We now show that modulo a convention-related unitary transformation $U$ given below in equation~(\ref{eq:Umatrix}), the $R({\boldmath\boldsymbol{\vartheta}})$  and $B({\boldmath\boldsymbol{\varphi}})$ are precisely the special relativistic transformations of the rotation and boosts. 

To establish this claim all we have to do is to obtain $\mathfrak{J}_i$ and $\mathfrak{K}_i$ defined in (\ref{eq:zeta-i}) and  (\ref{eq:kappa-i}) and show that a unitary transformation $U$ exists such that the $U$-transformed $\mathfrak{J}_i$ and $\mathfrak{K}_i$ reduce to $J_i$ and $K_i$ of special relativity. The $U$ simply implements the convention that the temporal coordinate is placed on top of the spatial co-ordinates in the standard sequence so that $x^\mu$ equals $(t,\x)$.

Implementing the definitions (\ref{eq:rl-generators})-(\ref{eq:kappa-i}) explicitly, and using Pauli matrices in the representation (\ref{eq:Pauli-Matrices}) we obtain
\begin{equation}
\mathfrak{J}_x= \frac{1}{2}\left(\begin{array}{cccc}
0 &1 & 1 &0 \\
1 & 0 & 0 & 1\\
1 & 0 & 0 &1\\
0 & 1 & 1 &0
\end{array}\right),
\end{equation}
\begin{equation}
\mathfrak{J}_y= \frac{1}{2}\left(\begin{array}{cccc}
0 &- i & -i  &0 \\
i & 0 & 0 & -i\\
i & 0 & 0 &-i\\
0 & i & i &0
\end{array}\right),\quad
\mathfrak{J}_z= \left(\begin{array}{cccc}
1 &0 & 0 &0 \\
0 & 0 & 0 & 0\\
0 & 0 & 0 &0\\
0 & 0 & 0 &-1
\end{array}\right)
\end{equation}
and 
\begin{equation}
\mathfrak{K}_x= \frac{1}{2} \left(\begin{array}{cccc}
0 & i & -i  &0 \\
i & 0 & 0 & -i \\
-i & 0 & 0 &i\\
0 & -i & i &0
\end{array}\right),
\end{equation}
\begin{equation}
\mathfrak{K}_y= \frac{1}{2}\left(\begin{array}{cccc}
0 & 1& - 1  &0 \\
-1 & 0 & 0 & -1\\
1& 0 & 0 &1\\
0 & 1 & -1 &0
\end{array}\right),\quad
\mathfrak{K}_z= \left(\begin{array}{cccc}
0 &0 & 0 &0 \\
0 & - i & 0 & 0\\
0 & 0 & i &0\\
0 & 0 & 0 & 0
\end{array}\right)
\end{equation}
A straight forward calculation now shows that with the $U$ defined as
\beq
U \stackrel{\textrm{def}}{=} \frac{1}{\sqrt{2}}
 \left(\begin{array}{cccc}
0 & i & -i  &0 \\
-i  & 0 & 0 & i \\
1 & 0 & 0 & 1\\
0 & i & i &0
\end{array}\right) \label{eq:Umatrix}
\eeq
we have the required result
\begin{align}
& U \mathfrak{J}_i U^{-1} \to J_i \\ & U\mathfrak{K}_i U^{-1} \to K_i
\end{align}
with $J_i$ and $K_i$ of special relativity given in (\ref{eq:kx}) to (\ref{eq:jyjz-new}). The usual $4\times 4$ transformation matrices of the special relativity are nothing but one or the other of the following
\begin{equation}
\left[{\Lambda^\mu}_\nu \right]= U \left\{
\begin{array}{l}
R({\boldmath\boldsymbol{\vartheta}}) \\
B({\boldmath\boldsymbol{\varphi}})
\end{array}\right\} U^{-1}
\end{equation}
The square brackets on the left hand side indicates that the expression is to be understood as  the matrix $\Lambda$ defined by the  elements enclosed.

The construction presented here may be interpreted as a parallel to  Atiyah's suggestion of a spinor  being a square root of geometry~\citep{Atiyah2013:ma}.
\index{Spinor!a square root of geometry}

\section{A few philosophic remarks}

It would become apparent as we open deeper into our discourse  that the $\mathcal{R}\oplus\mathcal{L}$ representation space, in contrast to the
$\mathcal{R}\otimes\mathcal{L}$ representation space just explored,  supports all the fermionic matter fields of the standard model of the high energy physics. So that which walks~\textendash~the matter fields~\textendash~and that in which it walks, that is spacetime, are determined by each other. The unifying theme being the underlying Lorentz algebra. From a philosophic point of view the theory that deals with the measurement of spatial and temporal distances through the proverbial rods and clocks in fact explores the universality of symmetries that determine the very substance those very rods and clocks are made of. I believe that this point of view is very similar to the one adopted by 
Harvey Brown in his monograph~\citep{Brown:2005hr}. 

It also assures us that the Planck-scale expected departures from the Lorentz algebra are likely to alter foundational principles that spring from the low-energy phenomena that we human are able to observe and study so far. Some of these principles and notions are those of Heisenberg algebra, wave particle duality, causality, particle-antiparticle symmetry, and the Lagrangians that govern matter and gauge fields in a Lorentz covariant framework. Any such departure may also be able to incorporate such notions as consciousness, and phenomena that are presently considered outside the realm of physics proper. 

We make this observation because one can easily argue that non-commutati\-vity of spacetime measurements already follows from the merger of quantum mechanics and the theory of general relativity~\citep{Ahluwalia:1993dd,Doplicher:1994zv,Kempf:1994su,Ahluwalia:2000iw}. In reference to the beautiful analysis of~\citep{Chryssomalakos:2004gk} we note that in their analysis relativistic symmetries and Heisenberg algebra are considered as two separate entities to be merged together whereas in the approach taken in this monograph Heisenberg algebra is seen as implicitly contained in the Lorentz algebra. 

However, incorporating the still deeper issues into the physics proper may require a unification of physics-related phenomena and consciousness-endowed biological systems~\citep{Ahluwalia:2017owu}.

\chapter{Discrete symmetries: Part 1 (Parity)}
\label{ch5}

Adapting the approach of \citep{Speranca:2013hqa}, this chapter presents a first-principle discussion of the parity operator. We will find that the 1928-Dirac equation expresses the fact  that the associated  spinors are the eigenspinors of the parity operator, and that $m^{-1}\gamma_\mu p^\mu$ is the covariant parity operator. We will show that for the Dirac spinors the right- and left- transforming components necessarily carry the same helicity, and bring to attention a phase factor that determines whether a spinor is a particle spinor, or an antiparticle spinor. Our results coincide with that of~\citep{Weinberg:1995mt}, while pointing out a flaw in the standard treatment of the subject in books such as~\citep{Ryder:1985wq,Hladik:1999tt}.

\section{Discrete symmetries}

Our description of the nature of reality divides one unified whole into parts. These are connected with each other by being one representation or the other of the Lorentz algebra.

An example of this broadbrush observation is that
each of the transformation for rotations and boosts  (\ref{eq:R-type-rb}) and (\ref{eq:L-type-rb}) contains in it a reference to two type of spaces: one from the familiar four-vector space of Minkowski through ${\boldmath\boldsymbol{\theta}}$ and 
${\boldmath\boldsymbol{\varphi}}$ (see section \ref{sec:constructing-xt}), and second to the spaces on which the generators of rotations and boosts act (this time, Minkowski being just one of them). The latter are  bifurcated in two types, the $\mathcal{R}$ and $\mathcal{L}$ type. These can transmute into each other if there exists a discrete transformation such that 
\begin{equation}
\Big\{ \exp[i\bmfj\cdot{\boldmath\boldsymbol{\vartheta}}] ,  \exp[  \pm \bmfj\cdot{\boldmath\boldsymbol{\varphi}}] \Big\} \leftrightarrows
\Big\{ \exp[i\bmfj\cdot{\boldmath\boldsymbol{\vartheta}}] ,  \exp[ \mp \bmfj\cdot{\boldmath\boldsymbol{\varphi}}] \Big\}  \label{eq:rl}
\end{equation}
Or, equivalently
\begin{equation}
\mathcal{R} \leftrightarrows \mathcal{L}
\end{equation}
There are two known ways in which this can be done. Each of these introduce a discrete symmetry. These merge seamlessly with the continuous kinematical symmetries and provide additional needed structure and understanding of reality. These are the symmetries of parity and charge conjugation:
\begin{itemize}
\item In the Minkowski four-vector space, parity is defined as the map:\index{Parity!in Minkowski space}
\begin{equation}
P: x^\mu = (t,\x) \to x^{\prime\mu}=(t,-\x). \label{eq:Definition-of-Parity-in-Four-Vectors-Space}
\end{equation}
Under it the $\bmfj$ and ${\boldmath\boldsymbol{\vartheta}}$ remain unaltered but the rapidity parameter $\vp$ changes sign.
 As a consequence parity interchanges the two representation spaces and implements (\ref{eq:rl}). In this chapter I start a discussion of the image of (\ref{eq:Definition-of-Parity-in-Four-Vectors-Space}) in the $\mathcal{R}\oplus\mathcal{L}$ representation space for $s=1/2$ and develop the associated mathematical structure that gives us two identities, leading to the 1928 Dirac equation in the momentum space.\footnote{The genesis of this chapter goes back to a pizza and hand written explanation by 
  Llohann 
Speran\c{c}a on a paper napkin one evening in Bar\~ao Geraldo in Campinas. This was during the two years I spent at IMECC, Unicamp, on a long term visit.}

\item A distinct discrete symmetry arises if the transmutation (\ref{eq:rl}) is \index{Charge conjugation} implemented by complex conjugating a spin $s$ representation space and acting it with the Wigner\index{Wigner time reversal operator} time reversal operator $\Theta$ defined as 
\beq\Theta \bmfj \Theta^{-1} \stackrel{\textrm{def}}= - \bmfj^\ast\eeq 
It introduces a new discrete symmetry, that of charge conjugation.\index{Charge conjugation} Again, taking $s=1/2$ as an example, I develop this discrete symmetry in detail in chapter~\ref{ch6}.  

\end{itemize}

\section{Weyl spinors\label{sec:notation}}\index{Weyl spinors}

To bring a sharper focus to spin one half we first gather together what we have learned so far in 
a compact form. 
 Under the Lorentz boosts  the right- and left-handed Weyl spinors transform as\footnote{The placing of $\mathcal{R}$ and $\mathcal{L}$ as a superscript or as a subscript is purely for convenience in a given context.} 
\begin{equation}
		\phi_{\mathcal{R}}(p^\mu) = \exp\left(+\frac{\s}{2}\cdot\vp\right) 
		\phi_\mathcal{R}(k^\mu),\quad
		\phi_\mathcal{L}(p^\mu) = \exp\left(-\frac{\s}{2}\cdot\vp\right) \phi_\mathcal{L}(k^\mu)\label{eq:w}
\end{equation}
where $\s$ represents the set  of Pauli matrices $\{\sigma_x,\sigma_y,\sigma_z\}$ in their standard representation given in (\ref{eq:Pauli-Matrices}); and the boost parameter $\vp$ is defined so that $\exp\left(i \K\cdot\vp\right)$ acting on the standard four momentum
 equals the general four momentum 
$
			p^\mu = (E, p \sin\theta\cos\phi,p\sin\theta\sin\phi,p\cos\theta).
$
This yields
$\cosh\varphi = E/m$, $\sinh\varphi= p/m$ with $\widehat{\vp} = \widehat {\p}$ [in agreement with
(\ref{eq:rapidity-parameter})], while
$\K$ are the $4\times 4$ matrices  for the generators of boosts in Minkowski space as per their definition in equations (\ref{eq:kx}) and (\ref{eq:kykz}).

The reader may recall that (\ref{eq:w})  follow from the fact that $- i \s/2$ are the generators of the boosts for the right-handed Weyl representation space, while $+ i \s/2$ are  for the left-handed Weyl representation space. 
For the direct sum of the right- and left-Weyl representation spaces, to be motivated below,
the boost and rotation generators thus read
\begin{equation}
				\kb  = \left[
				\begin{array}{cc}
				- i \s/2 & \0\\
					\0 & + i \s/2
					\end{array}
					\right], \quad
 \bz = \left[
\begin{array}{cc}
 \s/2 & \0\\
\0 &  \s/2
\end{array}
\right].  \label{eq:pi}
\end{equation}

The set of generators $\{\K,\J\}$ and $\{\kb,\bz\}$, separately, satisfy the same unifying algebra, the Lorentz algebra~\textendash~given in equations~(\ref{eq:LorentzAlgebra})~\textendash~and are simply its different representations. 

The $\left[{\Lambda^\mu}_\nu\right] \stackrel{\textrm{def}}{=}\Lambda$ representing boosts and rotations in Minkowski space is thus given by
\begin{equation}
\Lambda \stackrel{\textrm{def}}{=}\left\{
\begin{array}{cl}
\exp\left(i\K\cdot\vp\right) & \mbox{for Lorentz boosts} \\
\exp\left(i\J\cdot\vt\right) & \mbox{for rotations}
\end{array}.\right.\label{eq:Minkowski-boost-rotation}
\end{equation}
 If $\{\K,\J\}$ are in their canonical form
the elements of $\Lambda$, $\K$, and $\J$ are of the form ${{a}^\mu}_\nu$, where the 
$4\times 4$ matrix $a$ stands generically for either one of them. Their indices may then be raised and lowered by the spacetime metric $\eta_{\mu\nu} =\textrm{diag}\{1,-1,-1,-1\}$.

\section{Parity operator for   the general four-component spinors
\label{sec:parity}}

To construct the image of parity operator for $s=1/2$ the discussion above suggests to define
 a four-component spinor in the  $\mathcal{R}\oplus\mathcal{L}$ representation space \begin{equation}
		\psi(p^\mu) = \left( \begin{array}{c}
		\phi_\mathcal{R}(p^\mu)\\
		\phi_\mathcal{L}(p^\mu)
		\end{array}
		\right) \label{eq:gen-4-comp-psi}
\end{equation}
with 
\beq
\psi(p^\mu) 
= \exp(i \kb\cdot\vp) \psi(k^\mu). \label{eq:psi}
\eeq
where $p^\mu = {\left[ \exp\left(i\K\cdot\vp\right)\right]^\mu}_\nu k^\nu$.

 The $\psi(k^\mu)$ are generally called  ``rest  spinors,'' while the $\psi(p^\mu)$ are often named ``boosted spinors.'' However, since no frame is a preferred frame, the $\psi(k^\mu)$ and  the infinitely many $\psi(p^\mu)$ reside in every frame. 
 
 These $\psi(p^\mu)$ spinor may be eigenspinors of the parity operator, or that of the  charge conjugation operator, or any other operator in the  
 $\left[\mathcal{R}\oplus\mathcal{L}\right]_{s=1/2}$ representation space.
 A physically important classification of spinors 
is by Lounesto~\citep[Chapter 12]{Lounesto:2001zz}.
For ready reference we remind that the rotation on $\psi(p^\mu)$  is implemented by
\begin{equation} 
\psi\left(p^{\prime\mu}\right) = \exp\left(i\bz\cdot\bm{\vartheta}\right) \psi\left(p^{\mu}\right)\label{eq:rotation-on-spinors}
\end{equation}
where $p^{\prime\mu} = {\left[\exp\left(i\J\cdot\bm{\vartheta}\right)\right]^\mu}_\nu p^\nu$.

Thus for massive particles  incorporating parity covariance  doubles the degrees of freedom for a spin one half representation space from $2$ to $4$. This doubling brings in a new degree of freedom, that of antiparticles. The argument has a natural extension for all spins.
\vspace{11pt}

\noindent
\textit{A parenthetic remark}~\textemdash

 Historically, antiparticles entered the theoretical physics scene through this doubling in the degrees of freedom. Later, it was realised that in a merger of quantum mechanical formalism and  the theory of special relativity, causality could only be preserved if one introduced antiparticles, for all spins. 
For massive particles of spin one half, it is done by doubling of the degrees of freedom inherent in the definition of $\psi(p^\mu)$, while for the massless particles the doubling happens through incorporating both helicities degrees of freedom~\textendash~one from the massless $\mathcal{R}$ and the opposite from the massless $\mathcal{L}$ representation space.  In a sense, the parity eigenvalues act as a charge under the charge conjugation operator, 
while helicity takes over the role of charge for massless particles and it gets interchanged by the charge conjugation operator. 

Parity need not be invoked to incorporate antiparticles in the massless case, helicity picks up that role in the $m\ne 0$ to the $m=0$ transition. In fact the $\mathcal{R}$ and 
$\mathcal{L}$ representation spaces get decoupled in the indicated transition.
It is natural to conjecture if helicity of massless particles, irrespective of spin, from neutrinos to photons to gravitons acts as a new type of charge. If so, does it couple to torsion, or a fundamentally new field? In a cosmological setting, it may suggest tiny difference in the cosmic background temperatures associated   with the different helicities and it could also lead to matter-antimatter asymmetry.
\vspace{11pt}

An examination of the question, ``How does $P$ affect $\psi(p^\mu)$?'' yields the Dirac operator~\citep{Speranca:2013hqa}.  The argument is as follows.

In accordance with the opening discussion of this chapter, the parity operator interchanges the right- and left- handed Weyl representation spaces.
 The effect of $P$ on $\psi(p^\mu)$ is thus realised by a $4\times 4$ matrix $\mathcal{P}$:
 
 \begin{itemize} 
 \item[a.]
which
 up to a global phase
must contain purely off-diagonal $2\times 2$ identity matrices $\openone$, and
\item[b.]
which in addition implements the action of $P$ on $p^\mu$.  
\end{itemize}
Up to a global phase, chosen to be $1$ (unless needed otherwise), the effect of $\mathcal{P}$ on $\psi(p^\mu)$ is therefore
given by
\beq
		\mathcal{P}\, \psi(p^\mu) =
		\underbrace{\left(\begin{array}{cc}
		\0 & \I\\
		\I & \0
		\end{array}
		\right)}_{ {\gamma_0}}\psi(p^{\prime\mu}) = \gamma_0 \psi(p^{\prime\mu}). \label{eq:psi2}
\eeq
Here, $p^{\prime\mu}$ is the $P$ transformed $p^\mu$ while $\0$ and $\I$ represent $2\times 2$ null and identity matrices, respectively. This is where the general textbook considerations on $\mathcal{P}$ stop. To be precise, the usual treatments arrive at $\mathcal{P}$ only after they introduce Dirac equation, and not at the primitive level we are pursuing. We will shortly see that the 1928 Dirac equation lies a short way ahead on the track.

 For a general spinor, Speran\c{c}a has noted that 
it provides a better understanding of $\mathcal{P}$ if in (\ref{eq:psi2}) we note that
 $ \psi(p^{\prime\mu}) $ may be related to  $ \psi(p^\mu)$ as follows~\citep{Speranca:2013hqa}
\begin{equation}
 	\psi\left(p^{\prime\mu}\right) = \exp\big[i \kb\cdot(-\vp)\big]\psi(k^\mu) 
\end{equation}
with  $\kb$ defined in (\ref{eq:pi}). But since from (\ref{eq:psi}), $\psi(k^\mu)  = \exp\left(-i\kb\cdot\vp\right)\psi(p^\mu)$ the above equation can be re-written as
\begin{equation}
\psi\left(p^{\prime\mu}\right)\,		 =  
\exp\left(-i \kb\cdot\vp\right)\exp\left(-i \kb\cdot\vp\right) \psi(p^\mu)
		  =  \exp(- 2 i \kb\cdot\vp) \psi(p^\mu). \label{eq:zimpok167} 
\end{equation} 
Substituting $\psi\left(p^{\prime\mu}\right)$ from the above equation in the $\mathcal{P}$-defining equation~(\ref{eq:psi2}), and on using the anti-commutativity of $\gamma_0$ with each of the generators of the boost ${\kb}$, 
\begin{equation}
\{\gamma_0,\kb_i\} = 0,\qquad  \mbox{with}\,\, i=x,y,z
\end{equation} 
equation (\ref{eq:psi2}) becomes
\begin{equation}
		\mathcal{P} \psi(p^\mu) = \exp(2 i \kb\cdot\vp) \gamma_0 \psi(p^\mu).\label{eq:psi3}
\end{equation}
Using the definition of $\kb$ given in equation~(\ref{eq:pi}) and recalling 
from (\ref{eq:rapidity-parameter}) that $\vp = \varphi\,\widehat \p$, the $\exp(2 i \kb\cdot\vp) \gamma_0$ factor in the above equation can be re-written as 
\begin{equation}
\exp(2 i \kb\cdot\vp)  \gamma_0   =
\exp\left[\left(
\begin{array}{cc}
{{\boldmath\boldsymbol{\sigma}}}\cdot\widehat{\p} &\0 \\
\0 & - {{\boldmath\boldsymbol{\sigma}}\cdot\widehat{\p}}
\end{array}\right) \varphi
\right]\gamma_0 \label{eq:expansion}
\end{equation}
Taking note of the identity 
\beq
\left(
\begin{array}{cc}
{{\boldmath\boldsymbol{\sigma}}}\cdot\widehat{\p} &\0 \\
\0 & - {{\boldmath\boldsymbol{\sigma}}\cdot\widehat{\p}}
\end{array}\right)^2 = \left(\begin{array}{cc}
\I &\0 \\
\0 &\I
\end{array}\right)
\eeq
we immediately obtain
\begin{align}
\exp & \left[\left(
\begin{array}{cc}
{{\boldmath\boldsymbol{\sigma}}}\cdot\widehat{\p} &\0 \\
\0 & - {{\boldmath\boldsymbol{\sigma}}\cdot\widehat{\p}}
\end{array}\right) \varphi
\right] \nonumber\\ & \hspace{29pt} = 
\left(\begin{array}{cc}
\I \cosh\varphi + {\boldmath\boldsymbol{\sigma}}\cdot\widehat\p \sinh\varphi & \0\\
\0 & \I \cosh\varphi - {\boldmath\boldsymbol{\sigma}}\cdot\widehat\p\sinh\varphi 
\end{array}\right) 
\end{align}
With $\gamma_0$ defined in (\ref{eq:psi2}), this result transforms (\ref{eq:expansion}) to
\begin{equation}
 \exp(2 i \kb\cdot\vp)  \gamma_0   =  \left(\begin{array}{cc}
\0 & \I \cosh\varphi + {\boldmath\boldsymbol{\sigma}}\cdot\widehat\p \sinh\varphi\\
\I \cosh\varphi - {\boldmath\boldsymbol{\sigma}}\cdot\widehat\p\sinh\varphi &\0
\end{array}\right) 
\end{equation}
Making the substitutions
\begin{align}
\cosh\varphi \to \frac{p^0}{m}\nonumber\\
\sinh\varphi \to \frac{p}{m}
\end{align}
and using the identifications
\begin{equation}
\gamma_0 \stackrel{\textrm{def}}{=} \left(\begin{array}{cc}
\0 & \openone \\
\openone & \0\end{array}\right),\quad
 \bm{\gamma} \stackrel{\textrm{def}}{=}  \left(\begin{array}{cc}
 \0 &\s \\
 -\s & \0
 \end{array}\right).\label{eq:diracgamma-lower} 
 \end{equation}
 where $\gamma_\mu=(\gamma_0,\bm{\gamma})$, $\bm{\gamma} \stackrel{\textrm{def}}{=}\gamma_1,\gamma_2,\gamma_3$, are the canonical Dirac matrices in the Weyl representation, yields\index{Dirac matrices}
 \begin{equation}
 		\exp\left( 2 i \kb\cdot\vp\right) \gamma_0 = m^{-1} \gamma_\mu p^\mu \label{eq:zimpok177}
 \end{equation}
 \vspace{7pt} 
 
 Thus (\ref{eq:psi3}) now takes the form
 \begin{equation}
		\mathcal{P} \, \psi(p^\mu) = m^{-1}\gamma_\mu p^\mu \, \psi(p^\mu).\label{eq:psi4}
\end{equation}

Up to a global phase this exercise yields the parity operator for the $\left[\mathcal{R}\oplus\mathcal{L}\right]_{s=1/2}$ representation space to be
\begin{equation}
		\mathcal{P} = m^{-1} \gamma_\mu p^\mu. \label{eq:P}
\end{equation}
\index{Parity!for $\left[\mathcal{R}\oplus\mathcal{L}\right]_{s=1/2}$ space}
In this manner the `recipe' contained in the discussion surrounding equation (\ref{eq:psi2}) transforms into a clear-cut mathematical operator and makes the physical content of the 1928 Dirac equation 
transparent.
The $\mathcal{P}$ applies to all  the $\left[\mathcal{R}\oplus\mathcal{L}\right]_{s=1/2}$  4-component spinors of the type $\psi(p^\mu)$. Only its eigenspinors, despite wide spread misconceptions to the contrary,   are 
Dirac spinors~\textendash~for, they alone satisfy the Dirac equation.  

The eigenvalues of $\mathcal{P}$ are
 $\pm 1$. Each of these has a  two fold degeneracy 
\begin{equation}
		\mathcal{P}\, \psi^S_\sigma(p^\mu) = + \psi^S_\sigma(p^\mu),\quad
		\mathcal{P}\, \psi^A_\sigma(p^\mu) = - \psi^A_\sigma(p^\mu). \label{eq:Dirac-p-new}
\end{equation}
The subscript $\sigma$ is the degeneracy index, while the 
superscripts refer to self and anti-self conjugacy of $\psi(p^\mu)$ under $\mathcal{P}$. With the help of Eq.~(\ref{eq:P}), 
Eq.~(\ref{eq:Dirac-p-new}) translates to
\begin{equation}
		\left(\gamma_\mu p^\mu -m \openone\right) \psi^S_\sigma(p^\mu) = 0,\quad
		\left(\gamma_\mu p^\mu +m\openone\right) \psi^A_\sigma(p^\mu) = 0.\label{eq:Dirac-pd}
\end{equation}\index{Dirac equation!in momentum space}

\noindent
\textit{Notational remark\textemdash}

The Dirac's~\citep{Dirac:1928hu} $u_\sigma(p^\mu)$ and $v_\sigma(p^\mu)$ spinors are thus seen as the eigenspinors of the parity operator, $\mathcal{P}$, with eigenvalues $+1$ and $-1$, respectively:
\begin{equation}
	\psi^S_\sigma(p^\mu) \to u_\sigma(p^\mu),\quad  \psi^A_\sigma(p^\mu)\to v_\sigma(p^\mu).
	\label{eq:dirac-sa-uv}
\end{equation}

At this stage equations (\ref{eq:Dirac-pd}) should not be given a higher status than that of identities.
This is not to undermine their eventual importance but that importance, history aside, resides in a quantum field theoretic context with $\psi^S_\sigma(p^\mu)$ and $\psi^A_\sigma(p^\mu)$ serving as expansion coefficients of the Dirac field~\citep{Weinberg:1995mt} and \textit{not} as wave functions in momentum space.

We end this part of the discussion on the spinorial parity operator by noting that $
\mathcal{P}^2 = \I _4$, 
and that the form of equation (\ref{eq:Dirac-pd}) is preserved under a global transformation
\begin{equation}
			\psi_\sigma(p^\mu) \to  \psi_\sigma^\prime(p^\mu)= \exp( i \alpha) \psi_\sigma(p^\mu),\quad
			 \forall \sigma
			\label{eq:remarks}
\end{equation}
with $\alpha\in\mathfrak{R}$. 

\section[The parity constraints on spinors, locality phases $\ldots$]{The parity constraint on spinors, locality phases, and constructing the Dirac spinors}

To obtain the explicit form of the Dirac spinors we use the following strategy
\begin{enumerate}[(a)] 
\renewcommand{\theenumi}{(\alph{enumi})}
\item  first we obtain the constraint put on a general four-component spinor for $p^\mu=k^\mu$, and construct $\psi(k^\mu)$
\item then act the boost operator 
\beq
 \exp(i \kb\cdot\vp)  
 = \sqrt{\frac{E + m }{2 m}}
						\left[
						\begin{array}{cc}
						\I + \frac{\boldsymbol{\sigma}\cdot\mathbf{p}}{E +m} & \0 \\
						\0 & \I - \frac{\boldsymbol{\sigma}\cdot\mathbf{p}}{E +m} 
						\end{array}
						\right]  \label{eq:sb-again}
\eeq 
on $\psi(k^\mu)$.
\end{enumerate}
To accomplish this calculation we re-write equations~(\ref{eq:Dirac-p-new}) into the form

\begin{align}
m^{-1}\left[
\left(
\begin{array}{cc}
\0 &\I\\
\I & \0
\end{array}
\right) p_0 + \left(
\begin{array}{cc}
\0 &\s\cdot\widehat\p\\
- \s\cdot\widehat\p & \0
\end{array}
\right) p
\right]& \left( \begin{array}{c}
		\phi_\mathcal{R}(p^\mu)\\
		\phi_\mathcal{L}(p^\mu)
		\end{array}
		\right) \nonumber\\
		&= \pm \left( \begin{array}{c}
		\phi_\mathcal{R}(p^\mu)\\
		\phi_\mathcal{L}(p^\mu)
		\end{array}
		\right) 
\end{align}
For $p^\mu = k^\mu$ this reduces to the constraint
\beq
 \left( \begin{array}{c}
		\phi_\mathcal{L}(k^\mu)\\
		\phi_\mathcal{R}(k^\mu)
		\end{array}
		\right) = \pm \left( \begin{array}{c}
		\phi_\mathcal{R}(k^\mu)\\
		\phi_\mathcal{L}(k^\mu)
		\end{array}
		\right) 
\eeq
Thus, keeping the notation introduced in (\ref{eq:dirac-sa-uv}) in mind, we have
\begin{equation}
\phi_\mathcal{R}(k^\mu) = \begin{cases} + \phi_\mathcal{L}(k^\mu), & \mbox{for}~u_\sigma(k^\mu)~ \mbox{spinors}
\\
- \phi_\mathcal{L}(k^\mu), & \mbox{for}~v_\sigma(k^\mu)~\mbox{spinors}
\end{cases}\label{eq:constraint12}
\end{equation}
That is,  for a four-component spinor $\psi(p^\mu)$ to be an eigenspinor of the $\mathcal{P}$ the relative phase between the left- and right- transforming Weyl spinors at rest must be the same for the self-conjugate spinors and opposite for the anti-self conjugate 
spinors. 

This result
 stands in conflict with the  well-known textbook treatments
of the subject~\citep{Ryder:1985wq,Hladik:1999tt} and must be 
corrected~\citep{Gaioli:1998ra,Ahluwalia:1998dv}. Christoph Burgard arrived at this result simply by examining the form of the rest spinors as found in most textbooks~\citep{Burgard:1989tamu}. Specifically, Ryder and Hladik assume that $\phi_\mathcal{R}(k^\mu) = \phi_\mathcal{L}(k^\mu)$ and claim that this in conjunction with the results (\ref{eq:w}) yields Dirac equation: $\left(i \gamma_\mu\partial^\mu - m\right) \psi(x) =0$. Their construct yields $\left(\gamma_\mu p^\mu -m \right)\psi(p^\mu) =0$, but misses $\left(\gamma_\mu p^\mu +m \right)\psi(p^\mu) =0$. With the latter missing one must make two mistakes to arrive at the Dirac equation -- this thread continues in 
 reference~\citep{Ahluwalia:2016jwz}.

The constraint (\ref{eq:constraint12}) is consistent with Weinberg's analysis~\citep{Weinberg:1995mt}. His analysis contains two additional elements: 
\begin{itemize}
\item one, in each of the $\psi(k^\mu)$   the spin projections -- or helicities -- for
$\phi_\mathcal{R}(k^\mu)$ and $\phi_\mathcal{L}(k^\mu)$ must be the same; and 
\item second, when these are used to construct the expansion coefficients of a quantum field the $\psi_\sigma(k^\mu)$ cannot have arbitrary global phases but must have specific values.\footnote{These observations, though not explicitly stated, can be easily read off from the mentioned analysis. }
\end{itemize}

The first of the mentioned results, in conjunction with~(\ref{eq:phasefactor}), implies that under rotation by an angle $\vartheta$ the Dirac spinors pick up a global phase $e^{\pm i\vartheta/2}$ depending on the helicity of the Weyl components involved. It ought to be emphasised that this 
result is specific to Dirac spinors. It does not hold for general four-component spinors. An example of this assertion is found in Chapter~\ref{ch10}.

To incorporate the phase factors constraints coded in~(\ref{eq:constraint12}) 
into the eigenspinors of the $\mathcal{P}$ we define the Weyl spinors at rest $\phi(k^\mu)$\index{Weyl spinors!at rest}
\begin{equation}
\s\cdot\widehat \p\, \phi_\pm(k^\mu) = \pm \phi_\pm(k^\mu)\label{eq:hd}
\end{equation}
and take
\begin{align}
\phi_+(k^\mu) & = \sqrt{m} \textrm{e}^{i\vartheta_1} \left(
									\begin{array}{c}
									\cos(\theta/2)\exp(- i \phi/2)\\
									\sin(\theta/2)\exp(+i \phi/2)
											\end{array}
									\right)
									 \label{eq:zimpok52-new} \\
\phi_-(k^\mu) & = \sqrt{m}  \textrm{e}^{i\vartheta_2} \left(
									\begin{array}{c}
									\sin(\theta/2)\exp(- i \phi/2)\\
									- \cos(\theta/2)\exp(+i \phi/2)
											\end{array}
									\right)  \label{eq:zimpok52-new-new}
\end{align}
with $\vartheta_1, \vartheta_2 \in \Re$. In contrast to our 2005 work~\citep{Ahluwalia:2004sz,Ahluwalia:2004ab} where we set 
 $\vartheta_1 = \vartheta_2 = 0$, we now take $\vartheta_1 = 0, \vartheta_2 = \pi$. These are part of the  locality phase factors,\index{Locality phase factors} the phase factors responsible for
 removing non-locality of the cited earlier works. With this choice of phase factors the $\phi(k^\mu)$ read\index{Locality phase factors}
\begin{align}
\phi_+(k^\mu) & = \sqrt{m}  \left(
									\begin{array}{c}
									\cos(\theta/2)\exp(- i \phi/2)\\
									\sin(\theta/2)\exp(+i \phi/2)
											\end{array}
									\right)
									 \label{eq:zimpok52-new-new2} \\
\phi_-(k^\mu) & = \sqrt{m}   \left(
									\begin{array}{c}
									- \sin(\theta/2)\exp(- i \phi/2)\\
									 \cos(\theta/2)\exp(+i \phi/2)
											\end{array}
									\right). 
									\label{eq:zimpok52-new-new-new}
\end{align} 
In terms of these two-component rest spinors, we define the four-component rest spinors \index{Dirac spinors!at rest}
\begin{align}
& u_+(k^\mu) = \textrm{e}^{i \lambda_1}\left(
\begin{array}{c}
\phi_+(k^\mu) \\
\phi_+(k^\mu) 
\end{array}
\right),\quad 
u_-(k^\mu) = \textrm{e}^{i \lambda_2}\left(
\begin{array}{l}
\phi_-(k^\mu) \\
\phi_-(k^\mu) 
\end{array}
\right)\\
& v_+(k^\mu) = \textrm{e}^{i \lambda_3}\left(
\begin{array}{c}
\phi_-(k^\mu) \\
- \phi_-(k^\mu) 
\end{array}
\right),\quad
v_-(k^\mu) 
= \textrm{e}^{i \lambda_4}  \left(\begin{array}{c}
 \phi_+(k^\mu) \\
-\phi_+(k^\mu) 
\end{array}
\right)
\end{align}
with $\lambda_1,\lambda_2,\lambda_3,\lambda_4\in \Re$. We set $\lambda_1=\lambda_2=\lambda_3 =0$ and $\lambda_4=\pi$, 
\begin{align}
& u_+(k^\mu)  = \left(
\begin{array}{c}
\phi_+(k^\mu) \\
\phi_+(k^\mu) 
\end{array}
\right),\quad
u_-(k^\mu) = \left(
\begin{array}{c}
\phi_-(k^\mu) \\
\phi_-(k^\mu) 
\end{array}
\right)\label{eq:Diracuk}\\
& v_+(k^\mu) = \left(
\begin{array}{c}
\phi_-(k^\mu) \\
- \phi_-(k^\mu) 
\end{array}
\right),\quad
v_-(k^\mu) 
=  \left(\begin{array}{c}
 - \phi_+(k^\mu) \\
\phi_+(k^\mu) 
\end{array}
\right).\label{eq:Diracvk}
\end{align}
to be consistent with equations (5.5.35) and (5.5.36) of Weinberg's analysis~\citep{Weinberg:1995mt}. 
That analysis ensures correct incorporation of Lorentz symmetries, parity covariance, and locality. Apart from the chosen phases, 
it is $\phi_+(k^\mu) $ which enters the definition of $v_-(k^\mu) $, and $\phi_-(k^\mu) $ which enters the definition of $v_+(k^\mu) $. In contrast,
it is $\phi_-(k^\mu) $ which enters the definition of $u_-(k^\mu) $, and $\phi_+(k^\mu) $ which enters the definition of $u_+(k^\mu) $.
This is important for the correct pairing of the creation and annihilation operators with $u_\pm(k^\mu)$ and $v_\pm(k^\mu)$ when constructing the Dirac quantum field.

Following  item (b) of the strategy that opened this section,
the  $u_\sigma(p^\mu)$ and $v_\sigma(p^\mu)$ for an arbitrary $p^\mu$
follow by acting the $\left[\mathcal{R}\oplus\mathcal{L}\right]_{s=1/2}$ boost given in equation~(\ref{eq:sb-again}) 
on the rest spinors
just enumerated. This exercise yields\index{Dirac spinors}
\begin{align} 
u_+(p^\mu)  & = \alpha \left(
\begin{array}{l}
\alpha_+ \,\phi_+(k^\mu)\\ 
\alpha_-\,\phi_+(k^\mu)\\ 
\end{array}
\right) \label{eq:upp}\\
u_-(p^\mu)  & = \alpha
\left(
\begin{array}{l}
\alpha_-\,\phi_-(k^\mu)\\ 
\alpha_+\,\phi_-(k^\mu)\
\end{array}
\right) \label{eq:uspinors}
\end{align}
and 
\begin{align}
v_+(p^\mu)  & = \alpha
\left(
\begin{array}{r}
\alpha_-\,\phi_-(k^\mu)\\ 
- \,\alpha_+\,\phi_-(k^\mu)\
\end{array}
\right) \label{eq:vp}\\
v_-(p^\mu)  &= \alpha \left(
\begin{array}{r}
-\,\alpha_+ \,\phi_+(k^\mu)\\ 
\alpha_-\,\phi_+(k^\mu)\
\end{array}
\right) \label{eq:vm}
\end{align}
where 
\begin{align}
\alpha = \sqrt{\frac{E+m}{2 m}},\quad \alpha_\pm = \left(1\pm\frac{p}{E+m}\right)\label{eq:apm}
\end{align}

The task is done: On the one hand we have constructed the constraint that a 
$\mathcal{R}\oplus\mathcal{L}$ spinor must satisfy for it to be an eigenspinor of the parity operator $m^{-1}\gamma_\mu p^\mu$ and unearthed how Dirac spinors follow when working through this approach. On the other hand we have also introduced certain phase factors  that affect the Lorentz and parity covariance, and locality,  of the quantum field these spinors allow to be constructed.

Dirac spinors are known since 1928. What we have added to the subject is its simple underlying structure, an act that leads us to look at the charge conjugation operator and its eigenspinors in the next chapter.  It is there that a reader to whom all this is fairly well known 
would encounter 
her folkloric wisdom  challenged, unexpectedly.

\chapter{Discrete symmetries: Part 2 (Charge conjugation)}
\label{ch6}

We now embark on developing a second discrete symmetry implied by the transmutation (\ref{eq:rl}). It is the symmetry of charge conjugation.\index{Charge conjugation} The reader is likely to have encountered it before, in particular, in the context of the 1928-Dirac equation. Here, we will see it arise in a manner that unifies its origin to the transmutation (\ref{eq:rl}) and the symmetry of parity discussed in the previous chapter.
It spawns
 by complex conjugating 
 $\mathcal{R}$, or  $\mathcal{L}$, representation spaces
 for spin one half followed by the operation of the Wigner time reversal operator. The extension of the argument to all spins -- or more precisely, to all representation spaces 
 of the type $\left[\mathcal{R}\oplus\mathcal{L}\right]_{j=\textrm{any}}$ --
 follows in a parallel manner. We show that the the charge conjugation operator transmutes the parity eigenvalues. This in turn leads to the anti-commutativity of the parity and charge conjugation operators.

\section{Magic of Wigner time reversal operator}\index{Magic of Wigner time reversal operator}

For spin $s$, given a $\bmfj$ the Wigner time reversal operator is defined as: \index{Wigner time reversal operator}
\begin{equation}
\Theta_{[s]} \bmfj \Theta_{[s]}^{-1} \stackrel{\textrm{def}}{=} - \bmfj^\ast
\end{equation} 
with $\Theta_{[s]} = (-1)^{s+\sigma}\delta_{\sigma^\prime,-\sigma}$ and $\Theta_{[s]}^\ast \Theta_{[s]} = (-1)^{2 s}$. We abbreviate 
$\Theta_{\left[1/2\right]}$ to $\Theta$. In $\Theta$ we mark the rows and columns in the order $\{-1/2,1/2\}$.   
  
For spin one half $
\bmfj = \s/2$. As such, the  Wigner time reversal operator 
acts on the Pauli matrices as follows
	\begin{equation}
		\Theta \s\Theta^{-1} = - \s^\ast\label{eq:wt}
	\end{equation}
with 
	\begin{equation}
			\Theta= \left(\begin{array}{cc} 0 & -1 \\ 1 & 0\end{array}\right), \quad
			\Theta^{-1} = - \Theta.
	\end{equation}
\vspace{11pt}

\noindent
It allows the following `magic'  to happen: 
\begin{quote}
\begin{itemize}
\renewcommand{\theenumi}{(\alph{enumi})}
\item 
If $\phi_R(p^\mu)$ transforms as a right-handed Weyl spinor then $\zeta_\rho \Theta \phi^\ast_R(p^\mu)$ -- with  $\zeta_\rho$  an arbitrary  phase factor --  transforms as a left-handed Weyl spinor.

\item 
If $\phi_L(p^\mu)$ transforms as a left-handed Weyl spinor then $\zeta_\lambda \Theta \phi^\ast_L(p^\mu)$ -- with  $\zeta_\lambda$ an arbitrary  phase factor --  transforms as a right-handed Weyl spinor.


\end{itemize}
\end{quote}

The way this comes about is as follows. 
First complex conjugate both the equations in (\ref{eq:w}), then multiply from the left by $\Theta$, and use the above defining feature of the Wigner time reversal operator. This sequence of 
manipulations~\textendash~after using the freedom to  multiply these equations by phase factors  $\zeta_\rho$ and  $\zeta_\lambda$ respectively~\textendash~ends up with the result
\begin{equation}		
\Big[\zeta_\rho\Theta \phi^\ast_R(p^\mu)\Big] = \exp\left( - \frac{\s}{2}\cdot\vp\right)\Big[\zeta_\rho\Theta \phi^\ast_R(k^\mu)\Big] 
\end{equation}
and
\begin{equation}\Big[\zeta_\lambda\Theta \phi^\ast_L(p^\mu)\Big] = \exp\left( + \frac{\s}{2}\cdot\vp\right)\Big[\zeta_\lambda\Theta \phi^\ast_L(k^\mu)\Big]	
\label{eq:boostL}  
\end{equation}
yielding the claimed magic of the Wigner time reversal operator. This crucial  observation motivates  the  introduction of  two sets of four-component spinors~\citep{Ahluwalia:1994uy,Ahluwalia:2004ab}
	\begin{equation}
	\rho(p^\mu) = \left(\begin{array}{c}
					\phi_R(p^\mu)\\
					\zeta_\rho \Theta \phi^\ast_R(p^\mu)
					\end{array}
					\right) , \quad
	\lambda(p^\mu) = \left(\begin{array}{c}
					\zeta_\lambda \Theta \phi^\ast_L(p^\mu)\\
					\phi_L(p^\mu)
					\end{array}
					\right).
					\label{eq:lambda}
	\end{equation}
The $\rho(p^\mu)$ do not provide an additional independent set of  spinors.  For that reason  we do not consider them further and confine our attention to $\lambda(p^\mu)$ only~\citep{Ahluwalia:1994uy}.

Confining to spin one half, Ramond~\citep{Ramond:1981pw} introduces this result as `magic of Pauli matrices'\index{Magic of Pauli matrices}  where $ \Theta$ gets concealed in Pauli's $\sigma_y$,  which  equals $i \Theta$. More importantly,  his analysis misses the full multiplicity of the phase factor $\zeta_\lambda$. Our argument  in terms of the Wigner time reversal operator $\Theta$ has the advantage that it immediately generalises  to higher spin $\mathcal{R}\oplus\mathcal{L}$ representation spaces.

In $\lambda(p^\mu)$ we  have four, rather than two, independent four-component spinors. The  helicities of the right and left transforming components of $\lambda(p^\mu)$ are opposite (see Section~\ref{Sec:sevenpointone}) --  in sharp contrast to the eigenspinors of the parity operator. 
This circumstance allows  the $\lambda(p^\mu)$ to escape their interpretation as Weyl spinors in a four-component disguise. This doubling is more than of an academic interest. It is required, as we shall see later in this monograph, to incorporate antiparticles when the $\lambda(p^\mu)$ are used as expansion coefficients of a quantum field.

\vspace{11pt}

\section{Charge conjugation operator for  the general four-component spinors}

With the just made observations at hand we are  led to entertain the possibility that in addition to the symmetry operator $\mathcal{P}$ there exists a second symmetry operator from~(\ref{eq:rl}), which up to a global phase factor, has the form
	\begin{equation}
	\mathcal{C} \stackrel{\textrm{def}}{=} \left(\begin{array}{cc}
					\0 & \alpha \Theta \\
					\beta\Theta & \0
					\end{array}
				\right) K \label{eq:cc}
	\end{equation}
where $K$ complex conjugates to its right.\index{Charge conjugation operator!for spin one half} The arguments that lead to~(\ref{eq:cc}) are similar to the ones that give~(\ref{eq:psi2}). Requiring $\mathcal{C}^2$ to be an identity operator determines $\alpha = i, \beta= -i$ (where we have used $K^2=1$). It results in
	\begin{equation}
	\mathcal{C} = \left(\begin{array}{cc}
					\0 & i\Theta \\
					-i \Theta & \0
					\end{array}
					\right) K = \gamma_2 K\label{eq:gamma2K}
	\end{equation}
	with
\begin{equation}
\gamma_2 =
 \left(\begin{array}{cc}
 \0 &\sigma_y \\
 -\sigma_y & \0
 \end{array}\right).
\end{equation}
There also exists a second solution with $\alpha = - i, \beta= i$. But this does not result in a physically different operator and in any case the additional minus sign can be absorbed in the indicated global phase. This is the same operator that appears in the particle-antiparticle symmetry associated with the 1928 Dirac equation. 

We have thus arrived at the charge conjugation operator from the analysis of the symmetries of the $4$-component representation space of spinors. This perspective has the advantage of immediate generalisation to any spin: if $\Theta_{[s]}$ is taken as Wigner time reversal operator for spin $s$ then the spin-$s$ charge conjugation operator in the $2(2s+1)$ dimensional representation space becomes~\citep[equation (A10)]{Lee:2012td}\index{Charge conjugation operator!for any spin} 
	\begin{equation}
		\mathcal{C} = \left[\begin{array}{cc}
		\0 & -i \Theta_{[s]}^{-1} \\
		-i \Theta_{[s]} & \0
		\end{array}
		\right] K. \label{eq:c}
	\end{equation}
For spin one half, $\Theta^{-1} = - \Theta$; consequently, the above expression 
coincides with the result for spin one half given in equation~(\ref{eq:gamma2K}).

Both $\mathcal{P}$ and $\mathcal{C}$ arise without reference to any wave equation, or equivalently without assuming a Lagrangian density. In fact, it will be apparent from our presentation that Lagrangian densities should be derived rather than assumed.

\section{Transmutation of $\mathcal{P}$ eigenvalues by  $\mathcal{C}$, and related results }

We immediately note that $\mathcal{C}$ defined above transmutes the eigenvalues of the 
$\mathcal{P}$ operator.  As a consequence they anti-commute. To see this we apply   $\mathcal{C}$ from the left on (\ref{eq:P}) 
\begin{align}
\mathcal{C}{\mathcal{P}} &  =
\gamma_2  K \big[ m^{-1} \gamma_\mu p^\mu\big] \nonumber \\
& = \gamma_2  K \big[m^{-1} \gamma_{(\mu\ne 2)} p^{(\mu\ne 2)} + 
 m^{-1} \gamma_2 p^2 \big]  \nonumber\\
 & =
\gamma_2  \big[m^{-1} \gamma_{(\mu\ne 2)} p^{(\mu\ne 2)} - 
 m^{-1} \gamma_2 p^2 \big] K \nonumber\\
 & =  \big[- m^{-1} \gamma_{(\mu\ne 2)} p^{(\mu\ne 2)} - 
 m^{-1} \gamma_2 p^2 \big] \gamma_2  K \nonumber\\
 & = - \big[ m^{-1} \gamma_\mu p^\mu\big] \gamma_2 K \nonumber\\
& = -\mathcal{P}\mathcal{C}.
\end{align}
Where we have successively used the facts that in Weyl representation $\gamma_2$ is imaginary (see
(\ref{eq:Pauli-Matrices}) and (\ref {eq:diracgamma-lower})), and that $\gamma_2$ anti-commutes with
$\gamma_{(\mu\ne 2)}$. This anti-commutativity of the parity and charge conjugation operators
\beq
\left\{ \mathcal{C}, \mathcal{P}\right\} = 0\label{eq:CP-formal}
\eeq
immediately yields the result that if $\psi(p^\mu)$ are eigenspinors of the parity operator
\beq
\mathcal{P} \psi(p^\mu) = \pm \psi(p^\mu)
\eeq
then the $\mathcal{C}$ transformed spinors $\mathcal{C} \psi(p^\mu)$ carry opposite parity eigenvalue
\beq
\mathcal{P} \left[ \mathcal{C} \psi\right] = \mp \left[ \mathcal{C} \psi\right]. 
\eeq      
Stated differently,  (\ref{eq:Dirac-pd}) changes as follows
\begin{align}
\underbrace{\left(\gamma_\mu p^\mu -m \openone\right) \psi^S_\sigma(p^\mu) = 0,\quad
		\left(\gamma_\mu p^\mu +m\openone\right) \psi^A_\sigma(p^\mu) = 0}_\downarrow~~~~~\nonumber\\
		\overbrace{\left(\gamma_\mu p^\mu + m \openone\right)\left[\mathcal{C}  \psi^S_\sigma(p^\mu)\right] = 0,\quad
		\left(\gamma_\mu p^\mu - m\openone\right) \left[\mathcal{C}\psi^A_\sigma(p^\mu)\right] = 0}.\label{eq:Dirac-pd-C}
\end{align}
In this sense, the relative eigenvalues of the $\mathcal{P}$ may be identified as a charge under the charge conjugation operator. 

The result on the anti-commutativity of the $\mathcal{C}$ and $\mathcal{P}$ operators arrived at
(\ref{eq:CP-formal}) may be obtained in another manner and it checks the internal consistency of various  choices of phase factors. For this analysis we note that (\ref{eq:dirac-sa-uv}) together with (\ref{eq:Dirac-pd-C}) imply
\beq
v_\sigma(p^\mu) \propto \mathcal{C} u_\sigma(p^\mu),\quad 
u_\sigma(p^\mu) \propto \mathcal{C} v_\sigma(p^\mu)
\eeq
where the proportionality constants may depend on $\sigma$. We wish to find the proportionality constant(s), so we act $\mathcal{C}$ on each of the $u_\sigma(p^\mu)$ and $v_\sigma(p^\mu)$. With 
expression for $\mathcal{C}$ given by (\ref{eq:gamma2K}) and for $u_+(p^\mu) $ given by
(\ref{eq:upp}) we work through first of the calculations as follows
\beq
\mathcal{C} u_+(p^\mu) =  i  \sqrt{\frac{E+m}{2 m}} 
\left(
\begin{array}{c}
\left(1-\frac{p}{E+m}\right)\Theta\phi_+^\ast(k^\mu) \\
- \left(1+\frac{p}{E+m}\right) \Theta\phi_+^\ast(k^\mu) 
\end{array}
\right)
\eeq
and since $ \Theta\phi^\ast_+ (k^\mu) =   \phi_-(k^\mu) $
\beq
\mathcal{C} u_+(p^\mu) =  i  \sqrt{\frac{E+m}{2 m}} 
\left(
\begin{array}{c}
\left(1-\frac{p}{E+m}\right)\phi_-(k^\mu) \\
- \left(1+\frac{p}{E+m}\right) \phi_-(k^\mu).
\end{array}
\right)
\eeq
Use of (\ref{eq:vp}) then yields\index{Charge conjugation!of Dirac spinors}
\beq
\mathcal{C} u_+(p^\mu) =  i v_+(p^\mu).
\eeq
The use of another identity  $ \Theta\phi^\ast_- (k^\mu) =  - \, \phi_+(k^\mu) $ and calculations similar to as above then tells us that 
\beq
\mathcal{C} u_\pm(p^\mu) =  i v_\pm(p^\mu),\quad \mathcal{C} v_\pm(p^\mu) =  i u_\pm(p^\mu).\label{eq:cuv}
\eeq
As a consequence, 
\beq
\mathcal{C}^2 = \I_4,\quad\{\mathcal{C},\mathcal{P}\} = 0\label{eq:CPanticommutator-P}
\eeq
for the eigenspinors of the parity operator. This is so because
\begin{align}
& \mathcal{C}\mathcal{P} \, u_-(p^\mu)  = \mathcal{C}  u_-(p^\mu) = i v_-(p^\mu)\\
& \mathcal{P}\mathcal{C} \, u_-(p^\mu)  = \mathcal{P}  \left(i v_-(p^\mu)\right) =  - i v_-(p^\mu).
\end{align}
Together, they yield the claimed anti-commutativity of the $\mathcal{C}$ and $\mathcal{P}$ for the eigenspinors of the parity operator $\mathcal{P}$.

Introducing the time reversal operator $\mathcal{T} = i \gamma \mathcal{C}$, with $\gamma$ given by equation (\ref{eq:gamma5}) below, an explicit calculation gives the result\index{Time reversal operator}
\begin{align}
\mathcal{T} u_+(p^\mu)  &= - u_-(p^\mu),\quad  \mathcal{T} u_-(p^\mu)  =  
u_+(p^\mu)\label{eq:tr12}\\
\mathcal{T} v_+(p^\mu)  &=  v_-(p^\mu),\quad  \mathcal{T} v_-(p^\mu)  =  
- v_+(p^\mu).\label{eq:tr34}
\end{align}
The introduced time reversal operator commutes with the charge conjugation operator and also with the parity operator
\begin{equation}
\left[\mathcal{C},\mathcal{T}\right]=0,\quad \left[\mathcal{P},\mathcal{T}\right]=0
\end{equation}
In conjunction with the facts that 
\begin{equation}
\mathcal{P}^2 = \I_4,\quad \mathcal{T}^2=-\I_4
\end{equation}
all the results obtained above combine to yield \index{Dirac spinors!CPT}
\begin{equation}
\left(\mathcal{C}\mathcal{P}\mathcal{T}\right)^2 = \I_4
\end{equation}
for the Dirac spinors.

\chapter{Eigenspinors of charge conjugation operator, Elko}
\label{ch7}

In the usual language, the 1937 Majorana field\index{Majorana field} starts with the Dirac field and identifies the 
$b^\dagger_\sigma(\p)$ with the $a^\dagger_\sigma(\p)$. In the process Majorana constructed a fundamentally neutral field. Later Majorana spinors\index{Majorana!spinors} were introduced as eigenspinors of the charge conjugation operator, see for example Ramond's primer~\citep{Ramond:1981pw}. But as soon these were motivated by the `magic of Pauli matrices,' they were elevated to Grassmann variables. Furthermore, they were simply considered as Weyl spinors in a four component form -- giving rise to the `disguise' argument. 
Nothing similar is required of the Dirac spinors --  at least, in the operator formalism of quantum field theory, which in our opinion is an unambiguous  conceptual continuation of the quantum formalism~\citep{Dirac:1930pam}. Finding the grassmann-isation of doubtful validity, at least for the reasons given in~\citep{Ramond:1981pw} and elsewhere~\citep{Aitchison:2004cs}, we here construct eigenspinors of the charge conjugation operator. We not only avoid the `disguise' argument, but we also refrain from grassmann-isation. The result is a set of four four-component spinors which stand at par with the Dirac spinors. Taken to their logical consequence the new spinors lead to  mass dimension one fermions.
To avoid a possible confusion from arising we call the new spinors as Elko (\underline{\textbf{E}}igenspinoren 
 des \underline{\textbf{L}}adungs\underline{\textbf{k}}on\-jugations\underline{\textbf{o}}perators) -- or, simply as eigenspinors of the charge conjugation operator.

\section{Elko}
 \label{Sec:sevenpointone}

We have at our disposal the charge conjugation operator from 
equation~(\ref{eq:gamma2K}) and the new spinors suggested by the magic of the Wigner time reversal operator from equation~(\ref{eq:lambda}):
\begin{align}
	& \mathcal{C} = \left(\begin{array}{cc}
					\0 & i \Theta \\
					-i \Theta & \0
					\end{array}
				\right) K, \label{eq:ccRepeat}\\
	& \lambda(p^\mu) = \left(\begin{array}{c}
					\zeta_\lambda \Theta \phi^\ast_L(p^\mu)\\
					\phi_L(p^\mu)
					\end{array}
					\right).
					\label{eq:lambdaRepeat}
	\end{align}
	We now establish the following two results:
\begin{itemize}

\item
The $\lambda(p^\mu)$ become eigenspinors of the charge conjugation operator with doubly degenerate eigenvalues, $\pm 1$, if the phase 
$\zeta_\lambda$ that appears in $\lambda(p^\mu)$ is set to $\pm i$. 

\item In contrast to the eigenspinors of the parity operator, the right- and left- transforming components of the new spinors have opposite helicities. 

\end{itemize}

\vspace{11pt}


To establish the first of the two enumerated results we act the charge conjugation operator on the new spinors. After the action of $K$ in $\mathcal{C}$, the result reads
	\begin{equation}
		\mathcal{C} \lambda(p^\mu)  = 
		\left(\begin{array}{cc}
					\0 & i\Theta \\
					-i \Theta & \0
					\end{array}
					\right) 
					\left(\begin{array}{c}
					\zeta^\ast_\lambda \Theta \phi_L(p^\mu)\\
					\phi^\ast_L(p^\mu)
					\end{array}
					\right) 
\end{equation}
Exploiting the property $\Theta^2 = -\I$ simplifies the above expression into
\begin{equation}		
\mathcal{C} \lambda(p^\mu) 			 =
					 \left(\begin{array}{cc}
					 i \Theta \phi^\ast_L(p^\mu)\\
					 i \zeta^\ast_\lambda \phi_L(p^\mu)
			\end{array}			 \right).	
		\label{eq:Cev}		
\end{equation}
The choice $\zeta_\lambda = \pm i$ makes $\lambda(p^\mu)$ become eigenspinors of $\mathcal{C}$ 
with doubly degenerate eigenvalues $\pm 1$:
\begin{equation}
	\mathcal{C}  \lambda^S(p^\mu) = +  \lambda^S(p^\mu),\quad
	\mathcal{C}  \lambda^A(p^\mu) = -  \lambda^A(p^\mu),    \label{eq:elko-cc}
	\end{equation} 
where	 
	 	\begin{equation}
  		\lambda(p^\mu) = \left\{
		\begin{array}{ll}
	 	\lambda^S(p^\mu) =
	\left(\begin{array}{c}  i \Theta \phi_L^\ast(p^\mu)\\
	                       \phi_L(p^\mu)
	                       \end{array} 	\right)	
		 & \mbox{for} ~ \zeta_\lambda = + i \\ \\  
	 	
		\lambda^A(p^\mu) =
		\left(\begin{array}{c}-  i \Theta \phi_L^\ast(p^\mu)\\
	                       \phi_L(p^\mu)
	                       \end{array} 	\right)	 &  \mbox{for} ~ \zeta_\lambda = - i   
		\end{array}\right.
		\label{eq:lsa}
 	 \end{equation}
 
 To establish the second of the two enumerated results we proceed as follows.
With the structure $\zeta_\lambda \Theta \phi^\ast_L(p^\mu)$ in mind, we complex conjugate (\ref{eq:hd}) to get (suppressing the subscript $\mathcal{L}$)
\begin{equation}
\s^\ast \cdot\widehat{\p}\, \phi^\ast_\pm(k^\mu) = \pm \phi^\ast_\pm(k^\mu)\label{eq:hdd}
\end{equation}
and then replace $\s^\ast$ by $- \Theta\s\Theta^{-1}$ in accordance with  (\ref{eq:wt})
\beq
\Theta\s\Theta^{-1}  \cdot\widehat{\p}\, \phi^\ast_\pm(k^\mu) = \mp \phi^\ast_\pm(k^\mu)
\eeq
Next we use the fact that  $\Theta^{-1} \leftrightharpoons -\Theta$ to replace
$\Theta\s\Theta^{-1}$ by $ \Theta^{-1}\s\Theta$
\beq
 \Theta^{-1}\s\Theta  \cdot\widehat{\p}\, \phi^\ast_\pm(k^\mu) = \mp \phi^\ast_\pm(k^\mu)
\eeq
A  left multiplication by $\Theta$ furnishes us the result
\beq
\s\cdot\widehat{\p} \left[ \Theta \phi_\pm^\ast(k^\mu) \right] =  \mp  \left[ \Theta \phi_\pm^\ast(k^\mu) \right] \label{eq:zimpok3}
\eeq
The result (\ref{eq:boostL}) immediately translates the validity of the above expression for all $p^\mu$
\beq
\s\cdot\widehat{\p} \left[ \Theta \phi_\pm^\ast(p^\mu) \right] =  \mp  \left[ \Theta \phi_\pm^\ast(p^\mu) \right] \label{eq:zimpok3new}
\eeq
Thus, the right- and left- transforming components of $\lambda(k^\mu)$  are constrained to have opposite helicities -- in sharp contrast to the constraints (\ref{eq:constraint12})  for the eigenspinors of the parity 
operator $\mathcal{P}$. We will see in Chapter~\ref{ch10} that this fact endows Elko with an unusual  property under rotation. Exploiting (\ref{eq:phasefactor}) it results in an unexpected cosmological effect.
 

\section{Restriction on local gauge symmetries}
\label{sec:Restriction}

Because of the presence of the operator $K$ in Eq.~(\ref{eq:gamma2K})
a global transformation of the type
\begin{equation}
			\lambda(p^\mu) \to  \lambda^\prime(p^\mu)= \exp( i \mathfrak{a} \alpha) 
			\lambda(p^\mu) \label{eq:counterpart}
\end{equation}
with $\mathfrak{a}^\dagger = \mathfrak{a}$, a $4\times 4$ matrix and $\alpha\in\mathfrak{R}$, does not preserve the self/anti-self conjugacy of $\lambda(p^\mu)$ under $\mathcal{C}$ unless
the matrix $\mathfrak{a}$ satisfies the condition
\begin{equation}
		\gamma_2 \mathfrak{a}^\ast +\mathfrak{a} \gamma_2 = 0
\end{equation}
The general form of $\mathfrak{a}$ satisfying this requirements is
\begin{align}
		\mathfrak{a} & = 
		\left[
		\begin{array}{cccc}
		 \epsilon  & \beta  & \lambda & 0 \\
		 \beta^\ast  & \delta  & 0 & \lambda  \\
 	\lambda^\ast  & 0 & -\delta  & \beta  \\
 	0 & \lambda^\ast  & \beta^\ast  & -\epsilon  \\
		\end{array}
		\right] 
		\label{eq:mathfraka}
\end{align}
with $ \epsilon,\delta \in \mathfrak{R}$ and $\beta,\lambda \in\mathfrak{C}$ (with no association with the same symbols used elsewhere in this work). 

For a field constructed with the eigenspinors of $\mathcal{C}$ as expansion coefficients,
the usual local $U(1)$ interaction is ruled out as 
 the form of $\mathfrak{a}$ given by (\ref{eq:mathfraka}) does not allow a solution with $\mathfrak{a}$ proportional to an identity matrix.
As remarked around equation (\ref{eq:remarks}), for the eigenspinors of the $\mathcal{P}$, defined by (\ref{eq:Dirac-p-new}), the counterpart of (\ref{eq:counterpart} ) is trivially satisfied. And thus the two fields, one based on $\mathcal{P}$ eigenspinors and the other constructed from the $\mathcal{C}$ eigenspinors, carry intrinsically different possibility for their interaction through local gauge fields. The simplest non-trivial choice consistent with (\ref{eq:mathfraka}) is given by 
\begin{equation}
\mathfrak{a} = \gamma = \frac{i}{4!} \epsilon_{\mu\nu\lambda\sigma}
\gamma^\mu\gamma^\nu\gamma^\lambda\gamma^\sigma = \left(\begin{array}{cc}
\I &\0\\
\0 & -\I
\end{array}\right)\label{eq:gamma5}
\end{equation}
where $\epsilon_{\mu\nu\lambda\sigma}$ is defined as
\begin{equation}
\epsilon_{\mu\nu\lambda\sigma} = \left\{
\begin{array}{cl} 
+1, &  \mbox{for  $\mu\nu\lambda\sigma$ even permutation of 0123}\\
- 1,&  \mbox{for  $\mu\nu\lambda\sigma$ odd permutation of 0123}\\\
0, & \mbox{if any two of the $\mu\nu\lambda\sigma$ are same}
\end{array}\right.
\end{equation}

\chapter{Construction of Elko}
\label{ch8}

With Elko defined in the previous chapter we now provide explicit construction of the new spinors and discuss the subtle departure from our 2005 publications~\citep{Ahluwalia:2004sz,Ahluwalia:2004ab}. These  departure, when coupled with the new dual introduced in Chapter~\ref{ch11}, lie at the heart of evaporating away the non-locality and a lack of full Lorentz covariance of our earlier works. The arguments that follow are adapted from~\citep{Ahluwalia:2016rwl,Ahluwalia:2016jwz}.

\section{Elko at rest}\index{Elko!at rest}

To obtain an explicit form of Elko requires the `rest' spinors $\lambda(k^\mu)$. That done, we then have for an arbitrary $p^\mu$\index{Elko!explicit construction of}
\begin{equation}
	\lambda(p^\mu) 
	= \sqrt{\frac{E + m }{2 m}}
						\left[
						\begin{array}{cc}
						\I + \frac{\boldsymbol{\sigma}\cdot\mathbf{p}}{E +m} & \0 \\
						\0 & \I - \frac{\boldsymbol{\sigma}\cdot\mathbf{p}}{E +m} 
						\end{array}
						\right] \lambda(k^\mu)  
	\label{eq:elkoboost-again}
\end{equation}
with the boost operator above the same as in ~(\ref{eq:sb-again}). 

To construct $\lambda(k^\mu)$ we have to choose global phases for each of the  two $\phi_\pm(k^\mu)$ as discussed after equations (\ref{eq:zimpok52-new}) and (\ref{eq:zimpok52-new-new}). With that choice made we are still left with the additional phase freedom
\begin{equation}
 \lambda^S_+(k^\mu)  = e^{i\xi_1} \left[
					\begin{array}{c}
					i \Theta\left[\phi_+(k^\mu)\right]^\ast\\
								\phi_+(k^\mu) 
					\end{array}
					\right],\hspace{7pt}  		 
\lambda^S_-(k^\mu)  = e^{i\xi_2} \left[
				\begin{array}{c}
				i \Theta\left[\phi_-(k^\mu)\right]^\ast\\
				\phi_-(k^\mu)
				\end{array}
				\right]  \label{eq:zimpok0}
\end{equation}
and
\begin{equation}
 \lambda^A_+(k^\mu)  = e^{i\xi_3} \left[
					\begin{array}{c}
					- i \Theta\left[\phi_-(k^\mu)\right]^\ast\\
								\phi_-(k^\mu)
					\end{array}
					\right], 
					\hspace{7pt}
	\lambda^A_-(k^\mu) = e^{i\xi_4} \left[
				\begin{array}{c}
				- i \Theta\left[\phi_+(k^\mu)\right]^\ast\\
					\phi_+(k^\mu)
				\end{array}
				\right]
				\label{eq:zimpok91}
\end{equation} 
with $\xi_1,\xi_2,\xi_3,\xi_4\in \Re$. We set $\xi_1=\xi_2=\xi_3 =0$ and 
$\xi_4=\pi$
\begin{align}
 \lambda^S_+(k^\mu)  & =  \left[
					\begin{array}{c}
					i \Theta\left[\phi_+(k^\mu)\right]^\ast\\
								\phi_+(k^\mu) 
					\end{array}
					\right],\hspace{7pt}  		 
\lambda^S_-(k^\mu)  =  \left[
				\begin{array}{c}
				i \Theta\left[\phi_-(k^\mu)\right]^\ast\\
				\phi_-(k^\mu)
				\end{array}
				\right]  \label{eq:zimpok0new}\\
 \lambda^A_+(k^\mu)  &=  \left[
					\begin{array}{c}
					- i \Theta\left[\phi_-(k^\mu)\right]^\ast\\
								\phi_-(k^\mu)
					\end{array}
					\right], 
					\hspace{7pt}
	\lambda^A_-(k^\mu) =  \left[
				\begin{array}{c}
				 i \Theta\left[\phi_+(k^\mu)\right]^\ast\\
					- \phi_+(k^\mu)
				\end{array}
				\right]
				\label{eq:zimpok91new}
				\end{align}

This choice of phases, coupled with advances in understanding duals and adjoints~\citep{Ahluwalia:2016rwl,Ahluwalia:2016jwz}, 
completely resolves the lingering problems with locality and Lorentz covariance encountered in~\citep{Ahluwalia:2004sz,Ahluwalia:2004ab,Ahluwalia:2008xi,Ahluwalia:2009rh,Ahluwalia:2010zn}. Table 7.1 tabulates the differences just mentioned explicitly. 

For the Dirac field, the counterpart of these observations follow seamlessly in the Weinberg's formalism. Since Weinberg does not note these details explicitly they seem to have escaped many of the recent textbook expositions on the theory of quantum fields. This has the consequence that the Dirac field, as presented, for example, in~\citep{Ryder:1985wq,Folland:2008zz,Schwartz:2014md}, hides violation of locality and Lorentz symmetry. The former can be seen by locality analysis of the those fields on Majorana-isation,\index{Majorana-isation} while the latter only becomes apparent on comparing the rest spinors with those of Weinberg's analysis. One way of understanding part of this story is to realise that  in a quantum field the complete set of spinors enter as expansion coefficients and the freedom of a global phase that each of these carried before -- by, say satisfying the Dirac equation -- is lost under the usual summation on the helicity degrees of freedom.

\begin{table}\label{tab:comparison}
\begin{minipage}{350pt}
\caption{A comparison of $\lambda(p^\mu)$ with those used in earlier work.}
\label{table2new}
\addtolength\tabcolsep{2pt}
\begin{tabular}{@{}c@{\hspace{91pt}}ccc@{\hspace{10pt}}}
\hline\hline
Here & In reference \citep{Ahluwalia:2004sz,Ahluwalia:2004ab}\footnote{The ${\bf 0}$ there carries the same meaning as $k^\mu$ here.} \\
\hline
$ \lambda^S_+(k^\mu)$ & $\lambda^S_{\{-,+\}}(k^\mu)$  \\ 
$\lambda^S_-(k^\mu)$ & $-\lambda^S_{\{+,-\}}(k^\mu) $\\
$ \lambda^A_+(k^\mu)$ & $- \lambda^A_{\{+,-\}}(k^\mu)$,  ${\textrm {and\; not\;}
				- \lambda^A_{\{-,+\}}(k^\mu)} $\\ 
$\lambda^A_-(k^\mu)$ & $-\lambda^A_{\{-,+\}}(k^\mu)$,  ${\textrm{ and\; not}\;} 
										- \lambda^A_{\{+,-\}}(k^\mu)$
										\\
\hline\hline
\end{tabular}
\end{minipage}
\end{table}\index{Elko!comparison with earlier work}

 \section{Elko are not Grassmann nor are they Weyl in disguise}
 
Under the Dirac dual, as would be seen in Chapters \ref{ch11} and \ref{ch12}, the naive mass term for Elko identically vanishes. So one would be tempted to treat Elko as grassmann numbers -- see, for example, \citep{Aitchison:2004cs}.\index{Elko!not grassmann numbers} As already remarked in  Chapter~\ref{ch7}, we find this transition from the complex valued Elko to their grassmann-isation \index{Grassmann-isation}as mathematically untenable. We would treat Elko as one treated the Dirac spinors in the operator formalism of quantum field theory.

Similarly, one should not be tempted to consider Elko as Weyl spinors in disguise. This is because I include the 
neglected $\zeta_\lambda = -1$ in our formalism. It would be seen as we proceed that just as the Dirac spinors span a four dimensional representation space, the Elko too are endowed with a completeness relation in the same four dimensional representation space (see equation (\ref{eq:completeness-li})). \index{Elko!not Weyl spinors  in disguise} 
Furthermore, we can exploit the algebraic identity 
\begin{equation} 
\Theta\phi^\ast_\pm (k^\mu) = \pm \,  \phi_\mp(k^\mu)   \label{eq:algebraic-identity}
\end{equation}
to rewrite the Elko at rest, given by equations (\ref{eq:zimpok0}) and (\ref{eq:zimpok91}), into\index{Elko!at rest}
\begin{align}
\lambda^S_+(k^\mu) = \left(\begin{array}{cc}
i\phi_-(k^\mu) \\
\phi_+(k^\mu)
\end{array}\right),\quad \lambda^S_-(k^\mu) = \left(\begin{array}{cc}
- i\phi_+(k^\mu) \\
\phi_-(k^\mu)
\end{array}\right) \\
\lambda^A_+(k^\mu) = \left(\begin{array}{cc}
i\phi_+(k^\mu) \\
\phi_-(k^\mu)
\end{array}\right),\quad \lambda^A_-(k^\mu) = \left(\begin{array}{cc}
i\phi_-(k^\mu) \\
-\phi_+(k^\mu)
\end{array}\right)
\end{align}
and compare these with the Dirac spinors at rest given in equations
(\ref{eq:Diracuk}) and (\ref{eq:Diracvk}) in the same degrees of freedom: $\phi_\pm(k^\mu)$. Written in this form all the phase choices, relative between the right- and left- transforming components, and global for each of the spinors, along with their helicities become manifest and stand in contrast to their Dirac counterpart.\index{Elko!comparison with Dirac spinors} 

\section{Elko for any momentum}\index{Elko!at any momentum}

The interplay of the result (\ref{eq:zimpok3}) with the boost
(\ref{eq:sb-again}) and the chosen form of  $\lambda(k^\mu)$ in (\ref{eq:zimpok0new}) to 
 (\ref{eq:zimpok91new}) results in the following form for 
 $\lambda(p^\mu)$\index{Elko!at any momentum}
\begin{align}
		\lambda^S_+(p^\mu) &= \sqrt{\frac{E+m}{2 m} }\left( 1-\frac{p}{E+m}\right)\lambda^S_+(k^\mu),\label{eq:lsp}\\		
		\lambda^S_-(p^\mu) &= \sqrt{\frac{E+m}{2 m} }\left( 1+\frac{p}{E+m}\right)\lambda^S_-(k^\mu)\label{eq:name}
		\end{align}
and
\begin{align}
		\lambda^A_+(p^\mu) &= \sqrt{\frac{E+m}{2 m} }\left( 1+\frac{p}{E+m}\right)\lambda^A_+(k^\mu),\label{eq:lap}\\
		\lambda^A_-(p^\mu) &= \sqrt{\frac{E+m}{2 m} }\left( 1-\frac{p}{E+m}\right)\lambda^A_-(k^\mu). 
		\label{eq:lam}
\end{align}
Or, in a more compact form
\begin{align}
\lambda^S_+(p^\mu) &= \beta_-\lambda^S_+(k^\mu),	
\quad	
		\lambda^S_-(p^\mu) = \beta_+\lambda^S_-(k^\mu)\\
		\lambda^A_+(p^\mu) &= \beta_+\lambda^A_+(k^\mu),
\quad
		\lambda^A_-(p^\mu) = \beta_-\lambda^A_-(k^\mu). 
\end{align}
where 
\begin{equation}
\beta_\pm = \alpha \alpha_\pm
\end{equation}
with $\alpha$ and $\alpha_\pm$ defined in equation~(\ref{eq:apm}).

In a sharp contrast to the eigenspinors of the parity operator the here-considered eigenspinors of the charge conjugation operator, $\lambda^{S,A}_\pm(p^\mu)$, are simply the rest spinors $\lambda(k^\mu)$ scaled by the indicated energy-dependent factors.  In particular, for the eigenspinors of $\mathcal{C}$ the boost does not mix various components of the `rest-frame' spinors. 

An inspection of equations (\ref{eq:lsp}) to (\ref{eq:lam}) suggests that for massless $\lambda(p^\mu)$ the number of degrees of freedom reduces to two, that is those associated with $\lambda^S_-(p^\mu)$ and $\lambda^A_+(p^\mu)$ while the  $\lambda^S_+(p^\mu)$ and $\lambda^A_-(p^\mu)$ vanish identically. 

We will see below that
the parity operator takes $\lambda_-^{S}(p^\mu) \to \lambda_+^{S}(p^\mu)$ and $\lambda_+^{A}(p^\mu) \to \lambda_-^{A}(p^\mu)$. Combining these two observations we conclude that in the massless limit $\lambda(p^\mu)$ have no reflection. \index{Elko!massless limit}

Strictly speaking for massless particles there is no rest frame, or `rest-frame' spinors. The theory must be constructed \textit{ab initio} except that the massless limit 
of certain massive representation spaces yields the massless theory. 
The $\mathcal{R}$ and $\mathcal{L}$ representation spaces belong to that class.
We refer the reader to Weinberg's 1964 work on the subject~\citep{Weinberg:1964ev}.
That such a limit may be taken for the $\mathcal{R}\oplus\mathcal{L}$ representation space is apparent from Weinberg's analysis.

The four four-component Elko, $\lambda^{S,A}_\pm(p^\mu)$, are the expansion coefficients of the new quantum field to be introduced below.
\vspace{11pt}

We parenthetically note that as soon as the first papers introducing Elko and mass dimension one fermions were 
published  da Rocha and Rodrigues Jr. noted that 
Elko belong to class 5 spinors~\citep{daRocha:2005ti}  in the Lounesto classification~\citep[Chapter 12]{Lounesto:2001zz}. \index{Elko!in Lounesto classification}

\chapter{A hint for mass dimension one fermions}

\label{ch9}

\label{sec:elko-do-not-satisfy-Dirac-equation}

A first hint towards mass dimension one fermions arises from the observation that the momentum space Dirac operators $(\gamma_\mu p^\mu \pm m \I)$ do not annihilate Elko
\begin{equation}
\left(\gamma_\mu p^\mu \pm m \I\right)\lambda(p^\mu) \ne 0\label{eq:DiracNot}
\end{equation}

To prove this we act $\gamma_\mu p^\mu$ on each of the Elko enumerated in equations~(\ref{eq:lsp})-(\ref{eq:lam}). We begin this exercise with 
$ \lambda^S_+(p^\mu)$
\begin{align}
\gamma_\mu p^\mu   \lambda^S_+(p^\mu)  = \sqrt{\frac{E+m}{2 m}} &\left[ 1 - \frac{p}{E+m}\right]\nonumber \\
  & \times  \underbrace{ \left[
  E  \gamma_0  + p \left(\begin{array}{cc}
 0 & \s\cdot\widehat{\p} \\
 -\s\cdot\widehat{\p} & 0
 \end{array}\right) 
  \right] }_{\gamma_\mu p^\mu}
 \lambda^S_+(k^\mu)\label{eq:zimpok4}
 \end{align}
and notice that 
\begin{equation}
\left(\begin{array}{cc}
 0 & \s\cdot\widehat{\p} \\
 -\s\cdot\widehat{\p} & 0
 \end{array}\right) \lambda^S_+(k^\mu) = \left(\begin{array}{cc}
 0 & \s\cdot\widehat{\p} \\
 -\s\cdot\widehat{\p} & 0
 \end{array}\right)   
 \left(
					\begin{array}{c}
					i \Theta\left[\phi_+(k^\mu)\right]^\ast\\
								\phi_+(k^\mu) 
					\end{array}
					\right).
\end{equation}
But 
\beq\s\cdot\widehat{\p} \,\phi_+(k^\mu) 
 = \phi_+(k^\mu) 
 \eeq
  while according to (\ref{eq:zimpok3})
 \begin{equation}\s\cdot\widehat{\p} \,\Big[\Theta\left[\phi_+(k^\mu)\right]^\ast\Big]
 = - \Theta\left[\phi_+(k^\mu)\right]^\ast
 \end{equation}
Therefore, we have the result
\begin{align}
\left(\begin{array}{cc}
 0 & \s\cdot\hat{\p} \\
 -\s\cdot\hat{\p} & 0
 \end{array}\right)& \lambda^S_+(k^\mu)  = 
\left(\begin{array}{c}
\phi_+(k^\mu)\\
i \Theta\left[\phi_+(k^\mu)\right]^\ast
\end{array}
\right)  \nonumber \\
 & = \left(\begin{array}{cc}
\0 & \openone \\
\openone & \0\end{array}\right) 
\left(\begin{array}{c}
i \Theta\left[\phi_+(k^\mu)\right]^\ast\\
\phi_+(k^\mu)
\end{array}
\right) 
\end{align}
That is
\begin{equation}
\left(\begin{array}{cc}
 0 & \s\cdot\hat{\p} \\
 -\s\cdot\hat{\p} & 0
 \end{array}\right) \lambda^S_+(k^\mu)  = 
 \gamma_0 \lambda^S_+(k^\mu).
\end{equation}

\noindent
As a consequence (\ref{eq:zimpok4}) simplifies   to 
\begin{equation}
  \gamma_\mu p^\mu  \lambda^S_+(p^\mu)   =    \sqrt{\frac{E+m}{2 m}} \left(1 - \frac{p}{E+m}\right)
\left(
  E    + p 
  \right)  \gamma_0\lambda^S_+(k^\mu).\label{eq:zimpok5}
 \end{equation}
The  standard dispersion relation allows for the replacement 
\begin{equation}
\left(1 - \frac{p}{E+m}\right)(E+p)
\rightarrow  m \left(1 + \frac{p}{E+m}\right)
\end{equation}
To evaluate $\gamma_0 \lambda^S_+(k^\mu) $ we first expand it as
\begin{equation}
\gamma_0 \lambda^S_+(k^\mu)  =
\left(
\begin{array}{c}
\phi_+(k^\mu) \\
i\Theta\phi^\ast_+(k^\mu)
\end{array}
\right)
\end{equation}
and exploit the algebraic result (\ref{eq:algebraic-identity})
 to make  the following substitutions
\begin{equation}
\phi_+(k^\mu) \to i \left(i \Theta\phi^\ast_-(k^\mu)  \right),\quad
i \Theta\phi^\ast_+(k^\mu) \to i\phi_-(k^\mu)
\end{equation}
to obtain the identity
	\begin{equation}
		\gamma_0 \lambda^S_+(k^\mu) = i \lambda^S_-(k^\mu) 
	\end{equation} 
Combined, these two observations reduce (\ref{eq:zimpok5}) to
	\begin{equation}
  	\gamma_\mu p^\mu  \lambda^S_+(p^\mu)  =   i m \sqrt{\frac{E+m}{2 m}} 	\left(1 + \frac{p}{E+m}\right) \lambda^S_-(k^\mu).\label{eq:zimpok6}
 	\end{equation}
Using  (\ref{eq:name}) in the right-hand side of (\ref{eq:zimpok6}) gives
 \begin{equation}
  \gamma_\mu p^\mu  \lambda^S_+(p^\mu) = i m  \lambda^S_-(p^\mu).    \label{eq:er-a1}
 \end{equation}
An exactly similar exercise complements (\ref{eq:er-a1}) with
\begin{align}
 \gamma_\mu p^\mu \lambda^S_-(p^\mu) & = - i m \lambda^S_+(p^\mu) \label{eq:er-a2}\\
 \gamma_\mu p^\mu \lambda^A_-(p^\mu) & = i m \lambda^A_+(p^\mu)  \label{eq:er-b1}\\
 \gamma_\mu p^\mu \lambda^A_+(p^\mu) & = - i m \lambda^A_-(p^\mu) . \label{eq:er-b2}
\end{align} 
These  results are consistent with 
those of~\citep{Dvoeglazov:1995eg,Dvoeglazov:1995kn}. Translated in words these equations combine to yield the following pivotal result: $(\gamma_\mu p^\mu \pm m)$ do not annihilate $\lambda(p^\mu)$.
\vspace{11pt}

Equations (\ref{eq:er-a1}) to  (\ref{eq:er-b2}), coupled with the discussion surrounding  (\ref{eq:counterpart}), contain the rudimentary seeds for the kinematical and dynamical content of the quantum field built upon $\lambda(p^\mu)$ as its expansion coefficients: first, $\lambda(p^\mu)$ are annihilated by the spinorial Klein-Gordon operator (and not by the Dirac operator), and second, 
the resulting kinematic structure cannot support the usual gauge symmetries of the standard model of the high energy physics.
To prove the  the former claim, we multiply (\ref{eq:er-a1}) from the left by $\gamma_\nu p^\nu $ and use (\ref{eq:er-a1}) and (\ref{eq:er-a2}) in succession 
\beq
\gamma_\nu p^\nu 
\gamma_\mu p^\mu  \lambda^S_+(p^\mu)   =  i m  \gamma_\nu p^\nu \lambda^S_-(p^\mu) 
 = im \left( - i m \lambda^S_+(p^\mu)\right) = m^2 \lambda^S_+(p^\mu)
\label{eq:zimpok1952}
\eeq
and then utilise the fact that  the left hand side of the above equation can be rewritten exploiting
$\{\gamma_\mu,\gamma_\nu\}= 2 \eta_{\mu\nu} \I_4$ -- where $\eta_{\mu\nu}$ is the space-time metric with signature $(+1,-1,-1,-1)$) -- as
\beq
\gamma_\nu p^\nu 
\gamma_\mu p^\mu =  \frac{1}{2}\{\gamma_\mu,\;\gamma_\nu\}   p^\mu p^\nu 
 = \eta_{\mu\nu}p^\mu  p^\nu \I_4 .
\eeq
Substituting this result in~(\ref{eq:zimpok1952}), and rearranging gives
\begin{equation}
(\eta_{\mu\nu}p^\mu  p^\nu \I_4 - m^2\I_4)\lambda^{S}_+(p^\mu) = 0.
\end{equation}
Repeating the same exercise with  (\ref{eq:er-a2}) to  (\ref{eq:er-b2}) as the starting point, yields
\begin{equation}
(\eta_{\mu\nu}p^\mu  p^\nu \I_4 - m^2\I_4)\lambda(p^\mu)= 0. \label{eq:skg}
\end{equation}
where $\lambda(p^\mu)$ stands for any of the four eigenspinors of the charge conjugation operator, $\lambda^{S,A}_\pm(p^\mu)$.
\vspace{11pt}

Lest the appearances betray, the $\lambda^{S,A}_\pm(p^\mu)$ are not a set of four scalars, disguised in four-component form repeated four times,  but spinors in the $\mathcal{R}\oplus\mathcal{L}$ representation space of spin one half. Under boosts these spinors contract or dilate by factors of
\begin{equation}
\beta_\pm = \sqrt{\frac{E+m}{2 m} }\left[ 1\pm  \frac{p}{E+m}\right]
\end{equation}
in accord with equations (\ref{eq:lsp}) to  (\ref{eq:lam}) and pick up a minus sign under
$2\pi$ rotation as dictated by (\ref{eq:rotation-on-spinors}).\footnote{See, Chapter~\ref{ch10} for departures from Dirac spinors.}
\vspace{11pt}

\chapter{CPT for Elko}

Since $\mathcal{P} = m^{-1}\gamma_\mu p^\mu$ the results obtained in the previous chapter also  translate to the action of $\mathcal{P}$ on the eigenspinors of the charge conjugation operator $\mathcal{C}$\index{Parity operator!action on Elko}
 \begin{align}
  \mathcal{P} \lambda^S_+(p^\mu) &= i   \lambda^S_-(p^\mu)   \label{eq:er-a1P}\\
    \mathcal{P} \lambda^S_-(p^\mu) & = - i  \lambda^S_+(p^\mu) \label{eq:er-a2P}\\
   \mathcal{P}  \lambda^A_-(p^\mu) & = i  \lambda^A_+(p^\mu)  \label{eq:er-b1P}\\
  \mathcal{P}  \lambda^A_+(p^\mu) & = - i  \lambda^A_-(p^\mu) . \label{eq:er-b2P}
\end{align}

The above can be compacted into the following
\begin{equation}
  \mathcal{P} \lambda^S_\pm (p^\mu) = \pm i   \lambda^S_\mp(p^\mu),\quad
 \mathcal{P}\lambda^A_\pm (p^\mu)  = \mp i  \lambda^A_\mp(p^\mu) \label{eq:parity-b2}
\end{equation}
and lead to 
\begin{equation}\mathcal{P}^2=\I_4\label{eq:psq}
\end{equation}
\vspace{11pt}
Acting $\mathcal{C}$ from the left on the first of the above equations gives
\begin{equation}
\mathcal{C} \mathcal{P} \lambda^S_+(p^\mu) = - i \mathcal{C}   \lambda^S_-(p^\mu) =  -i \lambda^S_-(p^\mu) . \label{eq:zimpok100}
\end{equation}
On the other hand
\begin{equation}
\mathcal{P} \mathcal{C}  \lambda^S_+(p^\mu) =
\mathcal{P}  \lambda^S_+(p^\mu) = i   \lambda^S_-(p^\mu) .
\end{equation}
Adding the above two results leads to anti-commutativity  for the $\mathcal{C}$ and $\mathcal{P}$ for $ \lambda^S_+(p^\mu) $.
Repeating the same exercise for $\lambda^S_-(p^\mu)$ and $\lambda^A_\pm(p^\mu) $
 establishes
that $\mathcal{C}$ and $\mathcal{P}$ anticommute for all the eigenspinors of the charge conjugation operator
\begin{equation}
\{\mathcal{C}, \mathcal{P}\} = 0\label{eq:cp-z}
\end{equation}
just as seen before in (\ref{eq:CPanticommutator-P}) for the eigenspinors of the parity operator $\mathcal{P}$.

 The action of the time reversal operator on $\lambda(p^\mu)$ is as follows (as an explicit calculation shows):
 \begin{align}
 \mathcal{T}\lambda^S_+(p^\mu) = i \lambda^A_-(p^\mu), \quad
   \mathcal{T}\lambda^S_-(p^\mu) = - i \lambda^A_+(p^\mu) \label{eq:timereversalS}\\
    \mathcal{T}\lambda^A_+(p^\mu) = i \lambda^S_-(p^\mu), \quad
   \mathcal{T}\lambda^A_-(p^\mu) = - i \lambda^S_+(p^\mu)
   \label{eq:timereversalA}
 \end{align}
\vspace{11pt} 
with the square of $\mathcal{T}$ acting on Elko giving, $-\I_4$
\begin{equation}
\mathcal{T}^2 = - \I_4\label{eq:tsq}
\end{equation}
In addition, we find that 
\begin{equation}
\left[\mathcal{C},\mathcal{T}\right]=0,\quad \left[\mathcal{P},\mathcal{T}\right]=0\label{eq:pt}
\end{equation}
hold for Elko, as for the Dirac spinors. In consequence, for Elko $\left(\mathcal{C}\mathcal{P}\mathcal{T}\right)^2=\I_4$. \index{Elko!CPT}This contrasts with contrary results reported in our own earlier work, and those of several other authors. The difference can be traced to an unambiguous treatment $\mathcal{P}$ for Elko, and to the differences tabulated in Table 7.1.

\chapter{Elko in Shirokov-Trautman, Wigner, and Lounesto classifications}

In the Shirokov-Trautman\index{Shirokov-Trautman classification} classification~\citep{Shirokov:1960ym,Trautman:2005qx}  a spinor is classified by the possible choices of signs $\lambda,\mu,\nu\in \{+,-\}$ in the relations
\begin{equation}
\mathcal{P} \mathcal{T} = \lambda\, \mathcal{T} \mathcal{P},\quad
\mathcal{P}^2=\mu\, \I_4,\quad\mathcal{T}^2 = \nu\, \I_4
\end{equation}
Commutativity of $\mathcal{P}$ and $\mathcal{T}$ as given in 
equation~(\ref{eq:pt}) gives $\lambda= +$, while results~(\ref{eq:psq}) and ~(\ref{eq:tsq}) give $\mu=+$ and $\nu=-$. Thus Elko belongs to 
$(\lambda,\mu,\nu)$ class identified as $(+,+,-)$.

In the Wigner classification\index{Wigner classification} one may go beyond the relative intrinsic parity of the particles and antiparticles to be same for bosons, and opposite for fermions~\citep{Wigner:1962ep}. The anti-commutativity of 
$\mathcal{C}$ and $\mathcal{P}$ reached in 
equation~(\ref{eq:cp-z}) places Elko in the same class as the Dirac spinors. This should not make the reader infer that all other properties of Elko and Dirac spinors are the same but only that both of these spinors have particles and antiparticles with the opposite relative intrinsic parity.

Lounesto classification\index{Lounesto classification} is based on the bilinear invariants associated with the spinors under the standard Dirac dual~\citep{Lounesto:2001zz}.
An early analysis given in reference~\citep{daRocha:2005ti} established that Elko belong to class 5 in this classification. The cited work has been extended in a series of later papers, see for example in~\citep{HoffdaSilva:2017waf,HoffdaSilva:2016ffx}, with additional new structure unearthed by the use of the dual appropriate for Elko. This classification may be particularly helpful for studying currents associated with Dirac spinors and Elko, and the gauge symmetries they 
arise from. However, Lounesto did not consider the issue of associated Lagrangian densities for each of the classes that he proposed.

\chapter{Rotation induced effects on Elko}\label{ch10}\index{Elko!rotation}


Despite its simple roots in the spin one half representation of the Lorentz algebra the celebrated minus sign that a $ 2 \pi$ rotation induces on  Dirac spinors has intrigued the physicist and the lay scholar alike~~\citep{Aharonov:1967zz,Dowker:1969ia,RAUCH1974369,Werner:1975wf,Silverman:1980mp,Horvathy:1984sm,Klein:1976qc}. 
For a general rotation about an axis, say $\widehat{\p}$, by an angle $\vartheta$ the effect of rotation is simply a multiplication  of the original spinor by a phase factor $\exp(\pm i \vartheta/2)$ -- the sign determined by the helicity of the spinor. 

In sharp contrast to the Dirac spinors,  Elko not only acquire a minus sign for the same $2 \pi$ rotation but the effect of a general rotation is to create a specific admixture of the self-conjugate and anti-self conjugate spinors -- the resulting spinors are still eigenspinors of the charge conjugation operator. This apparent paradox  is resolved.
This property is a direct consequence of the fact that the right and left transforming components of Elko are endowed with opposite helicities (see equation (\ref{eq:zimpok3new}), and each of these picks up an opposite phase in accordance with 
the result~(\ref{eq:phasefactor}).\footnote{This chapter is an adapted version of~\citep{Ahluwalia:2018hfm}}

\section{Setting up an orthonormal cartesian coordinate system with 
$\widehat{\textbf{p}}$ as one of its axis}

Understanding the new spinors vastly simplifies in the basis we have chosen for them. Their right- and left- transforming components have their spin projections assigned not with respect an external $\widehat{z}$-axis but to a self referential unit vector associated with their motion: that is $\widehat{\textbf{p}}$. To  establish the result outlined above we find it convenient to erect an orthonormal cartesian coordinate system with $\widehat{\textbf{p}}$ as one of its axis.
With $\widehat{\textbf{p}}$ chosen as\footnote{Where $\sin[\theta]$ is abbreviated as $s_\theta$ with obvious extension to other trigonometric function.} 
\begin{equation}
\left(s_\theta c_\phi, s_\theta s_\phi,c_\theta\right)
\end{equation}
 we introduce two more unit vectors 
 \begin{align}
\widehat{\eta}_+ & \stackrel{\textrm{def}}{=}\frac{1}{\sqrt{2+a^2}}\left(1,1,a\right), \quad a \in \Re \\
\widehat{\eta}_-  &  \stackrel{\textrm{def}}{=} 
\frac{\widehat{\textbf{p}}\times\widehat{\eta}_+ }
{\sqrt{\left(\widehat{\textbf{p}}\times\widehat{\eta}_+ \right)
\cdot 
\left(\widehat{\textbf{p}}\times\widehat{\eta}_+ \right)}}
\label{eq:etam}
\end{align}
and impose the requirement
\begin{align}
& \widehat{\eta}_+\cdot \widehat{\eta}_- = 0,\quad
\widehat{\eta}_+\cdot \widehat{\textbf{p}} = 0,\quad
\widehat{\eta}_-\cdot \widehat{\textbf{p}} = 0\\
& \widehat{\eta}_+\cdot \widehat{\eta}_+ = 1,\quad
\widehat{\eta}_-\cdot \widehat{\eta}_- = 1.
\end{align}
Requiring $\widehat{\eta}_+\cdot \widehat{\textbf{p}} $ to vanish 
reduces 
$\widehat{\eta}_+ $ to
\begin{equation}
\widehat{\eta}_+ = \frac{1}{\sqrt{2 + (1+s_{2\phi})t^2_\theta}}\Big( 1,1,-(c_\phi + s_\phi)t_\theta\Big).
\end{equation}
Definition (\ref{eq:etam}) then immediately yields
\begin{align}
\widehat{\eta}_- = \frac{1}{\sqrt{2 + (1+s_{2\phi})t^2_\theta}}\Big(  -c_\theta - s_\theta s_\phi(c_\phi+s_\phi)t_\theta, \\
 c_\theta + s_\theta c_\phi(c_\phi+s_\phi)t_\theta,
 s_\theta(c_\phi-s_\phi) \Big).
\end{align}

In the limit when both the $\theta$ and $\phi$ tend to zero, the above-defined unit vectors take the form
\begin{align}
\widehat{\eta}_+\big\vert_{\theta\to 0,\phi\to 0} &= \frac{1}{\sqrt{2}}\left(1,1,0\right) \label{eq:etaplus}\\
\widehat{\eta}_-\big\vert_{\theta\to 0,\phi\to 0} &= \frac{1}{\sqrt{2}}\left(-1,1,0\right)\label{eq:etaminus}
\end{align}
and orthonormal system $\left(\widehat{\eta}_-,\widehat{\eta}_+,\widehat{\textbf{p}}\right)$
 does not reduce to the standard cartesian coordinate system  
$(\widehat{\textbf{x}},\widehat{\textbf{y}},\widehat{\textbf{z}})$. For this reason, without destroying the orthonormality of the introduced unit vectors and guided by (\ref{eq:etaplus}) and (\ref{eq:etaminus}), 
 we exploit the freedom of a rotation about the $ \widehat{\textbf{p}}$ axis (in the plane defined by $\widehat{\eta}_+$ and $\widehat{\eta}_-$) and introduce
 \begin{equation}
 \widehat{\textbf{p}}_\pm \stackrel{\textrm{def}}{=} \frac{1}{\sqrt{2}}\left(\widehat{\eta}_+ \pm \widehat{\eta}_-\right)
 \end{equation}
The set of axes $\left(\widehat{\textbf{p}}_-,\widehat{\textbf{p}}_+,\widehat{\textbf{p}}\right)$ do indeed form a right-handed coordinate system that reduces to the standard cartesian system in the limit  both the $\theta$ and the $\phi$ tend to zero.

\section{Generators of the rotation in the new coordinate system}
 
We now introduce three generators of rotations about each of the three axes
\begin{equation}
\texttt{J}_- \stackrel{\textrm{def}}{=} \frac{\s}{2}\cdot \widehat{\textbf{p}}_-
,\quad 
\texttt{J}_+ \stackrel{\textrm{def}}{=} \frac{\s}{2}\cdot \widehat{\textbf{p}}_+
,\quad 
\texttt{J}_p \stackrel{\textrm{def}}{=} \frac{\s}{2}\cdot \widehat{\textbf{p}}
\end{equation}
$\texttt{J}_p $ coincides with the helicity operator $\mathfrak{h} $ defined in (\ref{eq:helicity}). As a check, a straightforward exercise shows that the three $\texttt{J}'s$ satisfy 
the $\mathfrak{s}\mathfrak{u}(2)$ algebra needed for generators of rotation
\begin{equation}
\left[\texttt{J}_-,\texttt{J}_+\right] = i  \texttt{J}_p,\quad
\left[\texttt{J}_p,\texttt{J}_-\right] = i  \texttt{J}_+,\quad
\left[\texttt{J}_+,\texttt{J}_p\right] = i  \texttt{J}_-
\end{equation}
Since Elko reside in the $\mathcal{R}\oplus\mathcal{L}$ representation space and as far as rotations are concerned both the $\mathcal{R}$ and
$\mathcal{L}$ spaces are served by the same generators of rotations. We therefore
introduce
\begin{equation}
\texttt{h}_ - =\left(\begin{array}{cc}
{\mathtt{J}_-} & \texttt{0}_2\\
\texttt{0}_2 &{\mathtt{J}_-} 
\end{array}
\right),\quad
\texttt{h}_ + =\left(\begin{array}{cc}
{\mathtt{J}_+} & \texttt{0}_2\\
\texttt{0}_2 &{\mathtt{J}_+} 
\end{array}
\right),\quad
\texttt{h}_ p =\left(\begin{array}{cc}
{\mathtt{J}_p} & \texttt{0}_2\\
\texttt{0}_2 &{\mathtt{J}_p} 
\end{array}
\right)
\end{equation}
where $\texttt{0}_2$ is a $2\times 2$ null matrix. A straight forward calculation then yields the result
\begin{align}
\textrm{h}_p \,\lambda^S_+(p^\mu) &= -\frac{1}{2}\lambda^A_-(p^\mu),\quad
\textrm{h}_p \,\lambda^S_-(p^\mu) = -\frac{1}{2}\lambda^A_+(p^\mu)\label{eq:hplamdaS}\\
\textrm{h}_p \,\lambda^A_+(p^\mu) & = -\frac{1}{2}\lambda^S_-(p^\mu),\quad
\textrm{h}_p \,\lambda^A_-(p^\mu)  = -\frac{1}{2}\lambda^S_+(p^\mu)\label{eq:hplamdaA}
\end{align}
with the consequence that each of the Elko is an eigenspinor of $\texttt{h}_p^2$
\begin{equation}
\textrm{h}_p^2 \,\lambda^{S,A}_\pm(p^\mu) = \frac{1}{4}\lambda^{S,A}_\pm(p^\mu)
\end{equation}
The action of $\texttt{h}_-$ and $\texttt{h}_+$ on Elko is more involved,  
for instance
\begin{equation}
\texttt{h}_+\,\lambda^S_+(p^\mu) =  - \frac{1}{2}\left(
\alpha \lambda^S_-(p^\mu) + \beta\lambda^A_+(p^\mu)\right)
\end{equation}
with
\begin{align}
\alpha =  - \frac{im (m+E)\big(\cos\phi(1+\sec\theta)+(-1+\sec\theta)\sin\phi\big)}
{\sqrt{4+2(1+\sin 2\phi)\tan^2\theta}}\\
\beta=  - \frac{m (m+E)\big(\cos\phi(-1+\sec\theta)+(1+\sec\theta)\sin\phi\big)}
{\sqrt{4+2(1+\sin 2\phi)\tan^2\theta}}
\end{align}
but the action of their squares is much simpler and reads
\begin{equation}
\textrm{h}_-^2 \,\lambda^{S,A}_\pm(p^\mu) = \frac{1}{4}\lambda^{S,A}_\pm(p^\mu),\quad
\textrm{h}_+^2 \,\lambda^{S,A}_\pm(p^\mu) = \frac{1}{4}
\lambda^{S,A}_\pm(p^\mu).
\end{equation}
This exercise thus establishes that 
\begin{equation}
\texttt{h}^2 \stackrel{\textrm{def}}{=} \textrm{h}_- ^2+ \textrm{h}_+^2 + \textrm{h}_p^2
\end{equation}
while acting on each of the Elko yields 
\begin{equation}
\texttt{h}^2 \lambda^{S,A}_\pm = \frac{3}{4}  \lambda^{S,A}_\pm =\frac{1}{2}
\left(1+ \frac{1}{2}\right) \lambda^{S,A}_\pm
\end{equation}
and confirms spin one half for Elko.

For ready reference, we note the counterpart of (\ref{eq:hplamdaS}) and 
 (\ref{eq:hplamdaA}) for the Dirac spinors: 
 \begin{align}
\textrm{h}_p \, u_+(p^\mu) &= \frac{1}{2} u_+(p^\mu),\quad
\textrm{h}_p \,u_-(p^\mu)  = -\frac{1}{2}u_-(p^\mu) 
\label{eq:hplamdau}\\
\textrm{h}_p \,v_+(p^\mu) & = -\frac{1}{2}v_+(p^\mu),\quad
\textrm{h}_p \,v_-(p^\mu)  = \frac{1}{2}v_-(p^\mu).\label{eq:hplamdav}
\end{align}

\section{The new effect}

The formalism developed in this chapter so far immediately establishes the opening claim of this chapter. For simplicity we consider a rotation by 
$\vartheta$ about $\widehat{\p}$ axis and find that  a $2\pi$ rotation induces the expected minus sign for Elko, but for a general rotation it mixes the self and antiself conjugate spinors:
\begin{equation}
\exp\left({i \texttt{h}_p \vartheta}\right)\, \lambda^S_\pm(p^\mu) = \cos\left({\vartheta}/{2}\right) \lambda^S_\pm(p^\mu) - i \sin\left({\vartheta}/{2}\right)
\lambda^A_\mp(p^\mu)\label{eq:22}
\end{equation}
 and
 \begin{equation}
\exp\left({i \texttt{h}_p \vartheta}\right)\, \lambda^A_\pm(p^\mu) = \cos\left({\vartheta}/{2}\right) \lambda^A_\pm(p^\mu) - i \sin\left({\vartheta}/{2}\right)
\lambda^S_\mp(p^\mu).\label{eq:23}
\end{equation}
In contrast, for the Dirac spinors we have the well-known result
\begin{equation}
\exp\left({i \texttt{h}_p \vartheta}\right) \psi_\pm(p^\mu) = \exp(\pm i\vartheta/2) \psi_\pm(p^\mu) 
\end{equation}
where $\psi_\pm(p^\mu)$ stands for any one of the four $u$ and $v$ spinors of Dirac given in equations (\ref{eq:upp}) to (\ref{eq:vm}).

\texttt{Do the results (\ref{eq:22}) and (\ref{eq:23}) imply that rotation induces loss of self/anti-self conjugacy under  charge conjugation $\mathcal{C}$?} The answer is: no. To see this we observe that
\begin{eqnarray}
\mathcal{C} \left[ i \lambda^A_{\pm}(\p)\right] &=& (-i)(-\lambda^A_{\pm}(\p))
=+ \left[ i \lambda^A_{\pm}(\p)\right] \label{eq:obs1}\\
\mathcal{C} \left[ i \lambda^S_{\pm}(\p)\right] &=& (-i)(\lambda^S_{\pm}(\p))
=- \left[ i \lambda^S_{\pm}(\p)\right].\label{eq:obs2}
\end{eqnarray}
We thus define a set of new self and anti-self conjugate spinors (see (\ref{eq:22}) and (\ref{eq:23}))
\begin{eqnarray}
 \lambda^s(\p) &\stackrel{\text{def}}{=}& \cos\left({\vartheta}/{2}\right) \lambda^S_\pm(\p) - i \sin\left({\vartheta}/{2}\right) \lambda^A_\mp(\p) \\
 \lambda^a(\p)& \stackrel{\text{def}}{=} &\cos\left({\vartheta}/{2}\right) \lambda^A_\pm(\p) - i \sin\left({\vartheta}/{2}\right) \lambda^S_\mp(\p)
\end{eqnarray}
and verify that 
 \begin{equation}\mathcal{C}\lambda^{s}_\pm (\p) = + \lambda^{s}_\pm (\p), \quad\mathcal{C}\lambda^{a}_\pm (\p) = - \lambda^{a}_\pm (\p).
\end{equation} 
As a result (\ref{eq:22}) and  (\ref{eq:23}) reduce to
\begin{eqnarray}
\exp\left({i \texttt{h}_p \vartheta}\right)\, \lambda^S_\pm(\p) = \lambda^s_\pm(\p) \\
\exp\left({i \texttt{h}_p \vartheta}\right)\, \lambda^A_\pm(\p) = \lambda^a_\pm(\p)
\end{eqnarray}
and confirm that rotation preserves $\mathcal{C}$-self/anti-self conjugacy of Elko. The explicit expressions for the new set of Elko are now readily obtained, and read:
\begin{equation}
\lambda^s_\pm(\p) = \varrho_\pm\lambda^S_\pm(\p),\quad
\lambda^a_\pm(\p) = \varrho_\mp\lambda^A_\pm(\p)
\end{equation}
where the $4\times 4$ matrices $\varrho_\pm$ are defined as
\begin{equation}
\varrho_\pm =\left(
\begin{array}{cc}
e^{\mp i\vartheta/2} \mathbb{I} & \mathbb{O}\\
\mathbb{O} & e^{\pm i\vartheta/2} \mathbb{I}
\end{array}
\right).
\end{equation}
In the above expression, $\mathbb{I}$ and $\mathbb{O}$ are $2\times2$ identity and null matrices respectively. Apart from the special 
values of $\vartheta$ for which
\begin{equation}
\varrho_\pm =\bigg\{\begin{array}{ll}
- 1, & \mbox{for $\theta = 2\pi$}\\
+ 1,& \mbox{for $\theta = 4\pi$}.
\end{array}
\end{equation}
 Elko and Dirac spinors behave differently.
 
 The next question thus arises: Since in all physical observables  Elko appear as bilinears, do the sets $\{\lambda^S_\pm(\p),
 \lambda^A_\pm(\p)\}$ and  $\{\lambda^s_\pm(\p),
 \lambda^a_\pm(\p)\}$ yield identical results? The answer is: yes. It comes about because
 \begin{equation}
 (\varrho_\pm)^\dagger \gamma_0 \varrho_\mp = \gamma_0.
 \end{equation}
 So from the orthonormality relations, to the completeness relations, to the spin sums, the two sets 
 $\{\lambda^S_\pm(\p),
 \lambda^A_\pm(\p)\}$ and  $\{\lambda^s_\pm(\p),
 \lambda^a_\pm(\p)\}$  carry identical results.

\chapter{Elko-Dirac interplay, a temptation and a departure}
\label{ch11}
\section{Null norm of massive Elko and Elko-Dirac interplay}

The result that the spin one-half eigenspinors of the charge conjugation operator satisfy only the Klein Gordon equation would suggest that at the `classical level' the Lagrangian density for the 
$\lambda(x)$ would simply be
\begin{equation}
\mathfrak{L}(x) = \partial^\mu{\overline{\lambda}(x)}\,\partial_\mu {{\lambda(x)}} - m^2 {\overline{\lambda}}(x) \lambda(x)\label{eq:wrong}
\end{equation}
where $\lambda(x)$ is a classical field with $\lambda^{S,A}(p^\mu)$ as its Fourier coefficients.
This  apparently natural choice is too naive. Its validity is challenged by  an explicit calculation which shows that under the Dirac dual
\begin{equation}
\overline{\lambda}^{S,A}_\alpha(p^\mu)\stackrel{\rm def}{=}\left[\lambda^{S,A}_\alpha(p^\mu)\right]^\dagger \gamma_0 
\end{equation}
the norm of each of the four $\lambda^{S,A}_\alpha(p^\mu)$ identically vanishes~\citep{Ahluwalia:1994uy}
\begin{equation}
\overline{\lambda}^S_\alpha(p^\mu) \lambda^S_\alpha(p^\mu) = 0,\quad
\overline{\lambda}^A_\alpha(p^\mu) \lambda^A_\alpha(p^\mu) = 0 .\label{eq:nullnorm}
\end{equation}
At the simplest, one may use the $\lambda(p^\mu)$ enumerated in equations (\ref{eq:lsp}) to (\ref{eq:lam}) and verify null norm (\ref{eq:nullnorm}) by a brute force algebraic calculation. Here we follow a more detailed route to arrive at the null norm of massive Elko. In the process we would not only see how precisely this comes about but also gain deeper insight into the relation between the Dirac spinors and Elko.

Since both the Dirac spinors and Elko reside in the same $\mathcal{R}\oplus\mathcal{L}\vert_{s=1/2}$ representation space, we can expand Elko in terms of the Dirac spinors as~\citep{Ahluwalia:1993xa,Ahluwalia:2004ab}
\begin{equation}
e_a = \sum_{b=1}^{4} \Omega_{ab} d_b,\quad a,b = 1,2,3,4
\end{equation}
where we have abbreviated Elko and Dirac spinors as
\begin{equation}
\left\{
\begin{array}{ll}
e_1 \\
e_2\\
e_3\\
e_4
\end{array}\right\}
\stackrel{\textrm{def}}{=}
\left\{
\begin{array}{ll}
\lambda^S_+(p^\mu) \\
\lambda^S_-(p^\mu)\\
\lambda^A_+(p^\mu)\\
\lambda^A_-(p^\mu)
\end{array}\right\},\quad
\left\{
\begin{array}{ll}
d_1 \\
d_2\\
d_3\\
d_4
\end{array}\right\}
\stackrel{\textrm{def}}{=}
\left\{
\begin{array}{ll}
u_+(p^\mu) \\
u_-(p^\mu)\\
v_+(p^\mu)\\
v_-(p^\mu)
\end{array}\right\}
\end{equation}
and 
\begin{equation}
\Omega_{ab} = \begin{cases}
 + (1/2 m)\overline{d}_b e_a,\quad \mbox{\textrm{for b=1,2}}\\
 - (1/2 m)\overline{d}_b e_a,\quad \mbox{\textrm{for b=3,4}}
\end{cases}
\end{equation}
An explicit calculation using equations (\ref{eq:lsp}) to (\ref{eq:lam}) for the Elko and equations (\ref{eq:upp}) to (\ref{eq:vm}) for the Dirac spinors yields
\begin{equation}
\Omega = \frac{1}{2} \left(
\begin{array}{cccc}
1 & i & i & 1\\
-i & 1 & -1 & i\\
i & 1 & -1 & -i\\
-1 & i & i & -1
\end{array}
\right)
\end{equation}\index{Elko!in terms of Dirac spinors}where rows are enumerated as $a=1,2,3,4$ and columns as $b=1,2,3,4$. With this formulation at hand we immediately see that we may explicitly write any of the Elko in terms of Dirac spinors. For instance
\begin{equation}
\lambda^S_+(p^\mu) = \frac{1}{2} \Big(  u_+(p^\mu) + i u_-(p^\mu) + i v_+(p^\mu) + v_-(p^\mu\Big)\label{eq:lambdasa}
\end{equation}
which yields under the Dirac dual reads
\begin{equation}
\overline{\lambda}^S_+(p^\mu) = \frac{1}{2} \Big(  \overline{u}_+(p^\mu) - i \overline{u}_-(p^\mu) -i \overline{v}_+(p^\mu) +\overline{v}_-(p^\mu)\Big).
\end{equation}
Thus, using the orthonormality of the Dirac spinors under the Dirac dual, we get two positive and two negative contributions -- each equaling $2 m$ in magnitude -- to the norm of 
$\lambda^S_+(p^\mu)$ under the Dirac dual. With the result that 
$\overline{\lambda}^S(p^\mu) \lambda^S(p^\mu)$ vanishes. Exactly the same pattern and result holds for the remaining three Elko.

So the mass matrix in the Lagrangian density (\ref{eq:wrong}) would  have vanishing diagonal elements. The off diagonal elements  would either vanish identically, or would be pure imaginary because, for instance
\begin{equation}
\overline{\lambda}^S_-(p^\mu) = \frac{1}{2} \Big( i \overline{u}_+(p^\mu) 
+ \overline{u}_-(p^\mu) 
-\overline{v}_+(p^\mu) -i \overline{v}_-(p^\mu)\Big)
\end{equation}
and it results in
\begin{equation}
\overline{\lambda}^S_-(p^\mu) \lambda^S_+(p^\mu) = i 2m\label{eq:i2m}
\end{equation}

\noindent
\section{Further on Elko-Dirac interplay}

The above discussed Elko-Dirac structure makes it clear that the inclusion of anti self conjugate spinors is necessary for a proper understanding of the eigenspinors of the charge conjugation operator.

The result that Elko cannot satisfy Dirac equation, as encoded in equations (\ref{eq:er-a1P}) to (\ref{eq:er-b2P}), can now be derived in a way which makes transparent inevitability of the conclusion.  To see this one 
may act the parity operator $\mathcal{P}$, $m^{-1}\gamma_\mu p^\mu$, on equation (\ref{eq:lambdasa})  to the effect
\begin{align}
m^{-1}\gamma_\mu p^\mu \,\lambda^S_+(p^\mu)
 &= \frac{1}{2}
\Big(  \underbrace{ m^{-1} \gamma_\mu p^\mu \,u_+(p^\mu)}_{ u_+(p^\mu)} + i \underbrace{m^{-1}\gamma_\mu p^\mu \,u_-(p^\mu) }_{u_-(p^\mu) }
\nonumber \\
& + i \underbrace{m^{-1}\gamma_\mu p^\mu \,v_+(p^\mu)}_{-v_+(p^\mu)}  + \underbrace{m^{-1}\gamma_\mu p^\mu\, v_-(p^\mu)}_{-v_-(p^\mu)}\Big)\nonumber\\
&= \frac{1}{2}\Big(u_+(p^\mu)+i u_-(p^\mu) - i v_+(p^\mu) - v_-(p^\mu))    \Big)\nonumber\\
& =i \underbrace{\frac{1}{2}\Big(-i u_+(p^\mu)+ u_-(p^\mu) -  v_+(p^\mu) + i v_-(p^\mu) }_{\lambda^S_-(p^\mu)}   \Big)\nonumber\\
& = i \lambda^S_-(p^\mu).
\end{align}
That is $\mathcal{P} \lambda^S_+(p^\mu) = i \lambda^S_-(p^\mu)$. Results (\ref{eq:er-a2P}) to  (\ref{eq:er-b2P}) follow similarly.

Finally, it is insightful to check how the results (\ref{eq:timereversalS}) and
 (\ref{eq:timereversalA}) on time reversal come from the Elko-Dirac relationship. Towards this end one may act the time reversal operator 
 $\mathcal{T}$, $i\gamma\mathcal{C}$, on $\lambda^S_+(p^\mu)$ expanded in terms of Dirac spinors in (\ref{eq:lambdasa}) to the effect
 \begin{align}
 \mathcal{T}\,\lambda^S_+(p^\mu) &= \frac{1}{2} i \gamma \mathcal{C} \Big(  u_+(p^\mu) + i u_-(p^\mu) + i v_+(p^\mu) + v_-(p^\mu)\Big)\nonumber\\
 &= \frac{1}{2} i \gamma  \Big(  \mathcal{C} u_+(p^\mu) - i \mathcal{C} u_-(p^\mu) - i \mathcal{C} v_+(p^\mu) + \mathcal{C} v_-(p^\mu)\Big).
 \end{align}
 The action of the charge conjugation operator $\mathcal{C}$ can now be implemented through equations (\ref{eq:cuv}) to obtain
\begin{equation}
 \mathcal{T}\,\lambda^S_+(p^\mu) = 
 \frac{1}{2} i \gamma  \Big(  i v_+(p^\mu) + v_-(p^\mu) +   u_+ (p^\mu) + i u_-(p^\mu)\Big).
\end{equation}
This requires us to note that $\gamma$ acts on the Dirac spinors as follows
\begin{equation}
\gamma\, u_\pm(p^\mu) = \mp\, v_\mp(p^\mu),\quad 
\gamma\, v_\pm(p^\mu) = \pm\, u_\mp(p^\mu).
\end{equation}
As a consequence we have
\begin{align}
 \mathcal{T}\,\lambda^S_+(p^\mu) & = 
 \frac{1}{2} i   \Big(  i u_-(p^\mu) - u_+(p^\mu) -   v_- (p^\mu) 
 + i v_+(p^\mu\Big)\nonumber\\
 & = i  \underbrace{\frac{1}{2}    \Big(  - u_+(p^\mu) +  i u_-(p^\mu) + i v_+(p^\mu) -   v_- (p^\mu) 
 \Big)}_{\lambda^A_-(p^\mu)}
\end{align}
That is
\begin{equation}
 \mathcal{T}\,\lambda^S_+(p^\mu) = i \lambda^A_-(p^\mu)
\end{equation}
in agreement with  (\ref{eq:timereversalS}).
The check for the remainder of the results in (\ref{eq:timereversalS}) and (\ref{eq:timereversalA}) follows exactly the same calculational flow. 

These calculations provide an intimate dependence of Elko on Dirac spinors, and of Dirac spinors on Elko. Since Elko are superposition of the particle  \textit{and}  antiparticle Dirac spinors, which in configuration space have different time evolutions, Elko and Dirac spinors cannot be governed by the same Lagrangian density.

\section{A temptation, and a departure\label{sec:temptation}}

One may thus be tempted to suggest that we introduce, instead, a Majorana mass term and treat the components of $\lambda(x)$ as anticommuting numbers (that is, as Grassmann numbers). This, in effect  --  but with doubtful mathematical justification -- would immediately promote a c-number classical field to the quantum field of Majorana and demand that the kinetic term be restored to that of Dirac. In the process we will be forced to abandon Elko as possible expansion coefficients of a quantum field, and return to a field with Dirac spinors as its expansion coefficients
\begin{align}
\psi(x) = & \int \frac{\text{d}^3p}{(2\pi)^3}  \frac{1}{\sqrt{2  E(\p)}}  \nonumber\\ &\times\sum_\sigma \Big[ a_\sigma(\p) u_\sigma(\p) \exp(- i p_\mu x^\mu)
+\, b^\dagger_\sigma(\p) v_\sigma(\p) \exp(i p_\mu x^\mu){\Big]}
\label{eq:DiracField}
\end{align}
which on setting \index{Majorana!condition}
\begin{equation}
b_\sigma^\dagger(\p) = a_\sigma^\dagger(\p),\qquad
{
\mbox{\textrm{\citep{Majorana:1937vz}}}}\label{eq:Majorana1937}
\end{equation}
and introducing $\mathcal{R}$ and $\mathcal{L}$ transforming components, $\psi_R(x)$ and $\psi_L(x)$, can be written as
\begin{equation}
\psi(x) = \left(
\begin{array}{c}
\psi_R(x)\\
\psi_L(x)
\end{array}
\right)
\end{equation}
with{\small
\begin{align}
\psi_R(x) = & \int \frac{\text{d}^3p}{(2\pi)^3}  \frac{1}{\sqrt{2  E(\p)}}  \nonumber\\ &\times\sum_\sigma \Big[ a_\sigma(\p) u^R_\sigma(\p) \exp(- i p_\mu x^\mu)
+\, a^\dagger_\sigma(\p) v^R_\sigma(\p) \exp(i p_\mu x^\mu){\Big]}\\
\psi_L(x) = & \int \frac{\text{d}^3p}{(2\pi)^3}  \frac{1}{\sqrt{2  E(\p)}}  \nonumber\\ &\times\sum_\sigma \Big[ a_\sigma(\p) u^L_\sigma(\p) \exp(- i p_\mu x^\mu)
+\, a^\dagger_\sigma(\p) v^L_\sigma(\p) \exp(i p_\mu x^\mu){\Big].}
\end{align}
}
On using the identities{\small
\begin{equation}
 u^R_\sigma(\p) = \Theta  \left[ v^L_\sigma(\p) \right]^\ast,\quad
 v^R_\sigma(\p) = \Theta  \left[ u^L_\sigma(\p) \right]^\ast
\end{equation}
}
the resulting field would then have the form of a Majorana field\index{Majorana!field}
	\begin{equation}
	\psi^M(x) =
	 \left(
	\begin{array}{c}
	 \Theta \psi^\ast_L(x)\\
	\psi_L(x)
	\end{array}
	\right) 
	\end{equation} 
where the superscript $M$ symbolises the Majorana condition (\ref{eq:Majorana1937}), and the $\ast$ symbol  on $\psi^\ast_L(x)$ complex conjugates without transposing, and takes $a_\sigma(\p)$ to $a^\dagger_\sigma(\p)$, and vice versa.  While it has a striking formal resemblance to Elko its physical and mathematical content is very different. Elko is a complex number valued four component spinor. It has certain transformation properties. It has null norm under the Dirac dual, and this fact is not enough to warrant to endow it with a Grassmann character. The $\psi^M(x)$ is a quantum field with complex number valued four component  Dirac spinors as its expansion coeeficients, it has its own transformation properties. Its Grassmann character arises from interpreting the creation and annihilation operators as anticommuting fermionic operators. For Elko the $\mathcal{R}$ and $\mathcal{L}$ transforming components have opposite helicities, no such interpretation can be associated to $\mathcal{R}$ and $\mathcal{L}$ transforming components of $\psi^M(x)$.

If we do not fall into this temptation for Elko an unexpected theoretical result follows that naturally leads us to a new class of fermions of spin one half. The question on the path of departure is deceptively simple: If the Dirac dual 
\begin{equation}\overline{\psi}(p^\mu)=\left[\psi(p^\mu)\right]^\dagger\gamma_0 \label{eq:Dirac-dual}
\end{equation}
was not given how shall we go about deciphering it? And, is
this a unique dual, or is there a freedom in its definition?, and what physics does it encode?

These questions are rarely asked in the physics literature. An  exception for the definition, called a ``convenience'' by Weinberg, is to note that the counterpart of $\Lambda$ in (\ref{eq:Minkowski-boost-rotation}) for the $\mathcal{R}\oplus\mathcal{L}\vert_{s=1/2}$ representation space
\begin{equation}
D(\Lambda) \stackrel{\textrm{def}}{=}\left\{
\begin{array}{cl}
\exp\left(i\kb\cdot\vp\right) & \mbox{for Lorentz boosts} \\
\exp\left(i\bz\cdot\vt\right) & \mbox{for rotations}
\end{array}\right.
\end{equation}
is not unitary, but pseudounitary\footnote{\label{fn:8}Strictly speaking, Weinberg's discussion is for a   spin one half quantum field
 but it readily adapts to the c-number spinors, $\psi(p^\mu)$, in the $\mathcal{R}\oplus\mathcal{L}\vert_{s=1/2}$ representation space.}
\begin{equation}
\gamma_0 D(\Lambda)^\dagger \gamma_0 = D(\Lambda)^{-1}.\label{eq:pseudounitarity}
\end{equation}
Weinberg uses this observation to motivate the definition (\ref{eq:Dirac-dual}).


For Elko the completion of (\ref{eq:nullnorm}) and (\ref{eq:i2m}) is given by
\begin{align}
	&\overline\lambda^S_\pm(p^\mu)  \lambda^S_\pm(p^\mu)  = 0,  \quad \overline\lambda^S_\pm(p^\mu)	\lambda^A_\pm(p^\mu) = 0,\quad 
	 \overline\lambda^S_\pm(p^\mu) \lambda^A_\mp(p^\mu) =0 \label{eq:norm-a}\\
	 &\overline\lambda^A_\pm(p^\mu) \lambda^A_\pm(p^\mu)  = 0,\quad \bar\lambda^A_\pm(p^\mu) \lambda^S_\pm(p^\mu) = 0,\quad
	 \overline\lambda^A_\pm(p^\mu) \lambda^S_\mp(p^\mu) =0 \label{eq:norm-b} 
\end{align}
and
\begin{equation}
\overline\lambda^S_\pm(p^\mu) \lambda^S_\mp(p^\mu) = \mp \,2 i m,\quad
	 \overline\lambda^A_\pm(p^\mu) \lambda^A_\mp(p^\mu)  = \pm \,2 i m. \label{eq:norm-c}
	 \end{equation}
Because of (\ref{eq:norm-c}), we can define -- see below --  a new dual and make it convenient  to formulate and calculate the physics of
quantum fields with Elko as their expansion coefficients. Once a lack of uniqueness of the Dirac dual is discovered, the new way to accommodate the pseudounitarity 
captured by (\ref{eq:pseudounitarity}) is introduced.  It opens up concrete new possibilities to go beyond the 
Dirac and Majorana fields for spin one half fermions without violating Lorentz covariance and without introducing non-locality. 
The programme now is as follows:
\begin{itemize}
\item Introduce a new dual so that not only the norm of Elko is  Lorentz invariant (and $\in \Re$), but also the spins sums.
\item Define a quantum field in terms of Elko, with the creation and annihilation operators satisfying the usual fermionic anticommutators.
\item Using the new dual, define a new adjoint for the quantum field.
Calculate the vacuum expectation value of the time ordered product of the field and its adjoint.
\item From the Feynman-Dyson propagator thus obtained, decipher the mass dimensionality of the new field, and check the Schwinger locality by calculating the relevant anticommutators.
\end{itemize}

With the exception of this chapter, we consciously refrained from introducing a dual of the spinors. The reason was simple. The Elko had to be obtained in a linear fashion. Its full poetry and beauty revealed without the shadows of the intricacies of the dual space. The inevitability of the final result hinted, and exposed, in its utter simplicity. That done we could then take our reader through the more amorphous cloud of duals and adjoints, to establishing the hint in its full maturing, knowing that others may refine it further but not alter its essence.

This shared with the reader, we proceed to developing the theory of dual spaces. Its flow holds as good for the representation space of our immediate interest as for any other representation space. But by confining to the indicated space we are able to make our argument less heavy, notationally. Beginning in the next chapter the reader would learn the deeper structure from which (\ref{eq:Dirac-dual}) arises, and why there is no proper metric for Weyl spinors, and for $\mathcal{R}$ and 
$\mathcal{L}$ representation spaces, separately, in general. We would also learn to construct the metric for many other representation spaces. In particular, we would easily see how the Lorentz algebra -- without reference to any other fact -- gives the metric for the Minkowski space, with the twist of a multiplicative phase factor. This phase may help us to understand four-vector spaces at a deeper quantum level. In the process we will discover how to construct the needed metric, with symmetries as the starting point.

\chapter{An \textit{ab initio} journey into duals}
\label{ch12}

\section{Motivation and a brief outline}

\noindent

A $n$ dimensional representation space may be spanned by $n$ independent vectors $\{\zeta_1,\zeta_2,\ldots\zeta_n\}$. These could be Weyl spinors, four-component spinors, four-vectors, or `vectors' spanning any representation. Our task is to introduce 
a unified approach to constructing dual spaces,\index{Dual spaces} and 
a metric that helps define bi-linear invariants.\index{Bi-linear invariants}

Our starting point is Lorentz algebra, and the representation spaces on which the symmetry transformations act through exponentiation of the generators in the sense defined in the opening chapters. 
Our motivation to look at the duals \textit{ab initio} resides in the null norm of Elko under the  Dirac dual.

While considering a representation space,
we  look at each of the basis vectors $\zeta_\alpha$, $\alpha=1,2, \ldots n$, as a complex number valued column vector. Our task is to define a dual that allows us to construct scalars (or more generally bilinears),  that are invariant (covariant) under a set of physically interesting symmetry transformations. These, besides the Lorentz transformations, would include the discrete transformations, of parity, of time reversal, and of charge conjugation.
For this task we generate a mechanism that transforms each of the $\zeta_\alpha$ 
into a row. We identify this mechanism, in part, with the complex conjugation of each of the $\zeta_\alpha$ and transposing it. In addition, motivated by the null norm of Elko under the Dirac dual, we introduce a pairing operator that uniquely pairs (with the freedom to multiply with an $\alpha$ dependent phase factor)  each of these newly generated rows in an  invertible manner with each of the  $\zeta_\alpha$. Say, the $\alpha$th column with the 
$\beta$th row and generate the set of row vectors, $\{{\zeta}^\prime_1,{\zeta}^\prime_2,\ldots{\zeta}^\prime_n\}$
\begin{equation}
{\zeta}^\prime_\alpha \stackrel{\textrm{def}}{=} \left(\Xi \,\zeta_\alpha\right)^\dagger\label{eq:dualdef}
\end{equation}
where we have taken the liberty of first pairing and then implementing the 
complex conjugation and transposition.  Of $\Xi$ we require that its inverse exists, and that its square is an identity operator, $\Xi^2 = \I_n$. 


To construct scalars from the ${\zeta}^\prime_\alpha$ and $\zeta_\alpha$
we introduce a $n\times n$ matrix $\eta$, and define
a dual vector, or just a dual
\begin{equation}
\dual{\zeta_\alpha}\stackrel{\textrm{def}}{=}  {\zeta}^\prime_\alpha \eta
\end{equation}
We call $\eta$ the  metric\index{Metric!its definition} for the representation space under consideration. 
It  is constrained to keep the bi-linear product  
\begin{equation}
\chi_\alpha= \dual{\zeta_\alpha}\,\zeta_\alpha
\end{equation}
invariant under a chosen set of transformation, or symmetries.
These constraints, as we will discover, generally
take the form of commutators/anticom\-mutators of $\eta$, with the relevant generators of the symmetry algebra, to vanish.


\noindent
\textit{Additional constraints~\textemdash}

It turns out that the dual as defined above still has an element of freedom. This freedom we shall discuss in detail.

 It is this general procedure that we now implement for the spinor 
 spaces. 

\section{The dual of spinors: constraints from the scalar invariants \label{sec:constraint-from-scalar-invariants}}

Consider a general set of $4$-component massive spinors $\varrho(p^\mu)$. We would like these to be orthonormal under the dual we are seeking. These do not have to be  eigenspinors of $\mathcal{P}$, that is, Dirac spinors, or the eigenspinors of $\mathcal{C}$ -- that is, Elko. As a specific implementation of the preceding discussion, we examine a general form of the dual defined  as \begin{equation}
\dual{\varrho}_\alpha(p^\mu) \stackrel{\rm def}{=}{\big[}\Xi(p^\mu) \, \varrho_\alpha(p^\mu){\big]}^\dagger \eta
\label{eq:gendual}
\end{equation}
where $\eta$ is a $4\times 4$ matrix, with its elements $\in \mathbb{C}$.  The task of $\Xi(p^\mu)$  is to take any one of the $\varrho_\alpha(p^\mu)$ and transform it, up to a phase,
 into one of the spinors $\varrho_{\alpha^\prime}(p^\mu)$ from the same set. It is not necessary that the indices $\alpha^\prime$ and $\alpha$ be the same. 
 We require $\Xi(p^\mu)$ to define an invertible map, with $\Xi^2 = \I_4$ (possibly, up to a phase). 
 \vspace{11pt}
 
 \section{The Dirac and Elko dual: a preview}
 
 The Dirac dual arises when we identify the pairing matrix with the identity matrix  
\begin{equation}\Xi(p^\mu) =\I_4\label{eq:Dirac-Xi}
\end{equation}
for which $\beta =\alpha$
 \begin{equation}
 \varrho_\alpha(p^\mu)\to\varrho_\alpha(p^\mu),\quad\forall\alpha
 \end{equation} 
and identifying $\varrho_\alpha(p^\mu)$ with the eigenspinors of the parity operator, that is with the $u_\sigma(p^\mu)$ and $v_\sigma(p^\mu)$.

The starting point for the Elko dual goes back to an early unpublished preprint before it was fully realised that these provide expansion coefficients for a  mass dimension one fermionic field of spin one half~\citep{Ahluwalia:2003jt}. The hint arose
from  the results found here in equations (\ref{eq:norm-a}) to (\ref{eq:norm-c}). That early  choice was shown by Speran\c{c}a to arise from the pairing matrix~\citep{Speranca:2013hqa} \index{Pairing matrix}
 \begin{align}
	\Xi(p^\mu)   \stackrel{\rm def}{=}  \frac{1}{2 m} \sum_{\alpha=\pm}	
			  \Big[\lambda^S_\alpha(p^\mu)\bar\lambda^S_\alpha(p^\mu) 	 - \lambda^A_\alpha(p^\mu)\bar\lambda^A_\alpha(p^\mu) 	\Big].\label{eq:Xi}
	\end{align}
It induces the needed map
\begin{align}
\lambda^S_+(p^\mu)& \to i \lambda^S_-(p^\mu)\label{eq:map1}\\
\lambda^S_-(p^\mu)& \to-  i \lambda^S_+(p^\mu)\label{eq:map2}\\
\lambda^A_+(p^\mu)& \to - i \lambda^A_-(p^\mu)\label{eq:map3}\\
\lambda^A_-(p^\mu)& \to+  i \lambda^S_+(p^\mu).\label{eq:map4}
\end{align}
 if $\varrho_\alpha(p^\mu)$ are identified with Elko, 
 $\lambda_\alpha(p^\mu)$. That is
\begin{align}
 \Xi(p^\mu) \lambda^S_\pm(p^\mu) &= \pm i  \lambda^S_\mp(p^\mu)\label{eq:mapS}\\
 \Xi(p^\mu) \lambda^A_\pm(p^\mu) &= \mp i  \lambda^A_\mp(p^\mu). \label{eq:mapA}\
\end{align}

\section{Constraints on the metric from Lorentz, and discrete, symmetries}

We now wish to determine the metric $\eta$, and $\Xi(p^\mu)$~\textendash~explicitly. We will see that the choices (\ref{eq:Dirac-Xi}),  (\ref{eq:mapS}) and (\ref{eq:mapA}),  are indeed allowed.
For the boosts, the requirement of a Lorentz invariant norm translates to
\begin{equation}
\underbrace{\Big[\Xi(k^\mu)\varrho(k^\mu)\Big]^\dagger  \eta \,\varrho(k^\mu) }_{{\chi(k^\mu)}}
= \underbrace{\Big[\Xi(p^\mu)\varrho(p^\mu)\Big]^\dagger \eta \,\varrho(p^\mu)}_{\chi(p^\mu)}
\label{eq:boostdemand}
\end{equation}
 with a similar expression for the rotations. 
Expressing $\varrho(p^\mu)$ as  $\exp(i \kb\cdot\vp) \varrho(k^\mu)$, and using $\kb^\dagger = -\kb$ (for an explicit form of  $\bm{\kappa}$ see Eq.~(\ref{eq:pi})),
the right-hand side of the above expression can be re-written as 
\begin{equation}
\Big[\Xi(p^\mu)\varrho(p^\mu)\Big]^\dagger \eta \,\varrho(p^\mu) 
= \varrho^\dagger(k^\mu)\, e^{i\boldsymbol{\kappa\cdot\varphi}} \, \Xi^\dagger(p^\mu)\, \eta\, e^{i\boldsymbol{\kappa\cdot\varphi}}\, \varrho(k^\mu).\label{eq:breq}
\end{equation}
Since 
\begin{equation}
\Xi(p^\mu) = e^{i\boldsymbol{\kappa\cdot\varphi}}\, \Xi(k^\mu)\, e^{-i\boldsymbol{\kappa\cdot\varphi}} \label{eq:XiTransformation-b}
\end{equation}
the $\Xi^\dagger(p^\mu)$ evaluates to
\begin{equation} \Xi^\dagger(p^\mu) \rightarrow e^{i\boldsymbol{\kappa^\dagger\cdot\varphi}}\, \Xi^\dagger(k^\mu)\, e^{-i\boldsymbol{\kappa^\dagger\cdot\varphi}} 
= e^{- i\boldsymbol{\kappa\cdot\varphi}}\, \Xi^\dagger(k^\mu)\, e^{i \boldsymbol{\kappa\cdot\varphi}} .
\end{equation}
Using this result in (\ref{eq:breq}) yields
 	\begin{align}
	\Big[\Xi(p^\mu)\varrho(p^\mu)\Big]^\dagger \eta \,\varrho(p^\mu) 
	 &=
 	 \varrho^\dagger(k^\mu) \Xi^\dagger(k^\mu) e^{i\boldsymbol{\kappa\cdot\varphi}} \eta \,e^{i\boldsymbol{\kappa\cdot\varphi}}\varrho(k^\mu) \nonumber \\
 &=
 	\Big[\Xi(k^\mu) \varrho(k^\mu)\Big]^\dagger e^{i\boldsymbol{\kappa\cdot\varphi}} \eta \,e^{i\boldsymbol{\kappa\cdot\varphi}}\varrho(k^\mu).
 	\end{align}
Using the just derived result into the right hand side of (\ref{eq:boostdemand}) gives
\begin{equation}
\Big[\Xi(k^\mu)\varrho(k^\mu)\Big]^\dagger \eta \,\varrho(k^\mu) 
	 =
 	\Big[\Xi(k^\mu) \varrho(k^\mu)\Big]^\dagger e^{i\boldsymbol{\kappa\cdot\varphi}} \eta \,e^{i\boldsymbol{\kappa\cdot\varphi}}\varrho(k^\mu).
\end{equation}
Up to  a caveat to be noted below in section~\ref{sec:Aremark}, it gives the constraint 
\begin{equation}
\eta  = e^{i\boldsymbol{\kappa\cdot\varphi}} \eta \,e^{i\boldsymbol{\kappa\cdot\varphi}} \label{eq:bc}
\end{equation}
That is, the metric $\eta$ must anticommute with \index{Metric!constraint from boost}
each of boost generators $\kb$ 	
\begin{equation}
	\{\kappa_i,\eta\}= 0,\qquad i = x,y,z.  	 \label{eq:eta-boost}
	\end{equation}
Implementing the above constraint with $i=z,y,z$ in succession reduces $\eta$ to have the form\index{Metric!its form after boost constraint}
\begin{equation}
\left(\begin{array}{cccc}
0 & 0 & a+i b & 0\\
0 & 0& 0 & a+ i b \\
c + i d & 0 & 0 & 0\\
0 & c+i d & 0 & 0
\end{array}\right)\label{eq:etaafterboost}
\end{equation}
where $a,b,c,d \in \Re$. This is valid for all four-component spinors.


\subsection{A freedom in the definition of the metric} \label{sec:Aremark}

Strictly speaking the equality in (\ref{eq:bc}) is up to a freedom of an operator which carries the $\varrho(p^\mu)$ as its eigenspinor with eigenvalue unity:
\begin{equation}
\eta  = e^{i\boldsymbol{\kappa\cdot\varphi}} \eta \,e^{i\boldsymbol{\kappa\cdot\varphi}} \Gamma
\end{equation}
with $\Gamma\varrho(p^\mu) = \varrho(p^\mu)$. Existence of $\Gamma$ operator(s) is an important freedom as it may reduce the symmetries under consideration, possibly to a subgroup of Lorentz. 

Apart from our Elko experience~\citep{Ahluwalia:2010zn}  that we shall further share with our reader in this chapter,
the above remark arises from a 2006 Cohen and Glashow 
observation~\citep{Cohen:2006ky} that, ``Indeed, invariance under HOM(2), rather than (as is often taught) the Lorentz group, is both necessary and sufficient to ensure that the speed of light is the same for all observers, and inter alia, to explain the null result to the Michelson-Morley experiment and its more sensitive successors.'' These subgroups are defined by the four algebras summarised in Table 11.1. The generators 
\begin{equation}
\mathfrak{T}_1\stackrel{\textrm{def}}{=}\mathfrak{K}_x + \mathfrak{J}_y, \quad \mathfrak{T}_2\stackrel{\textrm{def}}{=}
\mathfrak{K}_y - \mathfrak{J}_x
\end{equation}
form the group T(2) -- a group isomorphic to the group of translations in a plane. 

Following  Cohen and Glashow a relativity formed by adjoining four spacetime translations with any of these groups is termed Very Special Relativity (VSR).\index{Very special relativity} While each of the  four VSRs have distinct physical characters  they all share the  remarkable defining property that incorporation of either $P$, $T$, $CP$, or $CT$ enlarges these subgroups to the full Lorentz group. At present, VSR does not violate any of the existing tests of Special Relativity and provides an unexpected realm for the investigation of spacetime symmetries including violation of discrete symmetries, and isotropy of space.

\begin{table}\label{tab:VSR}
\begin{minipage}{350pt}
\caption{Algebras associated with the four VSR subgroups}
\label{table2new}
\addtolength\tabcolsep{2pt}
\begin{tabular}{@{}l@{\hspace{49pt}}lll@{\hspace{10pt}}}
\hline\hline
Name & Generators & Sub algebra\\
\hline
$ \mathfrak{t}(2)$ & $\mathfrak{T}_1,\mathfrak{T}_2$  &$\left[\mathfrak{T}_1,\mathfrak{T}_2\right]=0$\\ 
$\mathfrak{e}(2)$ & $\mathfrak{T}_1,\mathfrak{T}_2,\mathfrak{J}_z$ 
&$\left[\mathfrak{T}_1,\mathfrak{T}_2\right]=0$, $\left[\mathfrak{T}_1,\mathfrak{J}_z\right]= -i \mathfrak{T}_2$,
$\left[\mathfrak{T}_2,\mathfrak{J}_z\right]= i \mathfrak{T}_1$\\
$\mathfrak{hom}(2)$ & $\mathfrak{T}_1,\mathfrak{T}_2,\mathfrak{K}_z $
&$\left[\mathfrak{T}_1,\mathfrak{T}_2\right]=0$, $\left[\mathfrak{T}_1,\mathfrak{K}_z\right]= i \mathfrak{T}_1$,
$\left[\mathfrak{T}_2,\mathfrak{K}_z\right]= i \mathfrak{T}_2$\\
$\mathfrak{sim}(2)$ & $\mathfrak{T}_1,\mathfrak{T}_2,\mathfrak{J}_z,\mathfrak{K}_z$ 
 & $\left[\mathfrak{T}_1,\mathfrak{T}_2\right]=0$, $\left[\mathfrak{T}_1,\mathfrak{K}_z\right]= i \mathfrak{T}_1$,
$\left[\mathfrak{T}_2,\mathfrak{K}_z\right]= i \mathfrak{T}_2$ \\	
\empty & \empty & $\left[\mathfrak{T}_1,\mathfrak{J}_z\right]= -i \mathfrak{T}_2$,
$\left[\mathfrak{T}_2,\mathfrak{J}_z\right]= i \mathfrak{T}_1$, $\left[\mathfrak{J}_z,\mathfrak{K}_z\right]=0$	\\
\hline\hline
\end{tabular}
\end{minipage}
\end{table}

A detailed discussion of VSR can be found in a Masters thesis by Gustavo Salinas De Souza~\citep{Gustavo:Thesis:2005} and in the context of Elko in~~\citep{Ahluwalia:2010zn}. Our analysis of Elko provides a concrete example of a VSR quantum field that exhibits non-locality governed by a VSR operator
\begin{equation}
\mathcal{G}(p^\mu)  \stackrel{\textrm{def}}{=}
\left(\begin{array}{cccc}
			0 & 0 & 0 & -i e^{-i\phi} \\
			0 & 0 & i  e^{i \phi} & 0 \\
			0 & -i  e^{-i \phi} & 0 & 0\\
			i  e^{i\phi} & 0 & 0 & 0
			\end{array}\right). \label{eq:Gdef}
\end{equation}
This result is consistent with the claim of Cohen and Glashow that a departure from SR to VSR introduces non-locality. In its character $\mathcal{G}(p^\mu) $ is similar to $\Gamma$ becuase
$
\mathcal{G}(p^\mu)\, \lambda^S(p_\mu) =  \, \lambda^S(p_\mu), \mbox{and}~
\mathcal{G}(p^\mu)\, \lambda^A(p_\mu) =  - \, \lambda^A(p_\mu)
$.

As a parenthetic remark we note that Nakayama shows a way to circumvent the no-go locality result of VSR~\citep{Nakayama:2018fib,Nakayama:2017eof}\index{Nakayama} while Ilderton\index{Ilderton} shows how non-locality and VSR symmetries appear on averaging observables over rapid field oscillations of a Lorentz covariant theory~\citep{Ilderton:2016rqk}. Cheng-Yang Lee\index{Lee, Cheng-Yang } circumvents the non-locality problem by introducing a configuration space 
$\mathcal{G}(p^\mu)$ at the cost of introducing fractional derivatives~\citep{Lee:2014opa}.

Given our experience with Elko we conjecture that $\Gamma$ encodes a theoretical mechanism of  Lorentz symmetry breaking in a generic manner to the subgroups summarised in Table 11.1. After we have developed 
the notion of Elko dual further,
we shall pick up this thread in section~\ref{Sec:IUCAA}

\subsection{The Dirac dual\label{sec:Dirac-dual}}

The simplest choice for $\Xi(p^\mu)$ is the  identity operator as 
in (\ref{eq:Dirac-Xi}). We will now show that the well-known Dirac dual corresponds to this choice. With  
$\Xi(p^\mu) = \openone_4$, the counterpart 
of (\ref{eq:boostdemand})
for rotations reads
\begin{equation}
\varrho^\dagger(p_\mu) \, \eta \,\varrho(p_\mu)  
= \varrho^\dagger(p^\prime_\mu)\, \eta \,\varrho(p^\prime_\mu) 
\label{eq:rotationdemand}
\end{equation}
where $\varrho(p^\prime_\mu) = e^{i\bm{\zeta}\cdot\bm{\theta}}\varrho(p_\mu)$. Using $\bm{\zeta}^\dagger
=\bm{\zeta}$ (for an explicit form of  $\bm{\zeta}$ see Eq.~(\ref{eq:pi})) translates 
 (\ref{eq:rotationdemand}) to
\begin{equation}
\varrho^\dagger(p_\mu) \, \eta \,\varrho(p_\mu)  
= 
\varrho^\dagger(p_\mu) \, e^{-i \bm{\zeta}\cdot{\bm{\theta}}}\, \eta 
e^{i \bm{\zeta}\cdot{\bm{\theta}}}\,\varrho(p_\mu).
\end{equation}
Modulo a freedom noted in section~\ref{sec:Aremark}, it gives the constraint
\begin{equation}
\eta=
e^{-i \bm{\zeta}\cdot{\bm{\theta}}}\, \eta 
e^{i \bm{\zeta}\cdot{\bm{\theta}}} .
\end{equation}
That is, the metric $\eta$ must commute with 
each of rotation generators $\bm{\zeta}$ 
\begin{equation}
\left[{\zeta}_i,\eta\right] =0,\qquad i = x,y,z.  	 \label{eq:eta-rotation}
\end{equation}
The constraint that $\eta$ must commute with 
$\bm{\zeta}_i$ does not reduce $\eta$ constrained by its anticommutativity with each of the boost generators, $\kappa_i$.

However, following an analysis paralleling the two previous calculations  the demand for the norm to be invariant under the parity transformation $\mathcal{P}$ is readily obtained to be
\begin{align}
\eta = m^{-2} \left[ \gamma_\mu p^\mu\right]^\dagger \eta \, \gamma_\mu p^\mu
\label{eq:parityconstraint}
\end{align}
A direct evaluation of the right hand side of the parity constraint (\ref{eq:parityconstraint}) with $\eta$ given by (\ref{eq:etaafterboost}) gives the result
\begin{equation}
m^{-2} \left[ \gamma_\mu p^\mu\right]^\dagger \eta \, \gamma_\mu p^\mu
 = \left(\begin{array}{cccc}
0 & 0 & c+i d & 0\\
0 & 0& 0 & c+ i d \\
a + i b & 0 & 0 & 0\\
0 & a+i b & 0 & 0
\end{array}\right)
\end{equation}
thus requiring $c=a$, and $d=b$. In consequence, the parity constraint further reduces the metric $\eta$ to\index{Metric!its form after parity constraint}
\begin{equation}
\eta= \left(\begin{array}{cccc}
0 & 0 & a+i b & 0\\
0 & 0& 0 & a+ i b\\
a + i b & 0 & 0 & 0\\
0 & a+i b & 0 & 0
\end{array}\right)
\end{equation}
Requiring the norm of the Dirac spinors to be real forces the choice $b=0$, and then $a$ is simply a scale factor.  It may be chosen to be unity
\begin{equation}
\eta= \left(\begin{array}{cccc}
0 & 0 & 1 & 0\\
0 & 0& 0 & 1\\
1& 0 & 0 & 0\\
0 & 1 & 0 & 0
\end{array}\right)
\end{equation} \label{eq:etacanonicalZimpok}
\index{Dirac dual}
We thus reproduce the canonical Dirac dual: it is defined by the choice of $\Xi(p^\mu)=\openone_4$, and the constraints on $\eta$ given by (\ref{eq:eta-boost}), (\ref{eq:eta-rotation}) and (\ref{eq:parityconstraint}).  

On the path of our departure we learn that the 
Dirac dual 
has additional underlying structure. In particular, it gives us a choice to violate parity, or to preserve it, depending on whether we choose $a/c$ in $\eta$ of (\ref{eq:etaafterboost}) to be unity, or different from unity. With  the former choice we reproduce the standard result (\ref{eq:Dirac-dual}), while the latter choice gives us a first-principle control on the extent to which parity may be violated in nature, or in a given physical process. 

 \section{The Elko dual\label{sec:elko-dual}}

The  dual for $\lambda(p^\mu)$ first  introduced in~\citep{Ahluwalia:2003jt}, and refined in~\citep{Ahluwalia:2008xi,Ahluwalia:2009rh} can now be more systematically understood by observing that those results correspond to the     
$\Xi(p^\mu)$ of the second example considered above (and given in equation (\ref{eq:Xi})).
Expression (\ref{eq:Xi}) can be evaluated to yield a compact form
	\begin{equation}
	\Xi(p^\mu) =
	m^{-1}  \mathcal{G}(p^\mu) \gamma_\mu p^\mu.
	\label{eq:XizZmpok}
	\end{equation}
where $\mathcal{G}(p^\mu)$ is defined in equation 
(\ref{eq:Gdef}).
Out of the four variables $m$, $p$, $\theta$ and $\phi$ that enter the definition 
\[p^\mu = (E, p \sin\theta\cos\phi,p\sin\theta\sin\phi,p\cos\theta)\]
$\mathcal{G}(p^\mu) $ depends only on $\phi$. 
The analysis of the boost constraint remains unaltered, with the result that we still have
	\begin{equation}
	\{\kb_i,\eta\}= 0,\quad i = x,y,z . 	 \label{eq:eta-boost-b}
	\end{equation}
The analysis for the rotation constraint changes. It begins as
\begin{equation}
\Big[m^{-1}  \mathcal{G}(p^\mu)  \gamma^\mu p_\mu  \, \lambda(p_\mu)\Big]^\dagger \,\eta\, \lambda(p_\mu)
 =
\Big[m^{-1}  \mathcal{G}(p^{\prime\mu}) \gamma^\mu p^\prime_\mu \,  \lambda(p^\prime_\mu)\Big]^\dagger \,\eta\, \lambda(p^\prime_\mu)
\end{equation}
where $\lambda(p_\mu)$ represents any of the four $\lambda^{S,A}_\pm(p^\mu)$, and the primed quantities refer to their rotation-induced counterparts. The above expression simplifies on using the following identities
\begin{equation}
\mathcal{G}(p^\mu)\, \lambda(p_\mu) = \pm \, \lambda(p_\mu),\qquad [\mathcal{G}(p^\mu),\,\gamma^\mu p_\mu ]=0,\label{eq:important}
\end{equation}
where the upper sign in the first equation above holds for $\lambda^S(p_\mu)$ and the lower sign is for $\lambda^A(p_\mu)$. The result is
\begin{equation}
\Big[ \gamma^\mu p_\mu  \, \lambda(p_\mu)\Big]^\dagger \,\eta\, \lambda(p_\mu)
 =
\Big[\gamma^\mu p^\prime_\mu \,  \lambda(p^\prime_\mu)\Big]^\dagger \,\eta\, \lambda(p^\prime_\mu).\label{eq:rotation-constraint-b}
\end{equation}
Expressing $\lambda(p^{\prime\mu})$ as  $e^{i \bm{\zeta}\cdot\bm{\theta}} \lambda(p^\mu)$, and using $\bm{\zeta}^\dagger
=\bm{\zeta}$,
the right-hand side of the above expression can be written as 
\begin{equation}
\Big[\gamma^\mu p^\prime_\mu \,  \lambda(p^\prime_\mu)\Big]^\dagger \,\eta\, \lambda(p^\prime_\mu) =
\lambda^\dagger(p_\mu)\,
e^{-i\bm{\zeta}\cdot\bm{\theta}} 
\left(\gamma^\mu p^\prime_\mu \right)^\dagger\,\eta\,
e^{i\bm{\zeta}\cdot\bm{\theta}} \,\lambda(p_\mu). \label{eq:extra}
\end{equation}
On taking note that
\begin{equation}
\gamma^\mu p^\prime_\mu = e^{i\bm{\zeta}\cdot\bm{\theta}}  \gamma^\mu p_\mu 
e^{-i\bm{\zeta}\cdot\bm{\theta}}
\end{equation}
equation (\ref{eq:extra}) becomes
\begin{equation}
\Big[\gamma^\mu p^\prime_\mu \,  \lambda(p^\prime_\mu)\Big]^\dagger \,\eta\, \lambda(p^\prime_\mu) =
\Big[\gamma^\mu p_\mu\,\lambda(p_\mu)\Big]^\dagger\,
e^{-i\bm{\zeta}\cdot\bm{\theta}} 
\eta\,
e^{i\bm{\zeta}\cdot\bm{\theta}} \,\lambda(p_\mu)
\end{equation}
Comparing the above expression with  (\ref{eq:rotation-constraint-b}) gives the constraint 
\begin{equation}
\eta = e^{-i\bm{\zeta}\cdot\bm{\theta}} 
\eta\,
e^{i\bm{\zeta}\cdot\bm{\theta}}  
\end{equation}
That is, the metric $\eta$ must commute with 
each of rotation generators $\bm{\zeta}$ 
\begin{equation}
  \left[\zeta_i,\eta\right] = 0 ,\quad i = x,y,z  	 \label{eq:eta-rotation-c}
\end{equation}
 Despite a non-trivial $\Xi(p^\mu)$, this is the same result as before.


It is readily seen that  $\left[\Xi(p^\mu)\right]^2=\I$ and $\left[\Xi(p^\mu)\right]^{-1}$ indeed  exists and equals $\Xi(p^\mu)$ itself. 
Thus, the
  dual for $\lambda(p^\mu)$ is defined by the choice of $\Xi(p^\mu)$ given by 
 (\ref{eq:XizZmpok}). 
 
 So far the constraints on $\eta$ turn out to be same as for the Dirac dual. 
 To distinguish it from other possibilities the new dual at the intermediate state of our calculations is represented by 
 \begin{equation}
\dual{\lambda}_\alpha(p^\mu)   \stackrel{\rm def}{=} {\big[}\Xi(p^\mu)\, \lambda_\alpha(p^\mu){\big]}^\dagger \eta
\label{eq:dual-b}
\end{equation}
with $\eta$ given by (\ref{eq:etacanonicalZimpok}). This choice of $\eta$ is purely for convenience at the moment. If the new particles to be introduced here are indeed an element of the physical reality then the ratio $a/c$ must not be set to unity but determined by appropriate observations/experiments.

Using (\ref{eq:mapS}) and  (\ref{eq:mapA}), (\ref{eq:dual-b}) translates to
\begin{align}
\dual\lambda^S_+(p^\mu) &= - i \left[\lambda^S_-(p^\mu)\right]^\dagger \eta  \nonumber
\\
\dual\lambda^S_-(p^\mu) &= i \left[\lambda^S_+(p^\mu)\right]^\dagger\eta\nonumber
\\
 \dual\lambda^A_+(p^\mu) &=  i \left[\lambda^A_-(p^\mu)\right]^\dagger\eta\nonumber
 \\
 \dual\lambda^A_-(p^\mu) &= - i \left[\lambda^A_+(p^\mu)\right]^\dagger\eta
\label{eq:otherd-3}
\end{align}

 We can thus rewrite the results (\ref{eq:norm-a}), (\ref{eq:norm-b}), and (\ref{eq:norm-c}) into the following orthonormality relations
\begin{align}
& \dual\lambda^S_\alpha(p^\mu) \lambda^S_{\alpha^\prime}(p^\mu)
 = 2 m \delta_{\alpha\alpha^\prime}\nonumber
 \\
&  \dual\lambda^A_\alpha(p^\mu) \lambda^A_{\alpha^\prime}(p^\mu)
 = - 2 m \delta_{\alpha\alpha^\prime} \nonumber
 \\
 & \dual\lambda^S_\alpha(p^\mu) \lambda^A_{\alpha^\prime}(p^\mu) = 0 =
 \dual\lambda^A_\alpha(p^\mu) \lambda^S_{\alpha^\prime}(p^\mu)\label{eq:zimpokJ9c}
\end{align}
with $\alpha$ and $\alpha^\prime = \pm$.

\section[The dual of spinors: constraint from the invariance \ldots]{The dual of spinors: constraint from the invariance of the Elko spin sums
\label{sec:constraints-from-the-spin-sums}}

The Dirac spin sums\index{Spin sums!Dirac}
\begin{align}
\sum_{\sigma=+,-} u_\sigma(p^\mu)\overline{u}_\sigma(p^\mu) & = \gamma_\mu p^\mu + m \I_4 \label{eq:spinsumsDiracu}\\
\sum_{\sigma=+,-} v_\sigma(p^\mu)\overline{v}_\sigma(p^\mu) & = \gamma_\mu p^\mu - m \I_4  \label{eq:spinsumsDiracv}
\end{align}
immediately follow from the definition of the Dirac dual and expressions for the $u_\pm(p^\mu)$ and
$v_\pm (p^\mu)$ given in equations (\ref{eq:upp}) to (\ref{eq:vm}). These are covariant under Lorentz symmetries and as such require no further scrutiny of the Dirac dual -- that is, as far as the construction of the dual is concerned.

The spin sums (\ref{eq:spinsumsDiracu}) and (\ref{eq:spinsumsDiracv}) yield the completeness relations for the Dirac spinors\index{Dirac spinors!completeness}
\begin{equation}
\frac{1}{2 m}\sum_{\sigma=+,-} 
\Big(
u_\sigma(p^\mu)\overline{u}_\sigma(p^\mu)  -
v_\sigma(p^\mu)\overline{u}_\sigma(p^\mu) \Big) = \I_4
\end{equation}

With $\lambda^S_\alpha(p^\mu)$ and
 $\lambda^A_\alpha(p^\mu)$ given by equations~(\ref{eq:lsp}) to \ref{eq:lam}), and their respective duals defined by equations $(\ref{eq:otherd-3})$,
the spin sums for Elko 
\begin{equation}
\sum_{\alpha} \lambda^S_\alpha(p^\mu) \dual \lambda^S_\alpha(p^\mu) \;\;{\rm and} \;\;
\sum_{\alpha}\lambda^A_\alpha(p^\mu) \dual \lambda^A_\alpha(p^\mu) 
\end{equation}
are completely defined.
The first of the two spin sums evaluates to
\begin{align}
 	 & i\; \underbrace{ \left[ \frac{E + m}{2 m}
 	 \left(1 - \frac{{p^2}}{(E+m)^2}\right)\right]}_{=1}   \nonumber\\ 		&\hspace{51pt}\times\underbrace{\bigg( - \lambda^S_+(k^\mu) \left[\lambda^S_-(k^\mu)		\right]^\dagger
  			+ \lambda^S_-(k^\mu) \left[\lambda^S_+(k^\mu)\right]^\dagger
			\bigg)\eta}_{= -i m \left[ \I_4 + \mathcal{G}(p^\mu) \right]}.	
\end{align}
The second of the spin sums can be evaluated in exactly the same manner. The combined result is \index{Spin sums!Elko}
\begin{align}\index{Spin sums!Elko}
\sum_{\alpha} \lambda^S_\alpha(p^\mu) \dual \lambda^S_\alpha(p^\mu) & = +
m \big[\I_4 + \mathcal{G}(p^\mu) \big] \label{eq:sss}\\
\sum_{\alpha} \lambda^A_\alpha(p^\mu) \dual \lambda^A_\alpha(p^\mu) & =
- m \big[\I_4 - \mathcal{G}(p^\mu)\big]  \label{eq:ssa}
\end{align}
with $\mathcal{G}(p^\mu)$ as in Eq.~(\ref{eq:Gdef}).
These spin sums have the eigenvalues $\{0,0,2m,2m\}$, and $\{0,0,-2m,-2m\}$, respectively. Since eigenvalues  of projectors  must be either zero or one~\citep[Section 3.3]{Weinberg:2012qm}, we define
\begin{align}
  P_S &\stackrel{\textrm{def}}{=}\frac{1}{2 m} \sum_{\alpha} \lambda^S_\alpha(p^\mu) \dual \lambda^S_\alpha(p^\mu)=  \frac{1}{2} \big[\I_4 + \mathcal{G}(p^\mu) \big]  \label{eq:s}
\\
P_{A}&\stackrel{\textrm{def}}{=} -\frac{1}{2 m} \sum_{\alpha} \lambda^A_\alpha(p^\mu) \dual \lambda^A_\alpha(p^\mu) =  \frac{1}{2} \big[\openone_4 -\mathcal{G}(p^\mu)\big]  \label{eq:a}
\end{align}
and confirm that indeed they are projectors and furnish the  completeness relation\index{Elko!completeness}
\begin{equation}
P_S^2 =P_S,\quad P_A^2 = P_A,\quad P_S+P_A = \openone_4. \label{eq:compl}
\end{equation}

Because $\mathcal{G}(p^\mu)$ is not Lorentz covariant its appearance in the spin sums 
violates Lorentz symmetry.

Till late 2015, all efforts to circumvent this 
problem had failed and gave rise to a suggestion that the formalism can only be covariant under a subgroup of the Lorentz group suggested by Cohen and Glashow~\citep{Cohen:2006ky,Ahluwalia:2010zn}. 

\section{The IUCAA breakthrough}\index{The IUCAA breakthrough}\label{Sec:IUCAA}

However, during a set of winter-2015 lectures I gave on mass dimension one fermions at the Inter-University Centre for Astronomy and Astrophysics (IUCAA)  
it became apparent that there is a freedom in the definition of the dual.\footnote{The concrete outline of the breakthrough passed through me at a conversation with Krishnamohan Parattu in the parking lot of Akashganga  guest house at IUCAA just as I was leaving for the Centre for the Studies of the Glass Bead Game that I was setting up in Bir, Himachal Pradesh.}
 It allows a re-definition of the dual in such a way that the 
 Lorentz invariance of the orthonormality relations remains intact, 
but it restores the Lorentz covariance of the spin sums. In fact it makes them invariant \citep{Ahluwalia:2016rwl}.

The basic idea is already discussed briefly in section~\ref{sec:Aremark}. Since the construction of Elko dual was not sufficiently developed yet, our discussion, of necessity, was framed in terms of Elko, rather than its dual. Now, it is possible to bypass that limitation and 
explore the solution to Lorentz symmetry breaking by considering the following re-definition of the Elko dual \index{Re-definition of Elko dual}
\begin{equation}
\dual{\lambda}^S_\alpha(p^\mu) \to \gdualn{\lambda}^S_\alpha(p^\mu) = \dual{\lambda}^S_\alpha(p^\mu) \mathcal{A},\quad
\dual{\lambda}^A_\alpha(p^\mu) \to \gdualn{\lambda}^A_\alpha(p^\mu) = \dual{\lambda}^A_\alpha(p^\mu) \mathcal{B}\label{eq:ab}
\end{equation}
with $\mathcal{A}$ and $\mathcal{B}$ constrained to have the 
following non-trivial properties: The $\lambda^S_\alpha(p^\mu)$ must be  eigenspinors of $\mathcal{A}$ with eigenvalue unity,  and similarly $\lambda^A_\alpha(p^\mu)$ must be  eigenspinors of $\mathcal{B}$ with eigenvalue unity\begin{equation}
\mathcal{A} \lambda^S_\alpha(p^\mu) = \lambda^S_\alpha(p^\mu),\quad
\mathcal{B} \lambda^A_\alpha(p^\mu) = \lambda^A_\alpha(p^\mu),\label{eq:4jan-a}\\
\end{equation}
and additionally $\mathcal{A}$ and $\mathcal{B}$ must be such that
\begin{equation}
\gdual{\lambda}^S_\alpha(p^\mu)\mathcal{A} \lambda^A_{\alpha^\prime}(p^\mu)=0,\quad \dual{\lambda}^A_\alpha(p^\mu)\mathcal{B} \lambda^S_{\alpha^\prime}(p^\mu) = 0.
\label{eq:4jan-b}
\end{equation}
 Under the new dual while the orthonormality relations (\ref{eq:zimpokJ9c}) remain unaltered in 
 form\index{Elko!orthonormality}
\begin{align}
& \gdualn\lambda^S_\alpha(p^\mu) \lambda^S_{\alpha^\prime}(p^\mu)
 = 2 m \delta_{\alpha\alpha^\prime}\label{eq:zimpokJ9an}\\
&  \gdualn\lambda^A_\alpha(p^\mu) \lambda^A_{\alpha^\prime}(p^\mu)
 = - 2 m \delta_{\alpha\alpha^\prime} \label{eq:zimpokJ9bn} \\
 &  \gdualn\lambda^S_\alpha(p^\mu) \lambda^A_{\alpha^\prime}(p^\mu) = 0 =
 \gdualn\lambda^A_\alpha(p^\mu) \lambda^S_{\alpha^\prime}(p^\mu)\label{eq:zimpokJ9cn}
\end{align}
the same very  re-definition alters the spin sums to
\begin{align}
\sum_{\alpha} \lambda^S_\alpha(p^\mu) \dualn \lambda^S_\alpha(p^\mu) & = 
m \big[\I_4 + \mathcal{G}(p^\mu) \big] \mathcal{A} \label{eq:sss-new}\\
\sum_{\alpha} \lambda^A_\alpha(p^\mu) \dualn \lambda^A_\alpha(p^\mu) & =
- m \big[\I_4 - \mathcal{G}(p^\mu)\big] \mathcal{B} \label{eq:ssa-new}
\end{align}
In what follow we will show that $\mathcal{A}$ and $\mathcal{B}$ exist that satisfy
the dual set of requirements encoded in (\ref{eq:4jan-a}) and (\ref{eq:4jan-b}) and at the same time find that a specific form of  $\mathcal{A}$ and $\mathcal{B}$ exists that renders the spin sums Lorentz invariant.

The spin sums determine the mass dimensionality of the quantum field that we will build from the here-constructed $\lambda(p^\mu)$  as its expansion coefficients. 
They, along with their duals,  enter the evaluation of the Feynman-Dyson propagator. For consistency with 
(\ref{eq:er-a1})-(\ref{eq:er-b2}) and (\ref{eq:skg}) this mass dimensionality must be one. This can be achieved in the formalism we are developing if the spin sums are Lorentz invariant, and are proportional to the identity.

Thus, up to a constant~\textendash~to be taken as $2$ to preserve orthonormality relations~\textendash~$\mathcal{A}$ and  $\mathcal{B}$ must  be inverses of $\big[\I_4 + \mathcal{G}(p^\mu) \big]$ and 
$\big[\I_4 - \mathcal{G}(p^\mu) \big]$ respectively. But since 
the determinants of $\big[\I_4 \pm \mathcal{G}(p^\mu)\big]$ identically vanish we proceed in a manner akin to that of Penrose~\citep{PSP:2043984} and Lee~\citep{Lee:2014opa}, and with $\tau \in \Re$ we introduce 
a $\tau$   deformation of the spin sums (\ref{eq:sss-new})  and  (\ref{eq:ssa-new})\footnote{Unlike Lee~\citep{Lee:2014opa} we refrain from introducing configuration counterpart of $\mathcal{G}(p^\mu)$ to avoid brining in the theory fractional derivatives.}
\begin{align}
\sum_{\alpha} \lambda^S_\alpha(p^\mu) \dualn \lambda^S_\alpha(p^\mu) & = 
m \big[\I_4 + \tau\mathcal{G}(p^\mu) \big] \mathcal{A}\Big\vert_{\tau \to 1} \label{eq:sss-newnew}\\
\sum_{\alpha} \lambda^A_\alpha(p^\mu) \dualn \lambda^A_\alpha(p^\mu) & =
- m \big[\I_4 - \tau\mathcal{G}(p^\mu)\big] \mathcal{B}\Big\vert_{\tau \to 1}. \label{eq:ssa-newnew}
\end{align}
We will see that the $\tau \to 1$ limit is non pathological 
in the infinitesimal small neighbourhood of $\tau = 1$ in the sense we shall make explicit.
We choose $\mathcal{A}$ and $\mathcal{B}$ 
to be 
\begin{align}
& \mathcal{A} = 2 \big[I_4 + \tau \mathcal{G}(p^\mu)\big]^{-1} = 2 \left(\frac{\I_4 - \tau \mathcal{G}(p^\mu)}{1-\tau^2}\right) \\
& \mathcal{B} = 2 \big[I_4 - \tau \mathcal{G}(p^\mu)\big]^{-1} = 2 \left(\frac{\I_4 + \tau \mathcal{G}(p^\mu)}{1-\tau^2}\right).
\end{align}
Making use of the identity $\mathcal{G}^2(p^\mu) = \I_4$, Eqs.~(\ref{eq:sss-newnew}) and (\ref{eq:ssa-newnew}) simplify to:
\begin{align}
\sum_{\alpha} \lambda^S_\alpha(p^\mu) \dualn \lambda^S_\alpha(p^\mu) & = 
2 m \big[\I_4 + \tau\mathcal{G}(p^\mu) \big]   \left(\frac{\I_4 - \tau \mathcal{G}(p^\mu)}{1-\tau^2}\right)  \bigg\vert_{\tau \to 1} \nonumber \\
&=2 m \I_4 \left(\frac{1-\tau^2}{1-\tau^2}\right)\bigg\vert_{\tau\to 1}  =
 2m \I_4\label{eq:sss-newnew-a}\\
 \sum_{\alpha} \lambda^A_\alpha(p^\mu) \dualn \lambda^A_\alpha(p^\mu) & = 
2 m \big[\I_4 - \tau\mathcal{G}(p^\mu) \big]   \left(\frac{\I_4 + \tau \mathcal{G}(p^\mu)}{1-\tau^2}\right)  \bigg\vert_{\tau \to 1} \nonumber \\
&=2 m \I_4 \left(\frac{1-\tau^2}{1-\tau^2}\right)\bigg\vert_{\tau\to 1}  =
- 2m \I_4
 \label{eq:ssa-newnew-b}
\end{align}
We now return to the orthonormality relations. Since from the first equation in (\ref{eq:important}), $\mathcal{G}(p^\mu) \lambda^S(p^\mu) = 
\lambda^S(p^\mu)$ while $\mathcal{G}(p^\mu) \lambda^A(p^\mu) = -
\lambda^A(p^\mu)$, we have the result demanded by the requirement (\ref{eq:4jan-a})
\begin{align}
\mathcal{A} \lambda^S_\alpha(p^\mu)  & = 2 \left(\frac{\I_4-\tau \mathcal{G}(p^\mu)}{1-\tau^2}\right) \lambda^S_\alpha(p^\mu)
 = 2 \left(\frac{1-\tau}{1-\tau^2}\right) \lambda^S_\alpha(p^\mu) 
 \nonumber\\
 &= \left(\frac{2}{1+\tau}\right)\bigg\vert_{\tau\to 1}  \lambda^S_\alpha(p^\mu) \nonumber\\
 & = \lambda^S_\alpha(p^\mu) 
 \end{align}
 and
 \begin{align}
\mathcal{B} \lambda^A_\alpha(p^\mu) &= 2 \left(\frac{\I_4+\tau \mathcal{G}(p^\mu)}{1-\tau^2}\right) \lambda^A_\alpha(p^\mu) 
= 2 \left(\frac{1-\tau}{1-\tau^2}\right) \lambda^A_\alpha(p^\mu) \nonumber\\
& = \left(\frac{2}{1+\tau}\right)\bigg\vert_{\tau\to 1}  \lambda^A_\alpha(p^\mu)\nonumber\\
&  = \lambda^A_\alpha(p^\mu)
\end{align}
where  in the first two terms on the right hand side of each of the above equations
the ${\tau\to 1}$ limit has been suppressed.

To examine the fulfilment of requirement (\ref{eq:4jan-b}) we note that
\begin{align}
\gdual{\lambda}^S_\alpha(p^\mu)\mathcal{A} \lambda^A_{\alpha^\prime}(p^\mu) = 
 2 \gdual{\lambda}^S_\alpha(p^\mu)\left(\frac{\I_4 - \tau \mathcal{G}(p^\mu)}{1-\tau^2}\right) \lambda^A_{\alpha^\prime}(p^\mu) \nonumber \\
=  2 \left(\frac{1}{1-\tau}\right)\bigg\vert_{\tau\to 1}
 \underbrace{\gdual{\lambda}^S_\alpha(p^\mu)\lambda^A_{\alpha^\prime}(p^\mu)}_{=\;0 ~\mbox{\small{(see eq. \ref{eq:zimpokJ9c})}}}
= \;0\\
\gdual{\lambda}^A_\alpha(p^\mu)\mathcal{B} \lambda^S_{\alpha^\prime}\alpha(p^\mu) = 
 2 \gdual{\lambda}^A_\alpha(p^\mu) \left(\frac{\I_4 + \tau \mathcal{G}(p^\mu)}{1-\tau^2}\right) \lambda^S_{\alpha^\prime}(p^\mu) \nonumber \\
 =  2 \left(\frac{1}{1-\tau}\right)\bigg\vert_{\tau\to 1}
 \underbrace{\gdual{\lambda}^A_\alpha(p^\mu)\lambda^S_{\alpha^\prime}(p^\mu)}
 _{=\;0 ~\mbox{\small{(see eq. \ref{eq:zimpokJ9c})}}}
= 0
\end{align}
where the  final equalities are to be understood as `in the infinitesimally  close neighbourhood of $\tau =1$, but not at $\tau=1$.' We will accept it as physically acceptable cost to be paid for the $\tau$ deformation forced upon us by the non-invertibility of $\big[\I_4 \pm \mathcal{G}(p^\mu)\big]$. With this caveat,
constraints (\ref{eq:4jan-a}) and (\ref{eq:4jan-b}) on $\mathcal{A}$ and 
$\mathcal{B}$ are satisfied resulting in the Lorentz invariant spin sums 
\begin{align}
&\sum_{\alpha} \lambda^S_\alpha(p^\mu) \dualn \lambda^S_\alpha(p^\mu) = 2m \I_4\label{eq:sss-newnew-a-new}\\
&\sum_{\alpha} \lambda^A_\alpha(p^\mu) \dualn \lambda^A_\alpha(p^\mu) = - 2m \I_4
 \label{eq:ssa-newnew-b-new}
\end{align}
without affecting the  Lorentz invariance of the  orthonormality relations (\ref{eq:zimpokJ9an})-(\ref{eq:zimpokJ9cn}).\footnote{
In~\citep{Rogerio:2016mxi}  Rogerio et al. provide additional support for the new dual introduced here.}\index{Spin sums!Elko}

The completeness relation that follows from the Lorentz invariant spin 
sums (\ref{eq:sss-newnew-a-new}) and (\ref{eq:ssa-newnew-b-new}) takes the form\index{Elko!completeness}
\begin{align}
\frac{1}{4 m} \sum_{\alpha}  \left( \lambda^S_\alpha(p^\mu) \dualn \lambda^S_\alpha(p^\mu) 
- \lambda^A_\alpha(p^\mu) \dualn \lambda^A_\alpha(p^\mu)\right) =  \I_4
 \label{eq:completeness-li}
\end{align}

We thus conclude that a systematic analysis of spinorial duals we have resolved a long-standing problem on the construction of a 
Lagrangian density for the c-number Majorana spinors, extended here into Elko. The sought after Lagrangian density is not as given in equation~(\ref{eq:wrong}), or as conjectured and found not to exist in~\citep[App. P]{Aitchison:2004cs},  but 
\begin{equation}
\mathfrak{L}(x) = \partial^\mu{\gdualn{\lambda}(x)}\,\partial_\mu {{\lambda(x)}} - m^2 {\gdualn{\lambda}}(x) \lambda(x).\label{eq:correct}
\end{equation}
Strictly speaking, this result should be taken as suggestive till it is fully established in its quantum field theoretic incarnation in the next chapter.

 \chapter{Mass dimension one fermions}
 \label{ch13}


\section{A quantum field with Elko  as its expansion coefficient}


We now use the eigenspinors of the charge conjugation operator, Elko: $\lambda^S_\alpha(p^\mu)$ and $\lambda^A_\alpha(p^\mu)$, as expansion coefficients to define a new quantum field of spin one half
\begin{align}
\mathfrak{f}(x)  \stackrel{\textrm{def}}{=} & \int \frac{\text{d}^3p}{(2\pi)^3}  \frac{1}{\sqrt{2 m E(\p)}} \nonumber \\  & \times \sum_\alpha \Big[ a_\alpha(\p)\lambda^S_\alpha(\p) e^{- i p\cdot x}
+\, b^\dagger_\alpha(\p)\lambda^A_\alpha(\p)e^{ i p\cdot x}
{\Big]}
\label{eq:newqf}
\end{align}
where we have taken the liberty to notationally replace the $\lambda(p^\mu)$ by $\lambda(\p)$.
To decipher the mass dimensionality of $\mathfrak{f}(x)$ and to develop a quantum field theoretic formalism for the new field we define its adjoint
 \begin{align}
\gdualn{\mathfrak{f}}(x) \stackrel{\textrm{def}}{=}  & \int \frac{\text{d}^3p}{(2\pi)^3}   \frac{1}{ \sqrt{2 m E(\p)}} \nonumber\\ &\times
\sum_\alpha \Big[ a^\dagger_\alpha(\p)\gdualn{\lambda}^S_\alpha(\p) e^{ i p\cdot x}
 + b_\alpha(\p)\gdualn{\lambda}^A_\alpha(\p) e^{-i p\cdot x}{\Big]}\label{eq:newadjoint}
\end{align}

The creation and annihilation operators,  at this stage, are left free to obey
fermionic
\begin{align}
& \left\{a_\alpha(\p),a^\dagger_{\alpha^\prime}(\p^\prime)\right\} = \left(2 \pi \right)^3 \delta^3\hspace{-2pt}\left(\p-\p^\prime\right) \delta_{\alpha\alpha^\prime} \label{eq:a-ad-zimpok}\\
& \left\{a_\alpha(\p),a_{\alpha^\prime}(\p^\prime)\right\} = 0,\quad \left\{a^\dagger_\alpha(\p),a^\dagger_{\alpha^\prime}(\p^\prime)\right\} =0\label{eq:aa-adad-zimpok}
\end{align}
or bosonic
\begin{align}
& \left[a_\alpha(\p),a^\dagger_{\alpha^\prime}(\p^\prime)\right] = \left(2 \pi \right)^3 \delta^3\hspace{-2pt}\left(\p-\p^\prime\right) \delta_{\alpha\alpha^\prime} \label{eq:a-ad-zimpok}\\
& \left[a_\alpha(\p),a_{\alpha^\prime}(\p^\prime)\right] = 0,\quad \left[a^\dagger_\alpha(\p),a^\dagger_{\alpha^\prime}(\p^\prime)\right] =0\label{eq:aa-adad-zimpok-zimpok}
\end{align}
statistics.
We assume similar anti-commutativity/commutativity for $b_\alpha(\p)$ and 
$b^\dagger_\alpha(\p)$.  Under the assumption that the vacuum state $\vert\hspace{3pt}\rangle$ is normalised to unity, they fix the normalisation of one particle states
\begin{equation}
\vert\p,\alpha,a\rangle \stackrel{\textrm{def}}{=}a^\dagger_\alpha(\p)\vert\hspace{3pt}\rangle,\quad \vert\p,\alpha,b\rangle \stackrel{\textrm{def}}{=}b^\dagger_\alpha(\p)\vert\hspace{3pt}\rangle
\end{equation}
to be
\begin{align}
\langle\p^{\prime},\alpha^\prime,a\vert\p,\alpha,a\rangle = \left(2 \pi \right)^3 \delta^3\hspace{-2pt}\left(\p-\p^\prime\right) \delta_{\alpha\alpha^\prime} \nonumber\\
\langle\p^{\prime},\alpha^\prime,b\vert\p,\alpha,b\rangle = \left(2 \pi \right)^3 \delta^3\hspace{-2pt}\left(\p-\p^\prime\right) \delta_{\alpha\alpha^\prime}
\label{eq:orthonormality-zimpok}
\end{align}
In the above expressions, we use the key
\begin{equation}
a = \textrm{particle}, \quad b=\textrm{antiparticle}
\end{equation}

\section{A hint that the new field is fermionic}

A Lie algebraically stable theoretical framework requires that position and momentum measurements do not commute: theories with special relativistic symmetries must 
be quantum in nature~\citep{Flato:1982yu,Faddeev:1989LD,VilelaMendes:1994zg,Chryssomalakos:2004gk}.\footnote{A weaker version of this argument based solely on the rotational symmetry is given in section~\ref{sec:quantum}.}
With the ensuing irreducible product of $\hbar/2$ between the
uncertainties in the position and the momentum measurements this necessity
forces non-vanishing amplitude for the propagation of a particle  from an emission event $x$ to a space-like separated absorption event $y$ -- that is to a classically forbidden region. But since time ordering of events is not preserved for space-like separations, causal paradoxes come to exist unless the same very process that was interpreted as a particle propagation for the set of observers with $y_0 > x_0$ is re-interpreted as propagation of an antiparticle for observers for whom $y_0 < x_0$~\citep[Ch. 2, Sec.13]{Weinberg:1972gc}.  In the quantum field theoretic framework -- only known way to merge quantum and relativistic realms --  the processes that connect space-like separated events are mediated by virtual particles\index{Virtual particles}, particles that are off shell, that is, with energy, momentum, and mass that deviate from: $E^2 = \p^2 +m^2$.  The time-energy uncertainty relation
intervenes to protect the  energy-momentum conservation for fluctuations off the mass shell. 

With this background, 
to decipher the statistics for the $\mathfrak{f}(x)$ and 
$\gdualn{\mathfrak{f}}(x)$ system we now consider two space-like separated events, $x$ and $y$, along the lines of~\citep{Ahluwalia:2015vea}.
Referring to the observation on the lack of time ordering preservation for such a set-up 
we recognise the existence of  two sets of inertial frames, 
ones in which $y_0 > x_0$  and the ones in which the reverse is true, 
$x_0 > y_0$. We call these sets of inertial frame as $\mathcal{O}$  and 
$\mathcal{O}^\prime$  respectively.
In $\mathcal{O}$, we calculate the amplitude for a particle to propagate from $x$ to $y$ and in $\mathcal{O}^\prime$ the amplitude for an antiparticle to propagate from $y$ to $x$. Causality requires that these two amplitudes may only differ, at most, by a phase factor:\footnote{For the smoothness of the discussion,
we suppress a normalisation factor with the dimensions of inverse length squared till equation (\ref{eq:normalisationAdded}).}
\begin{equation}
\textrm{Amp}(x\to y, \textrm{particle})\big\vert_\mathcal{O}
=e^{i \theta}  \textrm{Amp}(y\to x, \textrm{antiparticle})\big\vert_{\mathcal{O}^\prime}\label{eq:phase}
\end{equation}
with $\theta\in\R$. The definition of $\mathfrak{f}(x)$ and its adjoint
$\gdualn{\mathfrak{f}}(x)$ given in equations (\ref{eq:newqf}) and (\ref{eq:newadjoint}), respectively, 
 allow us to obtain the following concrete results for the needed amplitudes:
 \begin{align}
\textrm{Amp}&(x\to y, \textrm{particle})\big\vert_\mathcal{O}  =
\langle\hspace{3pt}\vert \mathfrak{f}(y)
\gdualn{\mathfrak{f}}(x)\vert\hspace{3pt}\rangle\nonumber\\
& =\int\frac{\text{d}^3p}{(2 \pi)^3}\left(\frac{1}{2 m E(\p)}\right)
 e^{-ip\cdot(y-x)}
 \sum_\alpha\lambda^S_\alpha(\p)
\dualn\lambda^S_\alpha(\p) \label{eq:amplitudeP}
\end{align}
and
\begin{align}
 \textrm{Amp}(y\to x, \textrm{antiparticle})\big\vert_{\mathcal{O}^\prime}
 &=\langle\hspace{3pt}\vert \gdualn{\mathfrak{f}}(x) \mathfrak{f}(y)\vert\hspace{3pt}\rangle\big\vert_{\mathcal{O}^\prime}\nonumber\\
 & = \left[\langle\hspace{3pt}\vert \gdualn{\mathfrak{f}}(x) \mathfrak{f}(y)\vert\hspace{3pt}\rangle\big\vert_{\mathcal{O}}\right]_{(x-y)\to (y-x)}\nonumber\\
  =\int\frac{\text{d}^3p}{(2 \pi)^3}\left(\frac{1}{2 m E(\p)}\right)
 & e^{-ip\cdot(y-x)}
 \sum_\alpha\lambda^A_\alpha(\p)
\dualn\lambda^A_\alpha(\p) 
\end{align}
The Elko spin sums (\ref{eq:sss-newnew-a-new}) and
(\ref{eq:ssa-newnew-b-new}) when substituted in the above calculated amplitudes,  yields the result that the phase factor in (\ref{eq:phase})
is minus one:
$e^{i \theta} = -1$.
Furthermore, a direct calculation shows that 
\begin{equation}
\textrm{Amp}(y\to x, \textrm{antiparticle})\big\vert_{\mathcal{O}^\prime}
= \textrm{Amp}(y\to x, \textrm{antiparticle})\big\vert_{\mathcal{O}}\
\end{equation}
Combined, the above two results translate the relation (\ref{eq:phase}) to
\begin{equation}
\textrm{Amp}(x\to y, \textrm{particle})\big\vert_\mathcal{O}
=  -  \textrm{Amp}(y\to x, \textrm{antiparticle})\big\vert_{\mathcal{O}}\label{eq:phase-b}
\end{equation}
That is
\begin{equation}
\left\{\gdualn{\mathfrak{f}}(x),\mathfrak{f}(y)\right\}=0
\end{equation}

This is a hint, a strong hint, that the new field must be fermionic.
We thus take  the choice (\ref{eq:aa-adad-zimpok}) as part of the definition of the $\mathfrak{f}(x)$.

In the standard S-matrix theory the events that we detect at `spatial infinity' -- that is, at distances far away from the interaction region -- contain contribution from virtual particles as well as from on-sell particles. This amplitude is the Feynman-Dyson propagator. It is considered in the next section. 

However, before we undertake that study we  
 explicitly substitute the spin sum from (\ref{eq:sss-newnew-a-new}) into (\ref{eq:amplitudeP}), do the resulting integration, and 
 introduce the normalisation constant alluded 
 to above\footnote{The far from trivial 
 integration was done by Sebastian Horvath in our collaborative 
 work~\citep{Ahluwalia:2011rg}.}
 \begin{align}
\textrm{Amp}(x\to y, \textrm{particle})\big\vert_\mathcal{O}   &=  \frac{i}{2} m^2
\int\frac{\text{d}^3p}{(2 \pi)^3}\left(\frac{1}{ E(\p)}\right)
 e^{-ip\cdot(y-x)}\,
\I_4 \nonumber\\
& =\frac{i}{4 \pi^2} \frac{m^3}{\sqrt{\epsilon^2-\tau^2} }K_1(m 
\sqrt{\epsilon^2-\tau^2}) \label{eq:normalisationAdded}
\end{align}
where $\epsilon\stackrel{\textrm{def}}{=}\vert\x^\prime - \x\vert$,  
$\tau\stackrel{\textrm{def}}{=}\vert\x^{0\prime} - x^0\vert$, $\epsilon > \tau$ so that $y$ and $x$ represent events separated by space-like interval,  and $K_\nu(\ldots)$  is the modified Bessel function of the second kind of order one, that is with $\nu=1$. 

For a historical thread (in the context of Dirac fermions), involving Dirac, Pauli, and Feynman, we refer our reader to the discussion following equation (2.13) of~\citep{Ahluwalia:2011rg}.

\section{Amplitude for propagation} 
We now study amplitude of propagation from $x$ to $x^\prime$ without spacetime interval being restricted to  space-like separations. It would reveal the  mass dimensionality of the new field to be one.
We would be mindful that causal paradoxes can only be avoided for contributions from space-like separations if we allow particles to be replaced by antiparticles whenever time ordering of the events is reversed. In the interaction region we do not measure spacetime coordinates, $x$ and $x^\prime$ -- all our measurements in the scattering/collision processes take place in `far away regions.' This creates an unavoidable  ambiguity in time ordering of the absorption and emission of the virtual particles. This is incorporated in our calculations, as in the standard quantum field theoretic formalism, by adding all possible amplitudes that connect $x$ and~$x^\prime$, at the same time we keep in mind the time ordering.

With this background, we note that:
\begin{itemize}
\item The fermionic statistics requires the amplitude for the $x \to x^\prime$ propagation to be antisymmetric under the exchange
$x\leftrightarrow x^\prime$.
\item
Causality requires that a particle, or an antiparticle, cannot be absorbed before it is emitted, and vice versa.
\end{itemize}
In the S-matrix formulation of quantum field theory,  amplitude for a particle to propagate from $x$ to $x^\prime$ 
incorporates all these facts by giving it 
the general form
\begin{align}
\mathcal{A}_{x\to x^\prime}  &= 
\textrm{Amp}   (x\to x^\prime,  \textrm{particle})\big\vert_{t^\prime>t} 
 - \textrm{Amp}(x^\prime\to x, \textrm{antiparticle})\big\vert_{t>t^\prime}
 \nonumber\\
& =  \xi \Big(\underbrace{\langle\hspace{3pt}\vert
\mathfrak{f}(x^\prime)\gdualn{\mathfrak{f}}(x)\vert\hspace{3pt}\rangle \theta(t^\prime-t)
-  \langle\hspace{3pt}\vert
\gdualn{\mathfrak{f}}(x) \mathfrak{f}(x^\prime)\vert\hspace{3pt}\rangle \theta(t-t^\prime)}_{\langle\hspace{4pt}\vert \mathfrak{T} ( \mathfrak{f}(x^\prime) \gdualn{f}(x)\vert\hspace{4pt}\rangle}\Big)
\end{align}
where $\xi\in\C$ is to be determined from the normalisation condition that
$
 \mathcal{A}_{x\to x^\prime} 
$
integrated over all possible separations $x-x^\prime$ be unity,\footnote{Or, more precisely  $e^{i\gamma}$, with $\gamma\in \R$.} and $\mathfrak{T}$ is the time ordering operator. The two vacuum expectation values that appear in  $\mathcal{A}_{x\to x^\prime}$ can be evaluated as before but with the care that  $(x-x^\prime)$ is no longer restricted to a space-like separation, the result is
 \begin{align}
\langle\hspace{3pt}\vert
\mathfrak{f}(x^\prime)\gdualn{\mathfrak{f}}(x)\vert\hspace{3pt}\rangle  & =\int\frac{\text{d}^3p}{(2 \pi)^3}\left(\frac{1}{2 m E(\p)}\right)
 e^{-ip\cdot(x^\prime-x)}
 \sum_\alpha\lambda^S_\alpha(\p)
\dualn\lambda^S_\alpha(\p) \label{eq:amplitudeP-newS}
\\
\langle\hspace{3pt}\vert
\gdualn{\mathfrak{f}}(x) \mathfrak{f}(x^\prime)\vert\hspace{3pt}\rangle 
  & = - \int\frac{\text{d}^3p}{(2 \pi)^3}\left(\frac{1}{2 m E(\p)}\right)
 e^{ip\cdot(x^\prime-x)}
 \sum_\alpha\lambda^A_\alpha(\p)
\dualn\lambda^A_\alpha(\p) .\label{eq:amplitudeP-newA}
\end{align}
The two Heaviside step function can be replaced by their integral representations
\begin{align}
\theta(t^\prime-t) &= \lim_{\epsilon\to 0^+} \int\frac{\text{d}\omega}{2\pi i}
\frac{e^{i \omega (t^\prime-t)}}{\omega- i \epsilon} \\
\theta(t-t^\prime) &= \lim_{\epsilon\to 0^+} \int\frac{\text{d}\omega}{2\pi i}
\frac{e^{i \omega (t-t^\prime)}}{\omega- i \epsilon}
\end{align}
where $\epsilon,\omega\in\R$. Using these results, and  
\begin{itemize}
\item substituting $\omega \to p_0 = -\omega+E(\p)$ in the first term and  $\omega \to p_0 = \omega- E(\p)$ in the second term
\item and using the results (\ref{eq:sss-newnew-a-new}) and 
(\ref{eq:ssa-newnew-b-new}) for the spin sums
\end{itemize}
we get
\begin{equation}
\mathcal{A}_{x\to x^\prime}  = i \,2 \xi \int\frac{\text{d}^4 p}{(2 \pi)^4}\,
e^{-i p_\mu(x^{\prime\mu}-x^\mu)}
\frac{\I_4}{p_\mu p^\mu -m^2 + i\epsilon}\label{eq:AmplitudeWithXi}
\end{equation}
We fix $\xi$ by the requirement that this amplitude when integrated over all possible separations $x^\prime-x$ yields unity. Towards that we take note of the integral representation of the delta function
\begin{equation}
\delta(x-a) = \frac{1}{2 \pi} \int_{-\infty}^{\infty}e^{i(x-a) t} \text{d}t
\end{equation}
and its symmetric aspect $\delta(x) = \delta(-x)$, and obtain
\begin{equation}
 i \,2 \xi \int\frac{\text{d}^4 p}{(2 \pi)^4}\,
\delta(p_\mu)
\frac{\I_4}{p_\mu p^\mu -m^2 + i\epsilon} = 
\I_4
\end{equation}
As such we have
\begin{equation}
\frac{ i \,2 \xi }{-m^2+i\epsilon} =1
\end{equation}
In the limit $\epsilon\to 0$
\begin{equation}
\xi = \frac{i m^2}{2}
\end{equation}
Up to a possible global phase $e^{i\gamma}$ mentioned earlier, the amplitude (\ref{eq:AmplitudeWithXi}) becomes\index{Amplitude for propagation}
\begin{equation}
\mathcal{A}_{x\to x^\prime}  = - m^2 \int\frac{\text{d}^4 p}{(2 \pi)^4}\,
e^{-i p_\mu(x^{\prime\mu}-x^\mu)}
\frac{\I_4}{p_\mu p^\mu -m^2 + i\epsilon}
\end{equation}
with $\epsilon\to 0^+$.

The $\xi$ differs from $\varpi$ of \citep{Ahluwalia:2004ab} by a factor of half:  $\xi = (1/2) \varpi $. The origin of this difference resides in the new spin sums  (\ref{eq:sss-newnew-a-new}) and 
(\ref{eq:ssa-newnew-b-new}).

\section{Mass dimension one fermions}
\index{Mass dimension one fermions}

To decipher the mass dimension of the new fermions we must know
the Feynman-Dyson propagator associated with the set 
$\mathfrak{f}(x)$ and $\gdualn{\mathfrak{f}}(x)$.

The Feynman-Dyson propagator\index{Feynman-Dyson propagator} is  defined to be proportional to 
$\mathcal{A}_{x\to x^\prime} $ in such a way that the proportionality constant is adjusted to make the Feynman-Dyson propagator  coincide with the Green function associated with the equation of motion for the field $\mathfrak{f}(x)$. To determine this proportionality constant we act
the spinorial Klein-Gordon operator on the amplitude $\mathcal{A}_{x\to x^\prime}$ 
\begin{equation}
\left(\partial_{\mu^\prime} \partial^{\mu^\prime} \I_4 + m^2\I_4\right)
\mathcal{A}_{x\to x^\prime} = m^2 \delta^4(x^\prime - x)
\end{equation}
So we define the Feynman-Dyson propagator 
\begin{align}
S_{\textrm{FD}}(x^\prime-x) & \stackrel{\textrm{def}}{=} - \frac{1}{m^2} 
\mathcal{A}_{x\to x^\prime}\nonumber\\
 &= \int\frac{\text{d}^4 p}{(2 \pi)^4}\,
e^{-i p_\mu(x^{\prime\mu}-x^\mu)}
\frac{\I_4}{p_\mu p^\mu -m^2 + i\epsilon}\label{eq:FD-prop-b}
\end{align}
With this definition
\begin{equation}
\left(\partial_{\mu^\prime} \partial^{\mu^\prime} \I_4 + m^2\I_4\right)
S_{\textrm{FD}}(x^\prime-x)  = -  \delta^4(x^\prime - x)\label{eq:KGDiracDelta}
\end{equation}
and the Feynman-Dyson propagator in terms of the new field and its adjoint takes the form
\begin{equation}
S_{\textrm{FD}}(x^\prime-x)= -\frac{i}{2}\left\langle\hspace{4pt}\left\vert \mathfrak{T} ( \mathfrak{f}(x^\prime) \gdualn{f}(x)\right\vert\hspace{4pt}\right\rangle
\end{equation}

Following the canonical discussion on the mass dimensionality of quantum fields given in~\citep[Section 12.1]{Weinberg:1995mt}
we find that mass dimension of the field $\mathfrak{f}(x)$ is  one \index{Lagrangian density}
\begin{equation}
\mathfrak{D}_{\mathfrak{f}} = 1 \label{eq:df1}
\end{equation}
and not three-half, as is the case for the Dirac field: for large momenta $p$,  $S_{\textrm{FD}} (x^\prime-x) \propto p^{-1}$ for the latter, while for the new field
$S_{\textrm{FD}} (x^\prime-x) \propto p^{-2}$. This result is precisely what was hinted at by our discussion in section~\ref{sec:elko-do-not-satisfy-Dirac-equation}.

The discussion of section~\ref{sec:elko-do-not-satisfy-Dirac-equation},
when coupled with the results encoded in equations~(\ref{eq:FD-prop-b}) and~(\ref{eq:KGDiracDelta}),
suggests the following  free field Lagrangian density for the new field 
\begin{equation}
\mathfrak{L}_0(x) = \frac{1}{2}\left(\partial^a\gdualn{\mathfrak{f}}\,\partial_a {\mathfrak{f}}(x) - m^2 \gdualn{\mathfrak{f}}(x) \mathfrak{f}(x)\right) \label{eq:fieldlagrangian}
\end{equation}
where we have taken liberty to let $a,b,\ldots$ take the flat space values 
$(0,1,2,3)$, and reserve $\mu,\nu,
\mbox{etc}$ to take on the values for the names of the co-ordinates in the gravitational context, for example $(t,x,y,z)$. The motivation to introduce the factor of half resides in our desire to obtain canonical form of the locality anticommutators. It does not alter the equation of motion.

Like the Dirac fermions, the new fermions, are of spin one half. Both satisfy Klein-Gordon equation. 
For the former the ``factorisation'' of the Klein-Gordon operator occurs through the Dirac operator, while for the latter it occurs 
through equations (\ref{eq:er-a1}) to (\ref{eq:er-b2}).
The expansion coefficients of the former are a complete set of eigenspinors of the parity operator, while for the latter  the expansion coefficients are a complete set of eigenspinors of the charge conjugation operator. Both the fields are local as we shall soon discover in section~\ref{sec:locality}.

Gravity enters the realm of Elko and mass dimension one fermions 
 by introducing tetrads $e^a_\mu$\index{Tetrads} through
$e^a_\mu e^b_\nu \eta_{ab} = g_{\mu\nu}$. Here $g_{\mu\nu}$ is the spacetime metric and $\eta_{ab}=\mbox{diag}(1,-1,-1,-1)$. Thus, for example, space-time  Dirac matrices are connected with their flat space counterpart by $\gamma^\mu = e^\mu _a \gamma^a$, which consequently satisfy
\begin{equation}
\left\{\gamma^\mu,\gamma^\nu\right \} = 2 g^{\mu\nu}
\end{equation}
The Latin indices  -- $a,b,\ldots = 0,1,\ldots$, called non-holonomic indices --  refer to a  local inertial frame. Greek indices  -- $\mu,\nu,\ldots = t,\ldots$, called holonomic indices --  refer to a generic non-inertial frame.

The Lagrangian density (\ref{eq:fieldlagrangian}) 
for the mass dimension one fermions in a torsion-free gravitational field follows
the prescription
\begin{align}
\partial_a \mathfrak{f}(x) & \to \nabla_\mu \mathfrak{f}(x)  \stackrel{\textrm{def}}{=} \partial_\mu \mathfrak{f}(x)- \Gamma_\mu \mathfrak{f}(x)\\
\partial_a\gdualn{\mathfrak{f}}(x) & \to \nabla_\mu\gdualn{\mathfrak{f}} (x)\stackrel{\textrm{def}}{=} \partial_\mu \gdualn{\mathfrak{f}} + 
\gdualn{\mathfrak{f}}(x) \Gamma_\mu
\end{align}
where
\begin{equation}
\Gamma_\mu = \frac{i}{2}\omega^{ab}_\mu\left(
\mathfrak{J}_{ab} \Big\vert_{(\mathcal{R}\oplus\mathcal{L})_{s=1/2}} \right)
\end{equation}
and the spin connection\index{Spin connection} is defined as
\begin{equation}
\omega^{ab}_\mu = e^a_\nu\left(\partial_\mu e^{\nu b}
+ e^{\sigma b}\Gamma^\nu_{\mu\sigma}\right)
\end{equation}
with $\Gamma^\nu_{\mu\sigma}$ denoting the Christoffel symbol associated with $g_{\mu\nu}$. In addition
\beq
\mathfrak{J}_{ab}\Big\vert_{(\mathcal{R}\oplus\mathcal{L})_{s=1/2}}  =  
\begin{cases}
\mathfrak{J}_{ij} =  - \mathfrak{J}_{ji} = \epsilon_{ijk} \zeta_k \\
\mathfrak{J}_{i0} =  -\mathfrak{J}_{0i}  = - \kappa_i
\end{cases}
\eeq
Equation (\ref{eq:pi}) provides explicit expressions for $\kb$ and $\bz$, respectively the boost and rotation generators in the $\mathcal{R}\oplus\mathcal{L}$ representations space for spin one half.

This prescription is as valid for Elko as for the Dirac spinors because the spinorial covariant derivative depends on the generators of the Lorentz algebra for spin one half in the indicated fashion. B\"ohmer has verified this observation independently, without invoking our argument, in~\citep{Boehmer:2007dh}.  The physics of Elko and torsion is discussed at length in
 \citep{Fabbri:2010ws,KOUWN:2013wza,Fabbri:2014foa,Pereira:2017efk,Pereira:2017bvq}. In the presence of torsion, denoted by a `tilde' above the symbols, the spinorial covariant derivatives no longer 
 commute
 \begin{equation}
 \left[\tilde{\nabla}_\mu,\tilde{\nabla}_\nu \right]\ne \0
 \end{equation}
 and (\ref{eq:fieldlagrangian}) is modified to
 \begin{equation}
\tilde{\mathfrak{L}}(x) = \frac{1}{2}\left(\tilde\nabla^\mu\gdualn{\mathfrak{f}}\,\tilde{\nabla}_\mu {\mathfrak{f}}(x) - m^2 \gdualn{\mathfrak{f}}(x) \mathfrak{f}(x)\right) \label{eq:fieldlagrangianB}
\end{equation}
allowing for such additional interactions as 
\begin{equation}
\frac{1}{2}  \tilde\nabla_\mu\gdualn{\mathfrak{f}}(x)\sigma^{\mu\nu}\tilde\nabla_\nu\mathfrak{f}(x)\end{equation}
 with
\begin{equation}\sigma^{\mu\nu}
\stackrel{\textrm{def}}{=}
\frac{1}{4}\Big[\gamma_\mu,\gamma_\nu\Big]
\end{equation}
This was first noted by Luca Fabbri in \citep{Fabbri:2010ws}.\index{Luca Fabbri}

Despite formal similarity, Elko and Dirac spinors experience gravity differently. From a physical point of view the $\mathcal{R}$ 
and $\mathcal{L}$ transforming components of Elko pick up different, and in certain circumstances opposite, gravitationally-induced phase factors. This is apparent from their helicity structure~\citep{Ahluwalia:2004kv}. 

Another distinguishing feature of Elko arises from  the contribution to the energy-momentum tensor  from the variation of spin connection. For the Dirac spinors this contribution identically vanishes. In contrast, for Elko it idoes not. This was first noted by Christian B\"ohmer in~\citep{Boehmer:2010ma}.\index{B\"ohmer}

\section{Locality structure of the new field\label{sec:locality}}

Now that we have $\mathfrak{L}_0(x)$ we can calculate the momentum conjugate to $\mathfrak{f}(x)$
\begin{equation}
\mathfrak{p}(x) = \frac{\partial \mathfrak{L}_0(x)}
{\partial {\dot{\mathfrak{f}}(x)}} =  \frac{1}{2} \frac{\partial}{\partial t}\gdualn{\mathfrak{f}}(x).
\end{equation}
To establish that the new field is local we calculate the standard equal-time anti-commutators. The first of the three anti-commutators we calculate is the `$\mathfrak{f}$-$\mathfrak{p}$' anti-commutator 
\begin{equation}
\left\{ \mathfrak{f}(t,\x),\;\mathfrak{p}(t,\x^\prime) \right\}.
\end{equation}
With the defintion
\begin{equation}
\sum_{\alpha,\bf{p}} \stackrel{\textrm{def}}{=}
 \int \frac{\text{d}^3p}{(2\pi)^3}  \frac{1}{\sqrt{2 m E(\p)}}
   \sum_\alpha 
\end{equation}
it expands to
\begin{align}
 \frac{1}{2} \sum_{\alpha,\bf{p}}\sum_{\alpha',\bf{p'}} 
 \left(iE(\p'\right) {\Big[} 
&  \left\{a_\alpha(\p),a^\dagger_{\alpha'} (\p')\right\}
 \lambda^S_\alpha(\p) \gdualn{\lambda}^S_{\alpha'}(\p')
 e^{-i p\cdot x+i p'\cdot x'}\nonumber\\
& -  \left\{b_\alpha(\p),b^\dagger_{\alpha'} (\p')\right\}
 \lambda^A_\alpha(\p) \gdualn{\lambda}^A_{\alpha'}(\p')
 e^{i p\cdot x- i p'\cdot x'}
 {\Big]}
\end{align}
Replacing each of the anticommutators by $\left(2 \pi \right)^3 \delta^3\hspace{-2pt}\left(\p-\p^\prime\right) 
\delta_{\alpha\alpha^\prime}$ and performing the $\p'$ integration, and $\alpha'$ summation, followed by (a) change of variables $\p\to -\p$ in the second integration, and (b) noting that the $t$ dependence in the exponentials cancels out, we get 
\begin{equation}
\frac{i}{4 m} \int \frac{\text{d}^3p}{(2\pi)^3} 
e^{i\bf{p}\cdot(\bf{x}-\bf{x'})}
\sum_\alpha\left(
 \lambda^S_\alpha(\p) \gdualn{\lambda}^S_{\alpha}(\p)
- \lambda^A_\alpha(- \p) \gdualn{\lambda}^A_{\alpha}(-\p)
\right)
\end{equation}
The first spin sum evaluates to $2 m\I$, while the second equals 
$-2 m\I$, giving the result $4 m \I$ for the summation over $\alpha$. This gives us
\begin{equation}
\left\{ \mathfrak{f}(t,\x),\;\mathfrak{p}(t,\x^\prime) \right\} = i \delta^3\left(\x-\x^\prime\right) \openone_4 .\label{eq:lac-1}
\end{equation}
Had we not included a factor $1/2$ in the definition of the Lagrangian density in (\ref{eq:fieldlagrangian}) we would have gotten an additional factor of $2$ multiplying the delta function in the above result.

A still simpler calculation shows that the remaining two, that is, `$\mathfrak{f}$-$\mathfrak{f}$' and  `$\mathfrak{p}$-$\mathfrak{p}$', equal time anti-commutators\index{Equal time anti-commutators} 
vanish
\begin{equation}
\left\{ \mathfrak{f}(t,\x),\;\mathfrak{f}(t,\x^\prime) \right\} = 0, \quad 
\left\{ \mathfrak{p}(t,\x),\;\mathfrak{p}(t,\x^\prime) \right\} = 0.\label{eq:lac-2and3}
\end{equation}
The field $\mathfrak{f}(x)$ is thus local in the sense of Schwinger~\citep[Sec. II, Eqs. 2.82]{PhysRev.82.914}. It is a much stronger condition of locality than that adopted by Schwartz~\citep[Sec. 24.4]{Schwartz:2013pla}.

\subsection{Majorana-isation of the new field\label{sec:Majorana-isation}}
\index{Majorana-isation}

Even though field $\mathfrak{f}(x)$ is uncharged under local U(1) supported by the Dirac fields of the standard model of high energy physics, it may carry a charge under a different  local U(1) gauge symmetry such as the one suggested in the discussion around (\ref{eq:mathfraka}). This gives rise to the  possibility of having a fundamentally neutral field in the sense of Majorana~\citep{Majorana:1937vz}
\begin{align}
\mathfrak{m}(x) &=   \int \frac{\text{d}^3p}{(2\pi)^3}  \frac{1}{ \sqrt{2 m E(\p)}} \nonumber\\ &\times \sum_\alpha \Big[ a_\alpha(\p)\lambda^S(\p) \exp(- i p_\mu x^\mu)
+ \, a^\dagger_\alpha(\p)\gdualn\lambda^A(\p) \exp(i p_\mu x^\mu){\Big]} 
\end{align}
with momentum conjugate 
\begin{equation}
\mathfrak{q} = \frac{1}{2} \frac{\partial}{\partial t}\gdualn{\mathfrak{m}}(x).
\end{equation}
The calculation for the `$\mathfrak{m}$-$\mathfrak{q}$' equal time anti-commutators goes through exactly as before and one gets
\begin{equation}
\left\{ \mathfrak{m}(t,\x),\;\mathfrak{q}(t,\x^\prime) \right\} = i \delta^3\left(\x-\x^\prime\right) \openone_4 .
\end{equation}
The calculation of the remaining two anti-commutators requires knowledge of the following 
`twisted' spin sums 
\begin{align}
&\sum_\alpha\left[ \lambda^S_\alpha(\p)\left[ {\lambda}^A_\alpha(\p)\right]^T  + 
 \lambda^A_\alpha(-\p) \left[{\lambda}^S_\alpha(-\p)\right]^T  \right]\\
 & \sum_\alpha\left[ \left[\gdualn{\lambda}^S_\alpha(\p)\right]^T {\gdualn{\lambda}}^A_\alpha(\p)  + 
 \left[\gdualn{\lambda}^A_\alpha(-\p)\right]^T \gdualn{\lambda}^S_\alpha(-\p)\right].
\end{align}
One finds that each of these vanishes. With this result at hand, we immediately decipher vanishing of the `$\mathfrak{m}$-$\mathfrak{m}$' and  `$\mathfrak{q}$-$\mathfrak{q}$', equal time anti-commutators 
\begin{equation}
\left\{ \mathfrak{m}(t,\x),\;\mathfrak{m}(t,\x^\prime) \right\} = 0, \quad 
\left\{ \mathfrak{q}(t,\x),\;\mathfrak{q}(t,\x^\prime) \right\} = 0.
\end{equation}
The field $\mathfrak{m}(x)$, like $\mathfrak{f}(x)$, is thus local in the sense of Schwinger~\citep[Sec. II, Eqs. 2.82]{PhysRev.82.914}.

\chapter{Mass dimension one fermions as a first principle dark matter}
\label{ch14}

\section{Mass dimension one fermions as dark matter}

\begin{itemize}
\item A mass dimension mismatch between the new fermions and the standard model matter fields --  one versus three halves --  forbids them to enter the standard model doublets.\index{Mass dimension mismatch} 
 In the process  their interaction with the standard model fields is
severely restricted. 
\vspace{7pt}

One exception is the dimension four operators
\begin{equation}
 \lambda_{\textrm\small \mathfrak{f}\phi} \gdualn{\mathfrak{f}}(x) \mathfrak{f}(x) \,\phi^\dagger(x)\phi(x),
\quad
 \lambda_{\textrm\small \mathfrak{m}\phi} \,\gdualn{\mathfrak{m}}(x) \mathfrak{m}(x) \,\phi^\dagger(x)\phi(x)
 \end{equation}
where  $ \lambda_{\textrm\small \mathfrak{f}\phi}$ and 
$ \lambda_{\textrm\small \mathfrak{m}\phi}$
are dimensionless coupling constant, and $\phi(x)$ is the Higgs doublet.  
For the Dirac field, a similar interaction is a dimension five operator. It is thus suppressed by one power of the unification/Planck scale. 

Therefore, mass dimension one fermions are natural dark matter candidate. Compared to bosonic dark matter a fermionic dark matter brings with it Pauli exclusion principle to affect structure formation.
\vspace{10pt}

\item
The system of mass dimension one fermions further supports perturbatively  re-normalisable 
dimen\-sion-four quartic self interactions\index{Quartic self interactions}
\begin{equation}
\lambda_\mathfrak{f} \left(\gdualn{\mathfrak{f}}(x) \mathfrak{f}(x)\right)^2,\quad\
\lambda_\mathfrak{m} \,\left(\gdualn{\mathfrak{m}}(x) \mathfrak{m}(x)\right)^2
\end{equation}
where $\lambda_\mathfrak{f}$ and $\lambda_\mathfrak{m}$ are dimensionless coupling constants. 
A similar quartic self interaction for the Dirac field, with or without Majorana-isation,  is a dimension six operator. It is thus suppressed by two power of the unification/Planck scale. 

\vspace{7pt}

Observational evidence that favours self interacting dark matter is reviewed in~\citep{Tulin:2017ara} and in a  Ph.D. thesis by Robertson~\citep{Robertson:2017igt}. An early reference 
is~\citep{Spergel:1999mh}. Following these references we simply note that self interacting dark matter provides a heat transport mechanism from the outer hotter to the cooler inner region of dark matter halos. It thermalises the inner halo and leads to a uniform velocity distribution as one moves outward in the halo.
 
\vspace{10pt}
\item The very definition of Elko does not allow covariance under local phase transformations of the standard model (see, section \ref{sec:Restriction} for local U(1)). This endows the new fermions with a natural darkness with respect to the standard model fields while leaving open the possibility of new gauge symmetries that apply to mass dimension one fermions.

\end{itemize}


Beyond the dimension-four interactions mentioned above one may also introduce the following Yukawa couplings\index{Yukawa couplings} of dimension
 three and half~\citep{ArnabDasgupta} 
\begin{align}
&\lambda_1\,
\varphi(x) \,\overline{\psi}(x)\mathfrak{f}(x),\quad
\lambda_2\,\varphi(x)  \gdualn{\mathfrak{f}} (x) {\psi}(x)\\
 &\lambda_3 \,
\varphi(x) \,\overline{\psi}(x)\mathfrak{m}(x),\quad
 \lambda_4\,\varphi(x)  \gdualn{\mathfrak{m}} (x) {\psi}(x)\label{eq:Yukawa}
\end{align}
where $\psi(x)$ is a Dirac/Majorana field, $\varphi(x)$ is a scalar field, and $\lambda's$  are dimensionfull coupling constants. These may be used to violate conservation of all three of the following: lepton number, electric charge, and dark charges\footnote{A similar mass dimension four coupling, without a quartic self interaction term, 
of Dirac/Major\-ana fermions
to a scalar can be found in~\citep{Kainulainen:2015sva}.}



The interactions of the form
\begin{equation}
\epsilon \gdualn{\mathfrak{f}}(x)\left[\gamma^a,\gamma^b\right]
\mathfrak{f}(x) F_{ab}(x),\quad
\varepsilon \,\gdualn{\mathfrak{m}}(x)\left[\gamma^a,\gamma^b\right]
\mathfrak{m}(x) F_{ab}(x)
\end{equation}
with dimensionless couplings $\epsilon$ and $\varepsilon$ are severely restricted due to stringent limits on photon mass\index{Photon mass}
\citep{PhysRevD.69.107501,PhysRevLett.90.081801,Bonetti:2016cpo}. Nevertheless, they could have significant astrophysical and cosmological implications where the smallness of the couplings may be compensated by huge dark matter densities. One crucial arena where such couplings may manifest are in 21-cm cosmology~\citep{Barkana:2018qrx}.




\section{A conjecture on a mass dimension transmuting symmetry}\label{Sec:conjecture}

A single component dark matter sector is likely to be an 
 oversimplified view of the cosmological reality with the possibility of distorting our intuitions.
We conjecture  that the realm of dark matter is populated by a set of
fields of  the $\mathfrak{f}(x)$ type and its Majorana-ised form 
$\mathfrak{m}(x)$.
 We envisage the possibility that a mass dimension transmuting symmetry\index{Mass dimension transmuting symmetry} exists. It takes mass dimension one fermions to mass dimension three halves fermions, and vice versa. Unlike supersymmetry the conjectured new symmetry does not alter the statistics, but only the mass dimensionality.
 
 In this possibility for every standard model fermion there exists a mass dimension one fermion, and vice versa. The possibility that Higgs is shared by both the sectors is the simplest unifying theme we can envisage. To complete our conjecture we suggest that dark gauge fields mirror the standard model gauge fields. The coupling constants and masses we leave as free parameters, but we would be surprised if they were not related to those of the standard model. 
 Echoing  the arguments of~\citep{Hardy:2014mqa,McDermott:2017vyk}:
 Combined, the dark sector may support some sort of dark fusion\index{Dark fusion} leading to dark nucleosynthesis.\index{Dark nucleosynthesis} 
 \vspace{11pt}
 
 \noindent
\textit{A parenthetic remark~\textemdash}

For reasons to be soon encountered the dark sector supported by Elko has  distinctive gravitational signatures with torsion revealing aspects that are absent with the Dirac spinors. Elko cosmology is still in its infancy, and given its first principle origins it holds promise to give us observational signatures not yet anticipated by cosmologists. Observations can be  easily misinterpreted in the absence of a systematic development of a cosmology that fully incorporates  the new fermions.
Such a process was begun by Christian B\"ohmer,\index{B\"ohmer} and is now an active realm of researches by a new generation of physicists mostly from South America, Europe, and Asia.

\section{Elko inflation and Elko dark energy}

Echoing the closing remarks of the last section, Christian B\"ohmer was  the first cosmologist to realise that:
\begin{itemize}

\item Elko not only help formulate  mass dimension one fermions for dark matter, as Daniel Grumiller\index{Grumiller} and I had argued, but also that Elko could serve as as a source of inflation and could drive the accelerated expansion of the universe~\citep{Boehmer:2006qq,Boehmer:2007dh,Boehmer:2008rz}. 

\item In a stark contrast to Dirac spinors, Elko  energy momentum tensor $T_{\mu\nu}$ has an important non-vanishing  contribution from the  variation of spin connection~\citep{Boehmer:2010ma}. \index{B\"ohmer}
\end{itemize}

These early works inspired a series of efforts to explore Elko as a source inflation, dark matter, and dark energy. 
Christian B\"ohmer was also the first to note that the spin angular momentum tensor associated with Elko cannot be entirely expressed as an axial torsion vector~\citep{Boehmer:2006qq}. He emphasised that this important difference from the Dirac spinors arises due to different helicity structures of the Elko and Dirac spinors. His groundbreaking paper also put forward a tiny coupling of Elko  to Yang-Mills fields and discussed its implications for consistently coupling massive spin one field to the Einstein-Cartan theory. Restricting to the  Einstein-Elko system  he constructed analytical ghost Elko solutions with the property of a vanishing energy-momentum tensor~\citep{Boehmer:2006qq}.
This was done to make the analytical calculations possible\footnote{This assumption was later placed on a more natural footing by~\citep{Chang:2015ufa} \textit{et al.}} and he showed that, ``the Elko \ldots are not only prime dark matter candidates but also prime candidates for inflation.'' With his collaborators 
B\"ohmer has placed  Elko cosmology on a firm footing with an eye on the available 
data. We refer the reader to references~\citep{Boehmer:2007ut,Boehmer:2008rz,Boehmer:2008ah,Boehmer:2009aw,Boehmer:2010ma} for details. While building Elko cosmology he has coined the term ``dark spinors'' for Elko.

The group of Julio Hoff da Silva\index{Hoff da Silva} and Saulo Pereira,\index{Pereira} focusing on exact analytical solutions, have taken Elko cosmology significantly beyond 
B\"ohmer's initial pioneering efforts. We refer the reader to their 
publications~\citep{daSilva:2014kfa,Pereira:2014wta,S:2014dja,Pereira:2014pqa}.
Concurrently 	extending the work of B\"ohmer, Gredat and Shankaranarayanan have considered an Elko-condensate driven inflation and shown that it is favoured by existing observational data~\citep{Gredat:2008qf}. This work has been followed by Basak and Shankaranarayanan to prove that, 
``Elko driven inflation can generate growing vector modes even in the first order." This allows them to generate vorticity during inflation to produce primordial magnetic field~\citep{Basak:2014qea}. 

Basak et al.\index{Basak} show that Elko cosmology provides two sets of attractor points. These correspond to slow and fast-roll inflation. The latter being unique to Elko~\citep{Basak:2012sn}. For earlier contribution to Elko cosmology from this group we refer  the reader \index{Shankaranarayanan} to~\citep{Shankaranarayanan:2010st,Shankaranarayanan:2009sz}. The cosmological coincidence problem in the context of Elko is discussed by Hao Wei \index{Hao Wei}in reference~\citep{Wei:2010ad}, and has been revisited 
in \citep[Sec. 7.1]{Bahamonde:2017ize}.
One of the early papers on stability of de Sitter solution in the context of Elko is ~\citep{Chee:2010ju}.

Elko cosmology has gained a significant and independent  boost through a recent study of phantom dark-energy Elko/dark-spinors  undertaken by Yu-Chiao Chang \textit{et al.}\index{Yu-Chiao Chang} In the context of Einstein-Cartan theory it resolves a host of problems with phantom dark energy models and predicts a final  de Sitter phase for our universe at late time with or without dark matter~\citep{Chang:2015ufa}. Their work not only makes Elko and mass dimension one fermions more physically viable but it also lends concrete physicality to torsion as an important possible element of reality.

\section{Darkness is relative, not self referential}

To avoid confusion, a clarifying remark seems necessary: The dark sector need not be self referentially dark. To dark matter, and to the all pervading field  -- dark 
energy -- responsible for the accelerated expansion of the universe, we are dark. But, darkness is relative, and the dark sector may not be self referentially dark. It may support its own gauge fields, and carry its own luminosity. Our luminosity arises from the gauge fields supported by the matter fields which use eigenspinors of the parity operator as its expansion coefficients. A self referentially luminous dark sector may arise  from the gauge fields that matter fields based on Elko support.


\chapter{Continuing the story}
\label{ch15}
In this closing chapter, I take the liberty of suggesting what in essence are research projects that a beginning students may pursue. Most of these have been developed in detail in my private notes. I would be happy to share the details with anyone interested in pursuing them further.

\section{Constructing the spacetime metric from Lorentz algebra}

Returning to section \ref{sec:constructing-xt} we continue an \textit{ab initio} examination of four vectors spanning the $\mathcal{R}\otimes\mathcal{L}\vert_{s=1/2}$ representation space. We denote these vectors by $\chi$, and take them in the form of four-component columns with their elements in $\C$. We shall assume that we have implemented the rotation defined by $U$ of equation (\ref{eq:Umatrix}), then if $\chi$ represents the position of an event then the elements  take values in $\Re$ but otherwise this restriction is not necessary. 

Following the discussion in chapter 
\ref{ch12}
the simplest dual for $\chi$ is
\begin{equation}
\overline{\chi} \stackrel{\textrm{def}}{=} \chi^\dagger\eta
\end{equation}
In order that $\overline{\chi} \chi$ is an invariant for observers connected by Lorentz transformations, the metric $\eta$ must satisfy
\begin{equation}
\{K_i,\eta\}=0,\quad \left[J_i,\eta\right]=0,\qquad i=x,y.z \label{eq:constraints}
\end{equation}
with the boost and rotation generators, $K_i$ and $J_i$, defined by equations (\ref{eq:kx})
 to (\ref{eq:jyjz-new}). 
 
We give $\eta$ the form
\begin{equation}
\eta = \left(
\begin{array}{cccc}
a+i \alpha & b+i \beta & c + i \sigma & d + i\delta\\
e +i \epsilon & f +  i\phi & g + i\gamma & h + i \lambda\\
j + i \zeta & k + i \kappa & m + i \mu & n + i\nu\\
p + i \omega & q + i\rho & r + i\chi & s + i \psi
\end{array}
\right)
\end{equation}
with $a \ldots s, \alpha\ldots\psi \in \Re$ and implement the 
constraints~(\ref{eq:constraints}). Its anti-commutativity with $K_z$ reduces it to the form
\begin{equation}
\eta = \left(
\begin{array}{cccc}
a+i \alpha & 0 &  0  & d + i\delta\\
0 & f +  i\phi & g + i\gamma &  0\\
0 & k + i \kappa & m + i \mu & 0\\
-d - i \delta & 0 & 0 &  -a - i\alpha
\end{array}
\right)
\end{equation}
while anti-commutativity with $K_y$ restricts it further to
\begin{equation}
\eta = \left(
\begin{array}{cccc}
a+i \alpha & 0 &  0  & 0\\
0 & f +  i\phi & 0 &  0\\
0 & 0 & -a - i \alpha& 0\\
0 & 0 & 0 &  -a - i\alpha
\end{array}
\right)
\end{equation}
Its final form is reached by implementing its anti-commutativity with $K_x$, and reads
\begin{equation}
\eta =   \delta e^{i\xi}\left(
\begin{array}{cccc}
1 & 0 &  0  & 0\\
0 & -1 & 0 &  0\\
0 & 0 & -1& 0\\
0 & 0 & 0 &  -1
\end{array}
\right)
\end{equation}
where we have  defined $a + i \alpha = \delta e^{i\xi}$, with $\delta,\xi\in\Re$. Since $\delta$ only sets the scale of the norm we may set it to unity, to the effect that 
\begin{equation}
\eta =   e^{i\xi}\left(
\begin{array}{cccc}
1 & 0 &  0  & 0\\
0 & -1 & 0 &  0\\
0 & 0 & -1& 0\\
0 & 0 & 0 &  -1
\end{array}
\right)\label{eq:SpacetimeMetricWithPhase}
\end{equation}
The commutativity of $\eta$ with the generators of rotations places no further restrictions on the metric. 

This exercise does two things for us. It algebraically constructs the metric for $\mathcal{R}\otimes\mathcal{L}\vert_{s=1/2}$ representation space, and unearths a phase that the Lorentz invariant norms allow.  Second, it provides a unified way of looking at spinors,
spacetime, and four vectors.

If we require a reality of the norm for  $\chi$, with all its four elements $\in\Re$, the phase angle $\xi$ can be restricted as follows
\begin{equation}
\xi= 
\begin{cases}
0, & \textrm{to yield $\eta$ in the West coast form}\\
\pi, & \textrm{to give the East coast version of $\eta$}
\end{cases}
\end{equation}
This is useful for spacetime vectors.
On the other hand, 
in a quantum context where $\chi$ may be a vector field, a non-vanishing $\xi$, $0 \le \xi  \le \pi$,
opens a discussion which was not possible hitherto: that is, before the discovery of the phase $e^{i\xi}$ in (\ref{eq:SpacetimeMetricWithPhase}).

\section{The $\mathbf {\left[\mathcal{R}\otimes\mathcal{L}\right]_{s=1/2}}$ representation space}


A field transforming according to $\left[\mathcal{R}\otimes\mathcal{L}\right]_{s=1/2}$ representations can be bifurcated into two Casimir sectors:  one, with eigenvalue $2=1(1+1)$, and the other with eigenvalue $0=0(0+1)$. The eigenvalue $2$ sector is three fold degenerate, which \textit{in the rest frame} can be distinguished with eigenvalues $+1,0,-1$  of spin one $J_z$. The Casimir sector with eigenvalue $0$ is non-degenerate.

 A quantum field defined with the three degrees of freedom associated with the Casimir sector of eigenvalue $2$ is known to violate unitarity at high energies.  The unitarity is restored to construct a renormalisable theory in the form of the standard model of high energy physics at the cost of introducing Higgs through spontaneous symmetry breaking.
 Historically, it served as a departure to develop gauge theories.


It is and was conventional to project out the Casimir sector with the vanishing eigenvalue. But, can one project out degrees of freedom from a representation space without violating something deep about the symmetries that gives birth to the very representation space on which a quantum field is built upon?

To cosnider this question we backtrack to our discussion
in
Section 2.2 of~\citep{Ahluwalia:2004ab} that a similar projecting out from the $\left[\mathcal{R}\oplus\mathcal{L}\right]_{s=1/2}$ representation space  would have excluded antiparticles from the Dirac's theory. 
Mass dimension one fermions come to exist because we do not ignore the two degrees of freedom associated with the anti-self conjugate spinors. Incorporating them in the theory brings about dark matter, or at least a first principle candidate for dark matter with quartic self interaction, and an unusual property associated with rotation of dark matter clouds based on Elko.

With that background we suspect that something similar may be happening in the conventional treatment of the $\left[\mathcal{R}\otimes\mathcal{L}\right]_{s=1/2}$ representation space. Inadvertently, we may be projecting out something that could morph into a Higgs.

One can now define a field that contains expansion coefficients corresponding to the three degrees of freedom associated with the Casimir invariant $2$ sector, and one from the Casimir invariant $0$ sector. The dual of the expansion coefficients is so defined that 
the two Casimir sectors have their own phase angle $\xi$. This modifies the adjoint of the field. As a result the  
the phase angles end up showing in the vacuum expectation value of the calculated Feynman-Dyson propagator and can be so adjusted as to obtain a well-behaved propagator that does not lead to unitarity violation at high energies, or to the negative norm states. 
The latter aspect is related to the fact that the creation and destruction operators satisfy the bosonic commutators.

The result  is that the resulting field contains, what in the rest frame, can be called spin one and spin zero. Such a scenario can serve as an \textit{ab initio} starting point to better understand  origin of the 
Higgs mechanism and spontaneous symmetry breaking.\footnote{The just outlined calculation  for the new field has been a subject of various discussions with Sebastian Horvath,\index{Horvath, Sebastian} Cheng-Yang Lee, \index{Lee, Cheng-Yang} and Dimitri Schritt 
 \index{Schritt, Dimitri} as is the extension to  higher spins. It must  be considered preliminary.} Extended to `spin $2$', the argument suggests `spin $1$' and `spin $0$' Higgs like bosons.

\section{Maxwell equations and beyond}\index{Maxwell equation}

When one extends the work of Chapter~\ref{ch5} to the $\mathcal{R}\oplus\mathcal{L}$ representation space of spin one and two, and takes the massless limit, one reaches a deeper understanding of Maxwell equations and Einstein's gravity in the weak field limit. A helpful reference is \citep{Weinberg:1964ev}, but care must be taken to check  the invertibility of $\mathfrak{J}\cdot \nabla$ operator~\citep{Ahluwalia:1993pj}.


\chapter*{Appendix: Further reading}	

			\label{ch17}
Recently Saulo Pereira\index{Pereira, Saulo} and R. C. Lima\index{Lima, R. C.} have shown that 
an asymptotically expanding universe creates
low-mass mass dimension one fermions much more copiously than Dirac fermions (of the same mass)~\citep{Pereira:2016eez}. If their preliminary results remain essentially unaffected by the new results presented here it would significantly help us to develop a first-principle cosmology based on Elko and mass dimension one fermions. 

Various Brazilian-Italian group of physicists  have examined such 
important topics as Hawking radiation of mass dimension one \index{Hawking radiation} fermions~\citep{daRocha:2014dla}, and continue to develop mathematical physics underlying  
Elko~\citep{daRocha:2007sd,daRocha:2007pz,daRocha:2008we,HoffdaSilva:2009is,daRocha:2011xb,daRocha:2011yr,Bernardini:2012sc,daSilva:2012wp,daRocha:2013qhu,Cavalcanti:2014uta,daRocha:2014dla,Bonora:2015ppa,daRocha:2016bil,Rogerio:2016grn,daSilva:2016htz,HoffdaSilva:2017vic,Neto:2017vgx}.
Of these we draw particular attention to mass-dimension transmuting operators considered in
~\citep{daRocha:2007pz,HoffdaSilva:2009is}. It would  help define a new symmetry between the Dirac field and the field associated with mass dimension one fermions if mass-dimension transmuting operators could be placed on a rigorous footing after incorporating locality and Lorentz covariance.\footnote{The need for a mass-dimension transmuting symmetry was first suggested by the present author to Rold\~ao da Rocha several years ago and is briefly mentioned in section~\ref{Sec:conjecture}.}

Max Chaves and Doug Singleton have suggested that mass dimension one fermions of spin one half  may have a possible connection with mass-dimension-one vector particles with fermionic statistics~\citep{Chaves:2008gd}. It may be worth examining if a new fundamental symmetry may be constructed that relates the works of~\citep{daRocha:2007pz,HoffdaSilva:2009is} with those of Chaves and Singleton.

Localisation of Elko in the brane has been considered in references~\citep{Liu:2011nb,Jardim:2014xla,Zhou:2017bbj}. Elko in the presence of torsion has been a subject of several insightful papers by Luca Fabbri. \index{Fabbri} We refer the reader to these and related publications~\citep{Fabbri:2009ka,Fabbri:2010qv,Fabbri:2010ws,Fabbri:2010va,Fabbri:2011mi,Fabbri:2012yg,Fabbri:2014foa}. Cosmological solutions of 5D Einstein equations with Elko condensates were obtained by Tae Hoon Lee where it was found that there exist exponentially expanding cosmological solution even in the absence of a cosmological constant~\citep{Lee:2012zze}.

\vspace{11pt}

We parenthetically note that all the works discussed so far remain essentially unchanged with the new developments reported here. In view of the new results on locality and Lorentz covariance it is important to revisit the analysis and claims of~\citep{Basak:2011wp} and also those calculations that use full apparatus of the theory of quantum fields, and not merely Elko.
In the same thread, given the interest in mass dimension one fermions a  number of $S$-matrix  calculations were done and published~\citep{Dias:2010aa,Lee:2015sqj,Alves:2014qua,Alves:2014kta,Agarwal:2014oaa,Lee:2015jpa,Alves:2017joy}. These need to be revisited also.

  \label{refs}

  \cleardoublepage

\printindex

\end{document}